\documentclass[%
 reprint, amsfonts,amssymb,amsmath,longbibliography,
superscriptaddress,aps,prb]{revtex4-2}

\usepackage{graphicx}
\usepackage{dcolumn}
\usepackage{bm}
\usepackage[colorlinks=true,citecolor=blue,linkcolor=blue,urlcolor=blue]{hyperref}
\usepackage{color}
\usepackage{xcolor}
\usepackage[normalem]{ulem} 

\begin{document}

\title{Lattice dynamics of the infinite-layer nickelate LaNiO$_2$}

\author{Shohei~Hayashida}
\email[]{s\_hayashida@cross.or.jp}
\affiliation{Neutron Science and Technology Center, Comprehensive Research Organization for Science and Society (CROSS), Tokai, Ibaraki 319-1106, Japan}
\affiliation{Max-Planck-Institute for Solid State Research, Heisenbergstra{\ss}e 1, 70569 Stuttgart, Germany}
\author{Vignesh~Sundaramurthy}
\affiliation{Max-Planck-Institute for Solid State Research, Heisenbergstra{\ss}e 1, 70569 Stuttgart, Germany}
\author{Wenfeng~Wu}
\affiliation{Key Laboratory of Materials Physics, Institute of Solid State Physics, HFIPS, Chinese Academy of Sciences, Hefei 230031, China}
\affiliation{Science Island Branch of Graduate School, University of Science and Technology of China, Hefei 230026, China}
\author{Pascal~Puphal}
\affiliation{Max-Planck-Institute for Solid State Research, Heisenbergstra{\ss}e 1, 70569 Stuttgart, Germany}
\author{Thomas~Keller}
\affiliation{Max-Planck-Institute for Solid State Research, Heisenbergstra{\ss}e 1, 70569 Stuttgart, Germany}
\affiliation{Max Planck Society Outstation at the Heinz Maier-Leibnitz Zentrum (MLZ), Lichtenbergstra{\ss}e 1, 85748 Garching, Germany}
\author{Björn~F{\aa}k}
\affiliation{Institut Laue-Langevin, 71 Avenue des Martyrs, 38042 Grenoble Cedex 9, France}
\author{Masahiko~Isobe}
\affiliation{Max-Planck-Institute for Solid State Research, Heisenbergstra{\ss}e 1, 70569 Stuttgart, Germany}
\author{Bernhard~Keimer}
\affiliation{Max-Planck-Institute for Solid State Research, Heisenbergstra{\ss}e 1, 70569 Stuttgart, Germany}
\author{Karsten~Held}
\affiliation{Institut für Festkörperphysik, Technische Universität Wien, 1040 Wien, Austria}
\author{Liang~Si}
\email[]{siliang@nwu.edu.cn}
\affiliation{School of Physics, Northwest University, Xi’an 710127, China}
\affiliation{Institut für Festkörperphysik, Technische Universität Wien, 1040 Wien, Austria}
\author{Matthias~Hepting}
\email[]{hepting@fkf.mpg.de}
\affiliation{Max-Planck-Institute for Solid State Research, Heisenbergstra{\ss}e 1, 70569 Stuttgart, Germany}
\affiliation{Max Planck Society Outstation at the Heinz Maier-Leibnitz Zentrum (MLZ), Lichtenbergstra{\ss}e 1, 85748 Garching, Germany}

\date{\today}

\begin{abstract}
Infinite-layer (IL) nickelates have rapidly emerged as a new class of superconductors. However, due to the technical challenges of their topotactic synthesis, they have so far been realized primarily as thin films or polycrystalline powder samples, limiting comprehensive investigations of fundamental physical properties such as the lattice dynamics. Here, we present a time-of-flight inelastic neutron scattering study on a sample composed of a large number of co-aligned bulk crystals of the IL nickelate LaNiO$_2$. We observe several dispersive phonon branches, which are in good agreement with lattice dynamical calculations based on density-functional perturbation theory. In addition, we compare the characteristics of selected LaNiO$_2$ phonon modes to those of isostructural cuprate superconductors. Our findings provide a reference point for future experimental and theoretical efforts aimed at understanding the interplay between lattice dynamics and electronic properties in IL nickelates.
\end{abstract}

\maketitle

\section{Introduction}
\label{sec:introduction}

The recent discovery of superconductivity in infinite-layer (IL) nickelates \cite{LiNat2019} has generated significant interest due to their structural and electronic similarities to high-temperature superconducting cuprates. In the latter, superconductivity emerges upon doping charge carriers into the CuO$_2$ planes, thereby suppressing the long-range antiferromagnetic order of the parent Mott-insulating state \cite{Keimer2015,Scalapino2012,Armitage2010,Lee2006}. Nevertheless, despite decades of intensive research, consensus on the microscopic mechanism underlying unconventional superconductivity in cuprates remains elusive. Therefore, the discovery and exploration of  related materials, such as IL nickelates, which might share key superconducting characteristics, offer promising opportunities to address this long-standing problem.

Striking similarities between IL nickelates of composition $R$NiO$_2$ ($R$: rare-earth ion) and cuprates were pointed out early in theoretical studies \cite{Anisimov1999}, even preceding the experimental confirmation of superconductivity in IL nickelate thin films \cite{LiNat2019,ZengPRL2020,LiPRL2020,OsadaPRM2020,GaoCPL2021,ZengSciAdv2022,Sahib2025}. In particular, both material classes exhibit NiO$_2$ (CuO$_2$) planes, wherein Ni$^{1+}$ (Cu$^{2+}$) ions nominally adopt a 3$d^9$ electronic configuration. However, later theoretical studies predicted distinct features of the electronic structure of IL nickelates \cite{Lee2004}, including a diminished 3$d$-2$p$ hybridization between Ni and the O ligands, while at the same time rare-earth 5$d$ Fermi surface pockets self-dope the Ni 3$d$ bands \cite{Nomura2019,Botana2020,Kitatani2020,Wu2020,Zhang2020E,Sakakibara2020,Lechermann2020,Wang2020E,Werner2020,Karp2020,Liu2021,Wan2021,Lang2021,Higashi2021,Held2022,Chen2022r}, which explains the metallic properties of the parent $R$NiO$_2$ compounds. Signatures of this distinct electronic structure of IL nickelates have been evidenced in recent photon and electron-spectroscopic measurements \cite{Hepting2020,Goodge2021,Ding2024,Sun2025}.

In cuprates, the electronic structure and lattice dynamics are strongly coupled, leading to pronounced anomalies in specific phonon modes \cite{Pintschovius1991}. Notably, the Cu–O bond-stretching (breathing) modes exhibit anomalous softening and broadening at characteristic wavevectors, indicative of strong electron-phonon interactions \cite{Fong1995,Reznik2008,Raichle2011,Ahmadova2020,Sterling2021,Reichardt1994,Pintschovius2005}. Some of these anomalies, revealed by phonon dispersion measurements using inelastic neutron and x-ray scattering across various cuprate families, have been discussed in the context of the microscopic mechanisms driving charge density wave formation \cite{Frano2020}. Phonons thus serve as a sensitive probe of the underlying charge correlations. Consensus holds that spin fluctuations \cite{LeTaconNatPhys2011,DeanNatMat2013} play a central role in the superconducting pairing mechanism in cuprates \cite{Lee2006,Scalapino2012,dai01,vilardi19}, whereas electron-phonon interactions are insufficient to account for the high superconducting transition temperatures, although phonons may still contribute to or interact with the pairing process \cite{cuk04,heid09,giustino08,reznik08,johnston10,Bauer2009}.


In superconducting IL nickelates, the pairing mechanism and symmetry remain under investigation. Initial spectroscopic studies on samples with potentially degraded surfaces yielded conflicting results regarding nodal versus fully gapped behavior \cite{Harvey2022,Chow2022Pairing,Gu2020,Wang2023STM,Choubey2021,Kreisel2022}, whereas recent ultrafast optical spectroscopy experiments suggest a weak-coupling $d$-wave pairing state \cite{Cheng2024}. Given their moderate superconducting transition temperature relative to cuprates \cite{Schilling1993,Lee2023}, the role of phonons and possible lattice instabilities remains an active field of inquiry \cite{Xia2022,Carrasco2022,Zhang2023Phonon,Meier2024,Sakakibara2024}. Theories diverge on this point, suggesting either strong electron-phonon coupling (EPC) \cite{Zhang2024EPC,Li2024Phonon,Alvarez2024} and phonon-driven $s$-wave gap scenarios \cite{Sui2023,Li2024Phonon,Craco2025}, or that EPC in IL nickelates is insufficient to explain superconductivity \cite{Nomura2019,DiCataldo2023}. 
In lieu thereof, spin fluctuations have been suggested as the dominant pairing mechanism \cite{Wu2020,Kitatani2023}, similarly to the cases of cuprates and iron-based superconductors \cite{Scalapino2012}.

\begin{figure}[tb]
\includegraphics[scale=1]{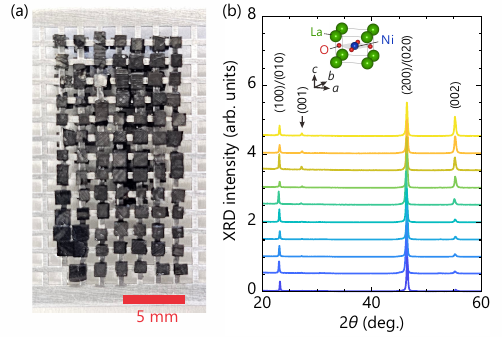}
\caption{(a) Sample array with co-aligned LaNiO$_{2}$ crystals on two sides of an Al grid. (b) Representative XRD patterns from the surface of individual LaNiO$_{2}$ crystals, acquired with Cu $K_\alpha$ radiation at 300 K. The Bragg peaks are indexed. The inset shows the tetragonal $P4/mmm$ unit cell of LaNiO$_{2}$.}
\label{fig:crystal}
\end{figure}

However, in spite of numerous theoretical studies on the lattice dynamics of IL nickelates and continued debate over the role of phonons \cite{Xia2022,Carrasco2022,Zhang2023Phonon,Meier2024,Sakakibara2024, Zhang2024EPC,Sui2023,Li2024Phonon,Adhikary2020,DiCataldo2023,Si2023,Sharma2024,Wang2025,Zhang2025EPC}, a basic experimental characterization of their phonon spectrum is lacking. This is likely due to the demanding nature of the topotactic synthesis process, which previously limited the availability of IL nickelate samples to thin films or polycrystalline powders, preventing the acquisition of highly resolved energy- and/or momentum-dependent spectroscopic data from bulk single-crystals, for instance by inelastic neutron scattering (INS). Additionally, optical spectroscopy has not been able to clearly distinguish phonons in IL nickelate films from substrate contributions \cite{Cervasio2023}, and no Raman-active phonon modes exist at the zone center for the $P4/mmm$ structure of $R$NiO$_2$ \cite{Burns1989,Puphal2021}.

In this work, we perform INS experiments on LaNiO$_2$ crystals, observing various branches of acoustic and optical phonons. We compare the measured phonon spectrum to lattice dynamical calculations obtained via density-functional perturbation theory (DFPT)~\cite{RevModPhys.73.515} and discuss the computed dispersions in the context of characteristic phonons in cuprates, including bond stretching, buckling, and bending modes as well as charge fluctuation-coupled modes. The results establish a basis for future studies exploring the interplay between lattice dynamics, EPC, and the electronic properties of IL nickelates.



\section{Materials and methods}
\label{sec:methods}

Cube-shaped LaNiO$_{2}$ crystals with dimensions of approximately 1~mm$^3$ were obtained through topotactic reduction of optical floating zone grown LaNiO$_3$ crystals \cite{Puphal2023,Puphal2023a}, using CaH$_2$ as the reducing agent. The details of the synthesis procedure are described in Ref.~\cite{Hayashida2024}. More than 100 LaNiO$_{2}$ crystals with a total mass of 870~mg were co-aligned on an Al grid for the INS experiment [Fig.~\ref{fig:crystal}(a)]. X-ray diffraction (XRD) from the surface of individual LaNiO$_{2}$ crystals [Fig.~\ref{fig:crystal}(b)] revealed the presence of $(H, 0, 0)$/$(0, K, 0)$ and $(0, 0, L)$-type Bragg peaks of the tetragonal $P4/mmm$ space group, suggesting that the crystals contain three twin domains, as described in detail in Ref.~\cite{Hayashida2024}. Different intensity ratios between the Bragg peaks in the XRD patterns of different crystals [Fig.~\ref{fig:crystal}(b)] indicate that each crystal might exhibit a distinct domain population across the probed surface region. In the INS experiment which picks up the bulk signal from the entire sample array consisting of more than 100 crystals, the varying domain populations are averaged out (see Supplementary Information~\cite{SM}) and we assume an equal population of the three domains for the INS data analyses.

\begin{figure}[tb]
\includegraphics[scale=1]{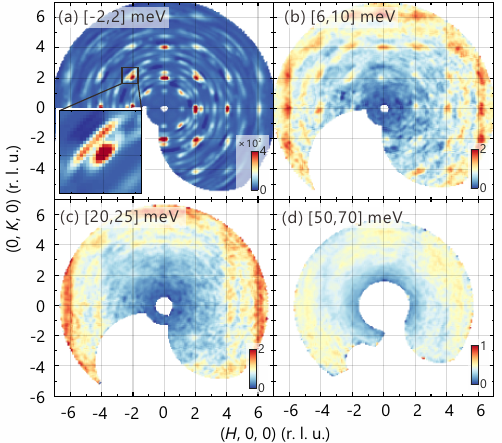}
\caption{(a)--(d) Constant energy slices of the INS spectra in the $(H,K,0)$ plane for incident neutron energy $E_{\rm i}=76$~meV. The transferred energies are integrated over the following ranges: (a) $-2\leq E \leq 2$~meV , (b) $6\leq E \leq 10$~meV, (c) $20\leq E \leq 25$~meV, and (d) $50\leq E \leq 70$~meV. The integration range along the out-of-plane direction is $\pm0.37$~{\AA}$^{-1}$. The inset in panel (a) highlights two satellite peaks from $P4/mmm$ twin domains around the $(-2,2,0)$ Bragg peak.}
\label{fig:constantE_Ei76meV}
\end{figure}

\begin{figure*}[tb]
\includegraphics[scale=1]{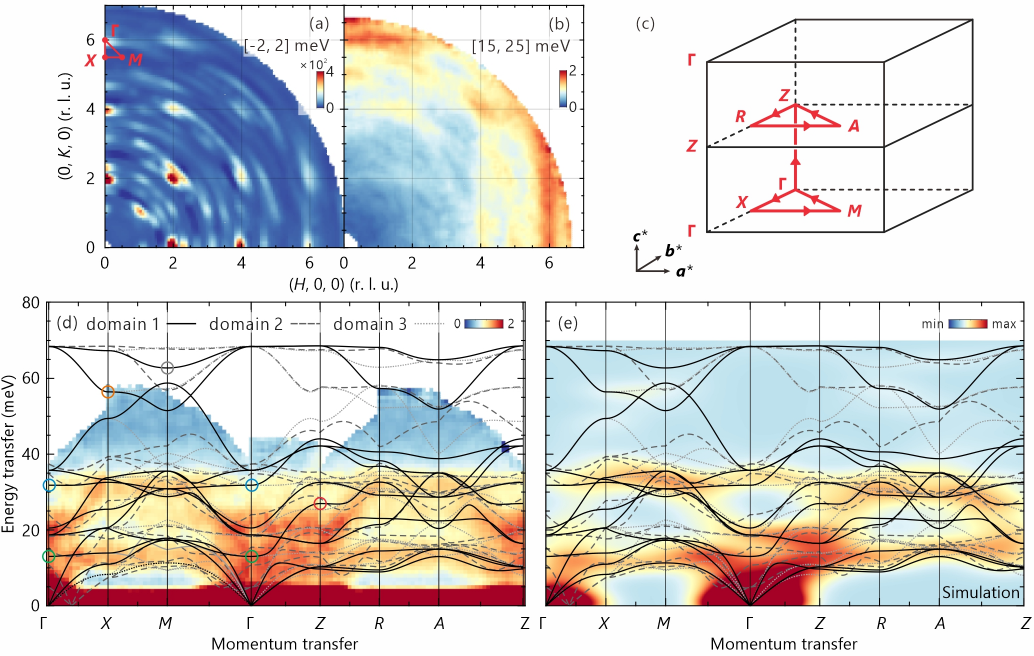}
\caption{(a),(b) Constant energy slices of folded INS spectra in the $(H,K,0)$ plane for incident energy $E_{\rm i}=76$~meV. The transferred energies are integrated over the following
ranges: (a) $-2\leq E \leq 2$~meV and (b) $15\leq E \leq 25$~meV. A path along high-symmetry points ($\Gamma$-$X$-$M$-$\Gamma$) is illustrated in red color around the (060) Bragg reflection in panel (a). (c) Schematic of a three-dimensional path moving along the high-symmetry points $\Gamma$-$X$-$M$-$\Gamma$-$Z$-$R$-$A$-$Z$ according to the tetragonal \textit{P4/mmm} unit cell of LaNiO$_2$. (d) INS map along a high-symmetry path from folded data around the (060) and (600) Bragg peaks, acquired with $E_{\rm i}=76$~meV. The integration range along the two orthogonal directions relative to the path is $\pm0.2$~{\AA}$^{-1}$. The DFPT computed phonon dispersions for the three twin domains of LaNiO$_2$ are superimposed as solid (domain 1), dashed (domain 2), and dotted lines (domain 3). The colored circles on the dispersion curves of domain 1 indicate specific phonon modes discussed in the text. (e) Phonon intensities calculated with the Euphonic software package, using the DFPT phonons as input. The intensities are averaged over the three domains, while the indexing of the high-symmetry path refers to domain 1. In addition, Gaussian broadening is applied along both the momentum ($\Delta Q=0.92$~{\AA}$^{-1}$) and energy ($\Delta E=4.9$~meV) transfer directions. Details about the domain averaging and broadening are given in the Supplementary Information~\cite{SM}.}
\label{fig:INS_spaghetti_Ei76meV}
\end{figure*}

The INS measurements were conducted at the thermal neutron time-of-flight spectrometer PANTHER (ILL, France)~\cite{Fak2022} 
at 1.5~K. The energy of the incident neutron beam was $E_{\rm i}=76$~meV, yielding a Gaussian energy resolution of $\Delta E=4.9$~meV at the elastic line. The scattering plane was chosen such that the crystallographic $ab$ plane was horizontal, although we note that this reference frame applies only to one out of the three orthogonal $P4/mmm$ twin domains in the sample. 
An Al grid sample holder without nickelate crystals was measured under the same conditions and the corresponding background signal was subtracted from all INS spectra shown in the following. 
Data reduction and background subtraction were performed using MANTID~\cite{Mantid}, and all data were analyzed using HORACE software~\cite{Horace}.

To determine the ground-state crystal structures of LaNiO$_{2}$, density-functional theory~\cite{PhysRev.136.B864,PhysRev.140.A1133} level structural relaxations were performed using the Vienna \textit{ab-initio} Simulation Package (\textsc{VASP}) \cite{kresse1996efficiency,PhysRevB.54.11169,PhysRevB.59.1758}. The Perdew-Burke-Ernzerhof version of the generalized gradient approximation~\cite{PhysRevLett.77.3865} was employed, sampling the  Brillouin zone (BZ) with a dense 13$\times$13$\times$15 Monkhorst–Pack $k$ mesh for a unit cell of LaNiO$_2$ and a 6$\times$6$\times$8 $k$ mesh for the 2$\times$2$\times$2 supercell of LaNiO$_{2}$. Energy and force convergence criteria were set to 10$^{-8}$\,eV and 0.01\,eV/\AA, respectively. The experimentally determined lattice parameters of LaNiO$_{2}$ were fixed during structural relaxations while atomic positions were fully relaxed. Phonon spectra for LaNiO$_{2}$ compounds were calculated using DFPT \cite{RevModPhys.73.515} and post-processed with the Phonopy code \cite{togo2015first}. The Euphonic software package~\cite{EuphonicWeb,EuphonicFair} was employed to simulate the INS phonon intensities, using the results from the DFPT calculation as an input.

\section{Results}
\label{sec:results}

Figures~\ref{fig:constantE_Ei76meV}(a)-\ref{fig:constantE_Ei76meV}(d) display  neutron scattering intensity maps in the $(H,K,0)$ plane acquired with $E_{\rm i}=76$~meV at the PANTHER spectrometer. The $H, K, L$ indexing refers to the LaNiO$_{2}$ twin domain with the crystallographic $ab$ plane lying in the scattering plane, denoted as \textit{domain 1} in the following. In the quasi-elastic scattering map in Fig.~\ref{fig:constantE_Ei76meV}(a) (energy integration range: $-2\leq E \leq 2$~meV), Bragg peaks emerge at integer $H$ and $K$ positions, exhibiting a rather smeared out intensity distribution. The peaks are particularly intense for even $H$ and $K$ values, which is consistent with the structure factor of LaNiO$_{2}$ in the  $P4/mmm$ space group. In addition, satellite peaks occur in proximity to the Bragg peaks at slightly deviating $H$ and $K$ values [see inset in Fig.~\ref{fig:constantE_Ei76meV}(a)]. These peaks originate from the other two twin domains of the tetragonal $P4/mmm$ space group of LaNiO$_{2}$, denoted as \textit{domain 2} and \textit{domain 3} in the following. In a previous XRD study on individual LaNiO$_{2}$ crystals, similarly broad Bragg peak intensity profiles and satellite peaks were observed in XRD maps of the $(H,K,0)$ plane~\cite{Puphal2023}, suggesting that the total mosaicity of the sample array  used for the present INS experiment is comparable to that of individual crystals. 

\begin{figure*}[tb]
\includegraphics[scale=1]{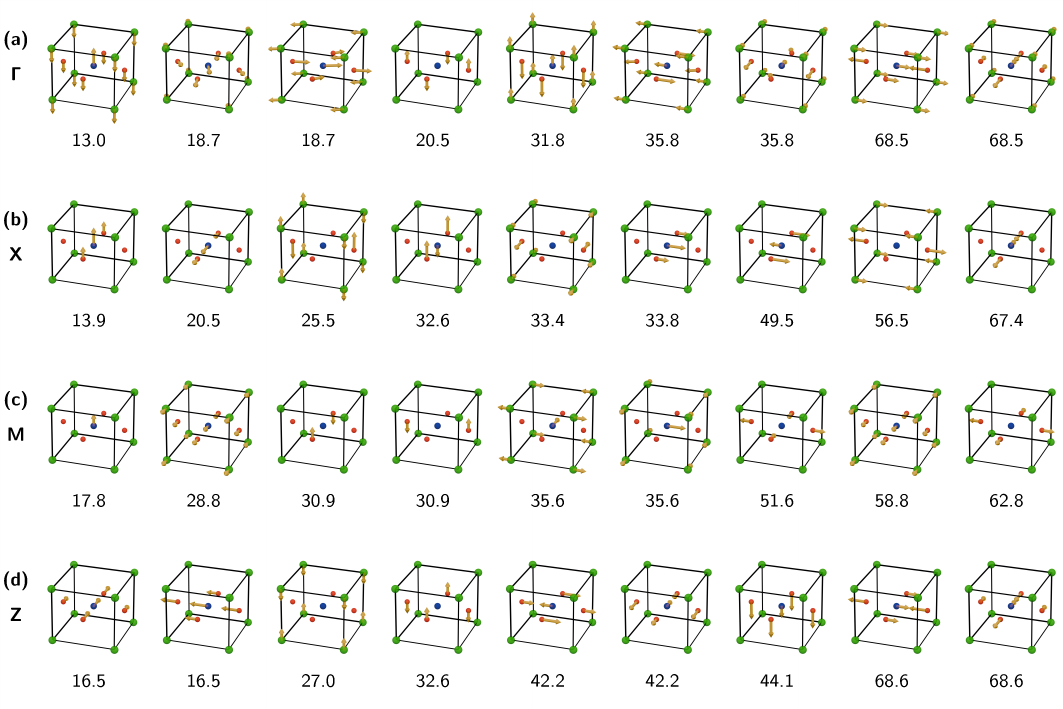}
\caption{(a) Atomic displacement patterns of selected phonons of LaNiO$_2$ at the $\Gamma$ point together with the computed phonon energies in units of meV. La atoms are shown in green, Ni in blue, and oxygen in red. The yellow arrows indicate the directions and amplitudes of the atomic vibrations. For clarity, the La, Ni, and O atoms are displayed with reduced atomic radii. (b)-(d) Displacement patterns of phonons at the $X$, $M$, and $Z$ points, respectively. 
}
\label{fig:phonons}
\end{figure*}

In the inelastic scattering map in Fig.~\ref{fig:constantE_Ei76meV}(b) with energy range $6\leq E \leq 10$~meV, pronounced accumulations of spectral weight occur around even $H$ and $K$ values, indicating the presence of characteristic low-energy excitations. The intensity of the spectral weight accumulations increases for larger $\mathbf{Q}$, which is opposite to the trend in the quasi-elastic scattering in Fig.~\ref{fig:constantE_Ei76meV}(a). This intensity increase for larger $\mathbf{Q}$ suggests that the excitations are likely phonons, due to the normal phonon polarization factor. For energy transfers above 20 meV [Figs.~\ref{fig:constantE_Ei76meV}(c) and \ref{fig:constantE_Ei76meV}(d)], the spectral weight accumulations become less localized and their intensity fades out. 


To further investigate the excitations revealed in Figs.~\ref{fig:constantE_Ei76meV}(b)-\ref{fig:constantE_Ei76meV}(d), we next inspect an INS intensity map along a high-symmetry path in the BZ, generated from folded data around the (060) and (600) Bragg peaks. Figures~\ref{fig:INS_spaghetti_Ei76meV}(a) and \ref{fig:INS_spaghetti_Ei76meV}(b) display the folded quasi-elastic and inelastic scattering maps in the $(H,K,0)$ plane, respectively, and Fig.~\ref{fig:INS_spaghetti_Ei76meV}(c) shows a schematic of the high-symmetry path within the $P4/mmm$ unit cell. In the corresponding INS map [Fig.~\ref{fig:INS_spaghetti_Ei76meV}(d)], we find that inelastic, dispersive spectral weight emanates from the $\Gamma$ point, while flat branches extend across wide paths in the BZ around 15 and 30 meV, connecting the $X$ and $M$ point as well as the $R$ and $A$ point. Additionally, a broad distribution of spectral weight occurs below 30 meV in the region between $\Gamma$ and $Z$. 

To corroborate that the observed spectral weight in Fig.~\ref{fig:INS_spaghetti_Ei76meV}(d) originates from phonons, we  compute the phonon dispersion in LaNiO$_{2}$ using DFPT. As an input for the DFPT calculation, we use the refined lattice parameters of the $P4/mmm$ unit cell from the INS experiment (see Supplementary Information~\cite{SM}). Furthermore, since the investigated sample contains three twin domains, we project the computed phonon dispersions of domains 2 and 3 onto the $\Gamma$–$X$–$M$–$\Gamma$–$Z$–$R$–$A$–$Z$ path of domain 1. Notably, the superimposed DFPT dispersions of acoustic and optical phonon branches closely match the INS spectral weight distribution in Fig.~\ref{fig:INS_spaghetti_Ei76meV}(d), including the steeply dispersing bands emanating from $\Gamma$ and the relatively flat bands between $X$ and $M$ as well as $R$ and $A$. However, while the experimentally observed spectral weight is sharply confined below $\sim$38 meV, the DFPT calculations predict several phonon branches extending up to $\sim$68 meV.

For a closer comparison between theory and experiment, we simulate the INS phonon intensity of LaNiO$_2$ using the Euphonic software package~\cite{EuphonicWeb,EuphonicFair}, with DFPT-derived eigenenergies and eigenvectors (see Fig.~\ref{fig:phonons} below) of the phonons as input. Figure~\ref{fig:INS_spaghetti_Ei76meV}(e) displays the resulting simulated intensity map, averaged over all three domains. In addition, adequate Gaussian broadenings were applied in both momentum and energy transfer directions in the map in Fig.~\ref{fig:INS_spaghetti_Ei76meV}(e) (for details see Supplementary Information~\cite{SM}). Notably, the simulation not only reproduces the absence of significant phonon intensity above $\sim$38 meV, but also qualitatively captures key features from Fig.~\ref{fig:INS_spaghetti_Ei76meV}(d), such as the spectral weight accumulation around the $\Gamma$–$Z$ region and the intense flat bands between $X$ and $M$ as well as $R$ and $A$.


In general, the diminishing INS intensity observed in LaNiO$_2$ above $\sim$38 meV can be attributed to several factors, including that the one-phonon scattering cross-section decreases with increasing phonon energy, scaling approximately as $1/\omega$, which naturally reduces intensity at higher energies. In addition, the high-energy phonons in LaNiO$_2$ predominantly involve oxygen bending and stretching modes with out-of-phase displacements between some of the neighboring oxygen atoms, reducing the net scattering amplitude.




\section{Discussion}
\label{sec:discussion}

The close agreement between the measured phonon signal and the DFPT results for modes below $\sim$38 meV suggests that this theoretical approach quite reliably captures the essential lattice dynamics of LaNiO$_2$. Hence, we next discuss selected DFPT calculated phonon modes in detail, also drawing comparisons to related modes in cuprates, which were intensively investigated by INS \cite{Fong1995,Reznik2008,Raichle2011,Ahmadova2020,Sterling2021,Reichardt1994,Pintschovius2005}.

A comprehensive overview of the phonon modes in LaNiO$_2$ along with the DFPT computed energies is presented in Fig.~\ref{fig:phonons}. At the $\Gamma$ point, the lowest-energy optical phonon exhibits a computed energy of 13.0 meV [green circle in Fig.~\ref{fig:INS_spaghetti_Ei76meV}(d)], involving an anti-phase motion between La and Ni ions along the $c$-axis direction [Fig.~\ref{fig:phonons}(a)]. The dispersion of the mode between $\Gamma$ and $X$ is nearly flat, which is reminiscent of the analogous mode in the isostructural IL cuprate SrCuO$_{2}$ \cite{klenner94}. A second  notable $\Gamma$-point mode  [blue circle in Fig.~\ref{fig:INS_spaghetti_Ei76meV}(d)] at 31.8 meV involves in-phase movements of La and Ni against the O atoms [Fig.~\ref{fig:phonons}(a)]. This mode also displays minimal dispersion towards $X$. In cuprates, {\it ab initio} rigid-ion model calculations suggest that such modes involving axial displacements of the O sublattice against the cation sublattices can split into longitudinal optical and transverse optical branches \cite{klenner94} and are associated with ferroelectric instabilities. However, this splitting is suppressed under metallic screening conditions \cite{klenner94}, consistent with the metallic ground state and absence of ferroelectric behavior in LaNiO$_2$ \cite{Hayashida2024}.

At the $X$ point, the 56.5-meV phonon [orange circle in Fig.~\ref{fig:INS_spaghetti_Ei76meV}(d)] corresponds to the oxygen half-breathing mode (HBM), characterized by the in-plane motion of the O ions along the Ni-O bond direction such that alternating oxygen atoms are displaced towards and away from a central Ni [Fig.~\ref{fig:phonons}(b)]. The analogous HBM in the IL cuprate SrCuO$_{2}$ was theoretically described in Ref.~\cite{klenner94} and experimentally observed by INS in related cuprates at comparable energies \cite{Reznik2008}, for instance, at 58.5 meV in HgBa$_{2}$CuO$_{4+\delta}$ \cite{Ahmadova2020}. Interestingly, this longitudinal zone boundary mode shows strong renormalization effects in doped cuprates below the superconducting transition, underlining the relevance of the HBM also for IL nickelates and warranting future experimental studies with techniques that can resolve the mode. 

At the $M$ point, the highest-energy phonon mode, located at 62.8~meV [gray circle in Fig.~\ref{fig:INS_spaghetti_Ei76meV}(d)], corresponds to the full-breathing mode (FBM), in which all oxygen ions move uniformly in-phase toward or away from the central Ni ion [Fig.~\ref{fig:phonons}(c)]. In cuprates, oxygen breathing distortions, such as the HBM and FBM, are known to be selectively sensitive to the on-site Coulomb repulsion $U_d$ of the localized Cu 3$d$ orbitals \cite{Falter1992,Falter1997,Falter2006}, with an increased $U_d$ suppressing charge fluctuations at the Cu site, resulting in a hardening of the breathing mode. In contrast, the smaller $U_d$ on Ni in IL nickelates generally leads to softer breathing modes compared to their cuprate counterparts. This trend was recently examined in LaNiO$_2$ by \textit{ab-initio} many-body methods \cite{Wang2025}, suggesting that the energy of oxygen breathing modes may serve as an indirect but sensitive probe of electronic correlation strength in this material class. 

Notably, the INS map in Fig.~\ref{fig:INS_spaghetti_Ei76meV}(d) shows no additional spectral weight around the $M$ point beyond that attributable to phonons. In particular, we find no indication of spin excitations at the $M$ point, also not in an analysis focused on small $\mathbf{Q}$ where the magnetic signal is expected to be strongest (see Supplementary Information~\cite{SM}). This absence is noteworthy given that the $M$ point corresponds to the antiferromagnetic zone center in square-lattice Heisenberg antiferromagnets, where a pronounced low-energy magnetic response in this region is well established for cuprates \cite{Birgeneau2006}. Moreover, resonant inelastic x-ray scattering experiments on IL nickelate thin films \cite{LuScience2021,TamNatMat2022,KriegerPRL2022} as well as on LaNiO$_2$ crystals from the same synthesis batch \cite{Hayashida2024} have revealed spin excitations consistent with paramagnons arising from $S=1/2$ square-lattice Heisenberg-type correlations. This can be explained by our estimate of the expected INS signal strength for spin excitations in IL nickelates (see Supplementary Information~\cite{SM}), which suggests that a significantly larger sample mass 
is required to resolve magnetic features in the present experiment.

At the $Z$ point, the 27.0-meV phonon [red circle in Fig.~\ref{fig:INS_spaghetti_Ei76meV}(d)] involves an axial breathing motion of the La ion against the NiO$_2$ planes [Fig.~\ref{fig:phonons}(d)]. In IL cuprates, the analogous mode has been reported to couple strongly to charge fluctuations \cite{klenner94}, typically manifesting as phonon linewidth broadening due to EPC, an effect that has been extensively studied in various cuprate compounds, offering important insights into the interplay between lattice dynamics and electronic correlations \cite{Zhang2007}. Dynamical interplane charge transfer due to the axial breathing mode is expected to be particularly enhanced in compounds with a short $c$-axis lattice parameter~\cite{klenner94}, 
such as the recently synthesized
Sm-based IL nickelates that might exhibit an enhanced $T_c$~\cite{Chow2025,Yang2025}.
In Fig.~\ref{fig:INS_spaghetti_Ei76meV}(a), in principle, the broad spectral weight distribution observed between $\Gamma$ and $Z$ could result from EPC-induced broadenings, consistent with the predicted presence of strong EPC in IL nickelates \cite{Zhang2024EPC,Sui2023,Li2024Phonon,Zhang2025EPC} and the coupling of certain modes with appropriate symmetry along the $c$-direction to charge fluctuations via EPC  \cite{klenner94}. However, the limited resolution and statistical quality of our current data preclude an unambiguous assessment of this effect in Fig.~\ref{fig:INS_spaghetti_Ei76meV}(d). Future high-resolution INS studies on LaNiO$_2$ crystals with higher crystalline quality and a possibly mono-domain character will be necessary to clarify the effect of EPC on certain modes and its general role in IL nickelates.

\section{Conclusion}
\label{sec:conclusion}


In summary, we have investigated the lattice dynamics of LaNiO$_2$ on a sample of co-aligned bulk crystals. Several branches of acoustic and optical phonons were observed below $\sim$38 meV, in good agreement with lattice dynamical calculations based on DFPT. Simulated INS intensity maps, incorporating DFPT mode-resolved eigenvectors and domain averaging, reproduced the main features of the experimental spectra and account for the lack of observable high-energy phonons that are dominated by oxygen displacements.

Based on our DFPT results, we have identified and discussed several characteristic phonon modes, including the HBM at the $X$ point and the FBM at the $M$ point, which are of particular relevance in the context of related cuprate superconductors. Furthermore, our computational results may also assist in the assignment of infrared-active phonons as well as phonon-induced features observed in other spectroscopic probes such as angle-resolved photoemission spectroscopy \cite{Ding2024,Sun2025}. While direct signatures of spin wave excitations and EPC-induced phonon broadening could not be resolved in the present data, such effects may become observable in future high-resolution INS studies on higher-quality and larger-mass LaNiO$_2$ crystals. Overall, our  results establish an experimental foundation for understanding the lattice dynamics of IL nickelates and lay the groundwork for future investigations addressing electron-phonon interactions, structural instabilities, and their possible relevance to superconductivity in this material class.

\begin{acknowledgments}
We thank C.~Falter and A.~Krajewska for insightful discussions. We acknowledge the Institut Laue Langevin, Grenoble, France for provision of neutron beamtime under Proposal No.~4-02-634 at PANTHER \cite{ILLdata}. We thank D.~Ueta and T.~Yokoo for a preliminary neutron scattering experiment  on a test sample performed at the polarized neutron spectrometer POLANO~\cite{Yokoo2013} with unpolarized neutrons in the  Material and Life Science Experiment Facility of the Japan Proton Accelerator Research Complex. The experiment on POLANO was approved by the Neutron Scattering Program Advisory Committee of the Institute of Materials Structure Science, High Energy Accelerator Research Organization, as the S-type research project (Project No.~2019S09). L.~S.~acknowledges support from the National Natural Science Foundation of China (Grant No.~12422407); K.~H. acknowledges support from  the Austrian Science Funds Grant No.~10.55776/I5398.   
\end{acknowledgments}

\bibliography{LaNiO2_INS}

\begin{thebibliography}{117}%
\makeatletter
\providecommand \@ifxundefined [1]{%
 \@ifx{#1\undefined}
}%
\providecommand \@ifnum [1]{%
 \ifnum #1\expandafter \@firstoftwo
 \else \expandafter \@secondoftwo
 \fi
}%
\providecommand \@ifx [1]{%
 \ifx #1\expandafter \@firstoftwo
 \else \expandafter \@secondoftwo
 \fi
}%
\providecommand \natexlab [1]{#1}%
\providecommand \enquote  [1]{``#1''}%
\providecommand \bibnamefont  [1]{#1}%
\providecommand \bibfnamefont [1]{#1}%
\providecommand \citenamefont [1]{#1}%
\providecommand \href@noop [0]{\@secondoftwo}%
\providecommand \href [0]{\begingroup \@sanitize@url \@href}%
\providecommand \@href[1]{\@@startlink{#1}\@@href}%
\providecommand \@@href[1]{\endgroup#1\@@endlink}%
\providecommand \@sanitize@url [0]{\catcode `\\12\catcode `\$12\catcode
  `\&12\catcode `\#12\catcode `\^12\catcode `\_12\catcode `\%12\relax}%
\providecommand \@@startlink[1]{}%
\providecommand \@@endlink[0]{}%
\providecommand \url  [0]{\begingroup\@sanitize@url \@url }%
\providecommand \@url [1]{\endgroup\@href {#1}{\urlprefix }}%
\providecommand \urlprefix  [0]{URL }%
\providecommand \Eprint [0]{\href }%
\providecommand \doibase [0]{https://doi.org/}%
\providecommand \selectlanguage [0]{\@gobble}%
\providecommand \bibinfo  [0]{\@secondoftwo}%
\providecommand \bibfield  [0]{\@secondoftwo}%
\providecommand \translation [1]{[#1]}%
\providecommand \BibitemOpen [0]{}%
\providecommand \bibitemStop [0]{}%
\providecommand \bibitemNoStop [0]{.\EOS\space}%
\providecommand \EOS [0]{\spacefactor3000\relax}%
\providecommand \BibitemShut  [1]{\csname bibitem#1\endcsname}%
\let\auto@bib@innerbib\@empty
\bibitem [{\citenamefont {Li}\ \emph {et~al.}(2019)\citenamefont {Li},
  \citenamefont {Lee}, \citenamefont {Wang}, \citenamefont {Osada},
  \citenamefont {Crossley}, \citenamefont {Lee}, \citenamefont {Cui},
  \citenamefont {Hikita},\ and\ \citenamefont {Hwang}}]{LiNat2019}%
  \BibitemOpen
  \bibfield  {author} {\bibinfo {author} {\bibfnamefont {D.}~\bibnamefont
  {Li}}, \bibinfo {author} {\bibfnamefont {K.}~\bibnamefont {Lee}}, \bibinfo
  {author} {\bibfnamefont {B.~Y.}\ \bibnamefont {Wang}}, \bibinfo {author}
  {\bibfnamefont {M.}~\bibnamefont {Osada}}, \bibinfo {author} {\bibfnamefont
  {S.}~\bibnamefont {Crossley}}, \bibinfo {author} {\bibfnamefont {H.~R.}\
  \bibnamefont {Lee}}, \bibinfo {author} {\bibfnamefont {Y.}~\bibnamefont
  {Cui}}, \bibinfo {author} {\bibfnamefont {Y.}~\bibnamefont {Hikita}},\ and\
  \bibinfo {author} {\bibfnamefont {H.~Y.}\ \bibnamefont {Hwang}},\ }\bibfield
  {title} {\bibinfo {title} {Superconductivity in an infinite-layer
  nickelate},\ }\href {https://doi.org/10.1038/s41586-019-1496-5} {\bibfield
  {journal} {\bibinfo  {journal} {Nature}\ }\textbf {\bibinfo {volume} {572}},\
  \bibinfo {pages} {624} (\bibinfo {year} {2019})}\BibitemShut {NoStop}%
\bibitem [{\citenamefont {Keimer}\ \emph {et~al.}(2015)\citenamefont {Keimer},
  \citenamefont {Kivelson}, \citenamefont {Norman}, \citenamefont {Uchida},\
  and\ \citenamefont {Zaanen}}]{Keimer2015}%
  \BibitemOpen
  \bibfield  {author} {\bibinfo {author} {\bibfnamefont {B.}~\bibnamefont
  {Keimer}}, \bibinfo {author} {\bibfnamefont {S.~A.}\ \bibnamefont
  {Kivelson}}, \bibinfo {author} {\bibfnamefont {M.~R.}\ \bibnamefont
  {Norman}}, \bibinfo {author} {\bibfnamefont {S.}~\bibnamefont {Uchida}},\
  and\ \bibinfo {author} {\bibfnamefont {J.}~\bibnamefont {Zaanen}},\
  }\bibfield  {title} {\bibinfo {title} {From quantum matter to
  high-temperature superconductivity in copper oxides},\ }\href
  {https://doi.org/10.1038/nature14165} {\bibfield  {journal} {\bibinfo
  {journal} {Nature}\ }\textbf {\bibinfo {volume} {518}},\ \bibinfo {pages}
  {179} (\bibinfo {year} {2015})}\BibitemShut {NoStop}%
\bibitem [{\citenamefont {Scalapino}(2012)}]{Scalapino2012}%
  \BibitemOpen
  \bibfield  {author} {\bibinfo {author} {\bibfnamefont {D.~J.}\ \bibnamefont
  {Scalapino}},\ }\bibfield  {title} {\bibinfo {title} {A common thread: The
  pairing interaction for unconventional superconductors},\ }\href
  {https://doi.org/10.1103/RevModPhys.84.1383} {\bibfield  {journal} {\bibinfo
  {journal} {Rev. Mod. Phys.}\ }\textbf {\bibinfo {volume} {84}},\ \bibinfo
  {pages} {1383} (\bibinfo {year} {2012})}\BibitemShut {NoStop}%
\bibitem [{\citenamefont {Armitage}\ \emph {et~al.}(2010)\citenamefont
  {Armitage}, \citenamefont {Fournier},\ and\ \citenamefont
  {Greene}}]{Armitage2010}%
  \BibitemOpen
  \bibfield  {author} {\bibinfo {author} {\bibfnamefont {N.~P.}\ \bibnamefont
  {Armitage}}, \bibinfo {author} {\bibfnamefont {P.}~\bibnamefont {Fournier}},\
  and\ \bibinfo {author} {\bibfnamefont {R.~L.}\ \bibnamefont {Greene}},\
  }\bibfield  {title} {\bibinfo {title} {Progress and perspectives on
  electron-doped cuprates},\ }\href
  {https://doi.org/10.1103/RevModPhys.82.2421} {\bibfield  {journal} {\bibinfo
  {journal} {Rev. Mod. Phys.}\ }\textbf {\bibinfo {volume} {82}},\ \bibinfo
  {pages} {2421} (\bibinfo {year} {2010})}\BibitemShut {NoStop}%
\bibitem [{\citenamefont {Lee}\ \emph {et~al.}(2006)\citenamefont {Lee},
  \citenamefont {Nagaosa},\ and\ \citenamefont {Wen}}]{Lee2006}%
  \BibitemOpen
  \bibfield  {author} {\bibinfo {author} {\bibfnamefont {P.~A.}\ \bibnamefont
  {Lee}}, \bibinfo {author} {\bibfnamefont {N.}~\bibnamefont {Nagaosa}},\ and\
  \bibinfo {author} {\bibfnamefont {X.-G.}\ \bibnamefont {Wen}},\ }\bibfield
  {title} {\bibinfo {title} {Doping a {Mott} insulator: Physics of
  high-temperature superconductivity},\ }\href
  {https://doi.org/10.1103/RevModPhys.78.17} {\bibfield  {journal} {\bibinfo
  {journal} {Rev. Mod. Phys.}\ }\textbf {\bibinfo {volume} {78}},\ \bibinfo
  {pages} {17} (\bibinfo {year} {2006})}\BibitemShut {NoStop}%
\bibitem [{\citenamefont {Anisimov}\ \emph {et~al.}(1999)\citenamefont
  {Anisimov}, \citenamefont {Bukhvalov},\ and\ \citenamefont
  {Rice}}]{Anisimov1999}%
  \BibitemOpen
  \bibfield  {author} {\bibinfo {author} {\bibfnamefont {V.~I.}\ \bibnamefont
  {Anisimov}}, \bibinfo {author} {\bibfnamefont {D.}~\bibnamefont
  {Bukhvalov}},\ and\ \bibinfo {author} {\bibfnamefont {T.~M.}\ \bibnamefont
  {Rice}},\ }\bibfield  {title} {\bibinfo {title} {Electronic structure of
  possible nickelate analogs to the cuprates},\ }\href
  {https://doi.org/10.1103/physrevb.59.7901} {\bibfield  {journal} {\bibinfo
  {journal} {Phys. Rev. B}\ }\textbf {\bibinfo {volume} {59}},\ \bibinfo
  {pages} {7901} (\bibinfo {year} {1999})}\BibitemShut {NoStop}%
\bibitem [{\citenamefont {Zeng}\ \emph {et~al.}(2020)\citenamefont {Zeng},
  \citenamefont {Tang}, \citenamefont {Yin}, \citenamefont {Li}, \citenamefont
  {Li}, \citenamefont {Huang}, \citenamefont {Hu}, \citenamefont {Liu},
  \citenamefont {Omar}, \citenamefont {Jani}, \citenamefont {Lim},
  \citenamefont {Han}, \citenamefont {Wan}, \citenamefont {Yang}, \citenamefont
  {Pennycook}, \citenamefont {Wee},\ and\ \citenamefont
  {Ariando}}]{ZengPRL2020}%
  \BibitemOpen
  \bibfield  {author} {\bibinfo {author} {\bibfnamefont {S.}~\bibnamefont
  {Zeng}}, \bibinfo {author} {\bibfnamefont {C.~S.}\ \bibnamefont {Tang}},
  \bibinfo {author} {\bibfnamefont {X.}~\bibnamefont {Yin}}, \bibinfo {author}
  {\bibfnamefont {C.}~\bibnamefont {Li}}, \bibinfo {author} {\bibfnamefont
  {M.}~\bibnamefont {Li}}, \bibinfo {author} {\bibfnamefont {Z.}~\bibnamefont
  {Huang}}, \bibinfo {author} {\bibfnamefont {J.}~\bibnamefont {Hu}}, \bibinfo
  {author} {\bibfnamefont {W.}~\bibnamefont {Liu}}, \bibinfo {author}
  {\bibfnamefont {G.~J.}\ \bibnamefont {Omar}}, \bibinfo {author}
  {\bibfnamefont {H.}~\bibnamefont {Jani}}, \bibinfo {author} {\bibfnamefont
  {Z.~S.}\ \bibnamefont {Lim}}, \bibinfo {author} {\bibfnamefont
  {K.}~\bibnamefont {Han}}, \bibinfo {author} {\bibfnamefont {D.}~\bibnamefont
  {Wan}}, \bibinfo {author} {\bibfnamefont {P.}~\bibnamefont {Yang}}, \bibinfo
  {author} {\bibfnamefont {S.~J.}\ \bibnamefont {Pennycook}}, \bibinfo {author}
  {\bibfnamefont {A.~T.~S.}\ \bibnamefont {Wee}},\ and\ \bibinfo {author}
  {\bibfnamefont {A.}~\bibnamefont {Ariando}},\ }\bibfield  {title} {\bibinfo
  {title} {{Phase Diagram and Superconducting Dome of Infinite-Layer
  ${\mathrm{Nd}}_{1\ensuremath{-}x}{\mathrm{Sr}}_{x}{\mathrm{NiO}}_{2}$ Thin
  Films}},\ }\href {https://doi.org/10.1103/PhysRevLett.125.147003} {\bibfield
  {journal} {\bibinfo  {journal} {Phys. Rev. Lett.}\ }\textbf {\bibinfo
  {volume} {125}},\ \bibinfo {pages} {147003} (\bibinfo {year}
  {2020})}\BibitemShut {NoStop}%
\bibitem [{\citenamefont {Li}\ \emph {et~al.}(2020)\citenamefont {Li},
  \citenamefont {Wang}, \citenamefont {Lee}, \citenamefont {Harvey},
  \citenamefont {Osada}, \citenamefont {Goodge}, \citenamefont {Kourkoutis},\
  and\ \citenamefont {Hwang}}]{LiPRL2020}%
  \BibitemOpen
  \bibfield  {author} {\bibinfo {author} {\bibfnamefont {D.}~\bibnamefont
  {Li}}, \bibinfo {author} {\bibfnamefont {B.~Y.}\ \bibnamefont {Wang}},
  \bibinfo {author} {\bibfnamefont {K.}~\bibnamefont {Lee}}, \bibinfo {author}
  {\bibfnamefont {S.~P.}\ \bibnamefont {Harvey}}, \bibinfo {author}
  {\bibfnamefont {M.}~\bibnamefont {Osada}}, \bibinfo {author} {\bibfnamefont
  {B.~H.}\ \bibnamefont {Goodge}}, \bibinfo {author} {\bibfnamefont {L.~F.}\
  \bibnamefont {Kourkoutis}},\ and\ \bibinfo {author} {\bibfnamefont {H.~Y.}\
  \bibnamefont {Hwang}},\ }\bibfield  {title} {\bibinfo {title}
  {{Superconducting Dome in
  ${\mathrm{Nd}}_{1\ensuremath{-}x}{\mathrm{Sr}}_{x}{\mathrm{NiO}}_{2}$
  Infinite Layer Films}},\ }\href
  {https://doi.org/10.1103/PhysRevLett.125.027001} {\bibfield  {journal}
  {\bibinfo  {journal} {Phys. Rev. Lett.}\ }\textbf {\bibinfo {volume} {125}},\
  \bibinfo {pages} {027001} (\bibinfo {year} {2020})}\BibitemShut {NoStop}%
\bibitem [{\citenamefont {Osada}\ \emph {et~al.}(2020)\citenamefont {Osada},
  \citenamefont {Wang}, \citenamefont {Lee}, \citenamefont {Li},\ and\
  \citenamefont {Hwang}}]{OsadaPRM2020}%
  \BibitemOpen
  \bibfield  {author} {\bibinfo {author} {\bibfnamefont {M.}~\bibnamefont
  {Osada}}, \bibinfo {author} {\bibfnamefont {B.~Y.}\ \bibnamefont {Wang}},
  \bibinfo {author} {\bibfnamefont {K.}~\bibnamefont {Lee}}, \bibinfo {author}
  {\bibfnamefont {D.}~\bibnamefont {Li}},\ and\ \bibinfo {author}
  {\bibfnamefont {H.~Y.}\ \bibnamefont {Hwang}},\ }\bibfield  {title} {\bibinfo
  {title} {{Phase diagram of infinite layer praseodymium nickelate
  ${\mathrm{Pr}}_{1\ensuremath{-}x}{\mathrm{Sr}}_{x}{\mathrm{NiO}}_{2}$ thin
  films}},\ }\href {https://doi.org/10.1103/PhysRevMaterials.4.121801}
  {\bibfield  {journal} {\bibinfo  {journal} {Phys. Rev. Mater.}\ }\textbf
  {\bibinfo {volume} {4}},\ \bibinfo {pages} {121801} (\bibinfo {year}
  {2020})}\BibitemShut {NoStop}%
\bibitem [{\citenamefont {Gao}\ \emph {et~al.}(2021)\citenamefont {Gao},
  \citenamefont {Zhao}, \citenamefont {Zhou},\ and\ \citenamefont
  {Zhu}}]{GaoCPL2021}%
  \BibitemOpen
  \bibfield  {author} {\bibinfo {author} {\bibfnamefont {Q.}~\bibnamefont
  {Gao}}, \bibinfo {author} {\bibfnamefont {Y.}~\bibnamefont {Zhao}}, \bibinfo
  {author} {\bibfnamefont {X.-J.}\ \bibnamefont {Zhou}},\ and\ \bibinfo
  {author} {\bibfnamefont {Z.}~\bibnamefont {Zhu}},\ }\bibfield  {title}
  {\bibinfo {title} {{Preparation of Superconducting Thin Films of
  Infinite-Layer Nickelate Nd$_{0.8}$Sr$_{0.2}$NiO$_{2}$}},\ }\href
  {https://doi.org/10.1088/0256-307X/38/7/077401} {\bibfield  {journal}
  {\bibinfo  {journal} {Chin. Phys. Lett.}\ }\textbf {\bibinfo {volume} {38}},\
  \bibinfo {pages} {077401} (\bibinfo {year} {2021})}\BibitemShut {NoStop}%
\bibitem [{\citenamefont {Zeng}\ \emph {et~al.}(2022)\citenamefont {Zeng},
  \citenamefont {Li}, \citenamefont {Chow}, \citenamefont {Cao}, \citenamefont
  {Zhang}, \citenamefont {Tang}, \citenamefont {Yin}, \citenamefont {Lim},
  \citenamefont {Hu}, \citenamefont {Yang},\ and\ \citenamefont
  {Ariando}}]{ZengSciAdv2022}%
  \BibitemOpen
  \bibfield  {author} {\bibinfo {author} {\bibfnamefont {S.}~\bibnamefont
  {Zeng}}, \bibinfo {author} {\bibfnamefont {C.}~\bibnamefont {Li}}, \bibinfo
  {author} {\bibfnamefont {L.~E.}\ \bibnamefont {Chow}}, \bibinfo {author}
  {\bibfnamefont {Y.}~\bibnamefont {Cao}}, \bibinfo {author} {\bibfnamefont
  {Z.}~\bibnamefont {Zhang}}, \bibinfo {author} {\bibfnamefont {C.~S.}\
  \bibnamefont {Tang}}, \bibinfo {author} {\bibfnamefont {X.}~\bibnamefont
  {Yin}}, \bibinfo {author} {\bibfnamefont {Z.~S.}\ \bibnamefont {Lim}},
  \bibinfo {author} {\bibfnamefont {J.}~\bibnamefont {Hu}}, \bibinfo {author}
  {\bibfnamefont {P.}~\bibnamefont {Yang}},\ and\ \bibinfo {author}
  {\bibfnamefont {A.}~\bibnamefont {Ariando}},\ }\bibfield  {title} {\bibinfo
  {title} {{Superconductivity in infinite-layer nickelate
  La$_{1-x}$Ca$_{x}$NiO$_{2}$ thin films}},\ }\href
  {https://doi.org/10.1126/sciadv.abl9927} {\bibfield  {journal} {\bibinfo
  {journal} {Sci. Adv.}\ }\textbf {\bibinfo {volume} {8}},\ \bibinfo {pages}
  {eabl9927} (\bibinfo {year} {2022})}\BibitemShut {NoStop}%
\bibitem [{\citenamefont {Sahib}\ \emph {et~al.}(2025)\citenamefont {Sahib},
  \citenamefont {Raji}, \citenamefont {Rosa}, \citenamefont {Merzoni},
  \citenamefont {Ghiringhelli}, \citenamefont {Salluzzo}, \citenamefont
  {Gloter}, \citenamefont {Viart},\ and\ \citenamefont {Preziosi}}]{Sahib2025}%
  \BibitemOpen
  \bibfield  {author} {\bibinfo {author} {\bibfnamefont {H.}~\bibnamefont
  {Sahib}}, \bibinfo {author} {\bibfnamefont {A.}~\bibnamefont {Raji}},
  \bibinfo {author} {\bibfnamefont {F.}~\bibnamefont {Rosa}}, \bibinfo {author}
  {\bibfnamefont {G.}~\bibnamefont {Merzoni}}, \bibinfo {author} {\bibfnamefont
  {G.}~\bibnamefont {Ghiringhelli}}, \bibinfo {author} {\bibfnamefont
  {M.}~\bibnamefont {Salluzzo}}, \bibinfo {author} {\bibfnamefont
  {A.}~\bibnamefont {Gloter}}, \bibinfo {author} {\bibfnamefont
  {N.}~\bibnamefont {Viart}},\ and\ \bibinfo {author} {\bibfnamefont
  {D.}~\bibnamefont {Preziosi}},\ }\bibfield  {title} {\bibinfo {title}
  {{Superconductivity in PrNiO$_2$ Infinite-Layer Nickelates}},\ }\href
  {https://doi.org/10.1002/adma.202416187} {\bibfield  {journal} {\bibinfo
  {journal} {Adv. Mater.}\ }\textbf {\bibinfo {volume} {37}},\ \bibinfo {pages}
  {202416187} (\bibinfo {year} {2025})}\BibitemShut {NoStop}%
\bibitem [{\citenamefont {Lee}\ and\ \citenamefont {Pickett}(2004)}]{Lee2004}%
  \BibitemOpen
  \bibfield  {author} {\bibinfo {author} {\bibfnamefont {K.-W.}\ \bibnamefont
  {Lee}}\ and\ \bibinfo {author} {\bibfnamefont {W.~E.}\ \bibnamefont
  {Pickett}},\ }\bibfield  {title} {\bibinfo {title} {Infinite-layer
  {LaNiO$_2$}: {Ni$^{1+}$} is not {Cu$^{2+}$}},\ }\href
  {https://doi.org/10.1103/physrevb.70.165109} {\bibfield  {journal} {\bibinfo
  {journal} {Phys. Rev. B}\ }\textbf {\bibinfo {volume} {70}},\ \bibinfo
  {pages} {165109} (\bibinfo {year} {2004})}\BibitemShut {NoStop}%
\bibitem [{\citenamefont {Nomura}\ \emph {et~al.}(2019)\citenamefont {Nomura},
  \citenamefont {Hirayama}, \citenamefont {Tadano}, \citenamefont {Yoshimoto},
  \citenamefont {Nakamura},\ and\ \citenamefont {Arita}}]{Nomura2019}%
  \BibitemOpen
  \bibfield  {author} {\bibinfo {author} {\bibfnamefont {Y.}~\bibnamefont
  {Nomura}}, \bibinfo {author} {\bibfnamefont {M.}~\bibnamefont {Hirayama}},
  \bibinfo {author} {\bibfnamefont {T.}~\bibnamefont {Tadano}}, \bibinfo
  {author} {\bibfnamefont {Y.}~\bibnamefont {Yoshimoto}}, \bibinfo {author}
  {\bibfnamefont {K.}~\bibnamefont {Nakamura}},\ and\ \bibinfo {author}
  {\bibfnamefont {R.}~\bibnamefont {Arita}},\ }\bibfield  {title} {\bibinfo
  {title} {{Formation of a two-dimensional single-component correlated electron
  system and band engineering in the nickelate superconductor
  ${\mathrm{NdNiO}}_{2}$}},\ }\href
  {https://doi.org/10.1103/PhysRevB.100.205138} {\bibfield  {journal} {\bibinfo
   {journal} {Phys. Rev. B}\ }\textbf {\bibinfo {volume} {100}},\ \bibinfo
  {pages} {205138} (\bibinfo {year} {2019})}\BibitemShut {NoStop}%
\bibitem [{\citenamefont {Botana}\ and\ \citenamefont
  {Norman}(2020)}]{Botana2020}%
  \BibitemOpen
  \bibfield  {author} {\bibinfo {author} {\bibfnamefont {A.~S.}\ \bibnamefont
  {Botana}}\ and\ \bibinfo {author} {\bibfnamefont {M.~R.}\ \bibnamefont
  {Norman}},\ }\bibfield  {title} {\bibinfo {title} {Similarities and
  differences between {LaNiO$_2$} and {CaCuO$_2$} and implications for
  superconductivity},\ }\href {https://doi.org/10.1103/physrevx.10.011024}
  {\bibfield  {journal} {\bibinfo  {journal} {Phys. Rev. X}\ }\textbf {\bibinfo
  {volume} {10}},\ \bibinfo {pages} {011024} (\bibinfo {year}
  {2020})}\BibitemShut {NoStop}%
\bibitem [{\citenamefont {Kitatani}\ \emph {et~al.}(2020)\citenamefont
  {Kitatani}, \citenamefont {Si}, \citenamefont {Janson}, \citenamefont
  {Arita}, \citenamefont {Zhong},\ and\ \citenamefont {Held}}]{Kitatani2020}%
  \BibitemOpen
  \bibfield  {author} {\bibinfo {author} {\bibfnamefont {M.}~\bibnamefont
  {Kitatani}}, \bibinfo {author} {\bibfnamefont {L.}~\bibnamefont {Si}},
  \bibinfo {author} {\bibfnamefont {O.}~\bibnamefont {Janson}}, \bibinfo
  {author} {\bibfnamefont {R.}~\bibnamefont {Arita}}, \bibinfo {author}
  {\bibfnamefont {Z.}~\bibnamefont {Zhong}},\ and\ \bibinfo {author}
  {\bibfnamefont {K.}~\bibnamefont {Held}},\ }\bibfield  {title} {\bibinfo
  {title} {{Nickelate superconductors—a renaissance of the one-band Hubbard
  model}},\ }\href {https://doi.org/10.1038/s41535-020-00260-y} {\bibfield
  {journal} {\bibinfo  {journal} {npj Quantum Mater.}\ }\textbf {\bibinfo
  {volume} {5}},\ \bibinfo {pages} {59} (\bibinfo {year} {2020})},\ \Eprint
  {https://arxiv.org/abs/2002.12230} {2002.12230} \BibitemShut {NoStop}%
\bibitem [{\citenamefont {Wu}\ \emph {et~al.}(2020)\citenamefont {Wu},
  \citenamefont {Di~Sante}, \citenamefont {Schwemmer}, \citenamefont {Hanke},
  \citenamefont {Hwang}, \citenamefont {Raghu},\ and\ \citenamefont
  {Thomale}}]{Wu2020}%
  \BibitemOpen
  \bibfield  {author} {\bibinfo {author} {\bibfnamefont {X.}~\bibnamefont
  {Wu}}, \bibinfo {author} {\bibfnamefont {D.}~\bibnamefont {Di~Sante}},
  \bibinfo {author} {\bibfnamefont {T.}~\bibnamefont {Schwemmer}}, \bibinfo
  {author} {\bibfnamefont {W.}~\bibnamefont {Hanke}}, \bibinfo {author}
  {\bibfnamefont {H.~Y.}\ \bibnamefont {Hwang}}, \bibinfo {author}
  {\bibfnamefont {S.}~\bibnamefont {Raghu}},\ and\ \bibinfo {author}
  {\bibfnamefont {R.}~\bibnamefont {Thomale}},\ }\bibfield  {title} {\bibinfo
  {title} {Robust ${d}_{{x}^{2}\ensuremath{-}{y}^{2}}$-wave superconductivity
  of infinite-layer nickelates},\ }\href
  {https://doi.org/10.1103/PhysRevB.101.060504} {\bibfield  {journal} {\bibinfo
   {journal} {Phys. Rev. B}\ }\textbf {\bibinfo {volume} {101}},\ \bibinfo
  {pages} {060504} (\bibinfo {year} {2020})}\BibitemShut {NoStop}%
\bibitem [{\citenamefont {Zhang}\ \emph {et~al.}(2020)\citenamefont {Zhang},
  \citenamefont {Jin}, \citenamefont {Wang}, \citenamefont {Xi}, \citenamefont
  {Shi}, \citenamefont {Ye},\ and\ \citenamefont {Mei}}]{Zhang2020E}%
  \BibitemOpen
  \bibfield  {author} {\bibinfo {author} {\bibfnamefont {H.}~\bibnamefont
  {Zhang}}, \bibinfo {author} {\bibfnamefont {L.}~\bibnamefont {Jin}}, \bibinfo
  {author} {\bibfnamefont {S.}~\bibnamefont {Wang}}, \bibinfo {author}
  {\bibfnamefont {B.}~\bibnamefont {Xi}}, \bibinfo {author} {\bibfnamefont
  {X.}~\bibnamefont {Shi}}, \bibinfo {author} {\bibfnamefont {F.}~\bibnamefont
  {Ye}},\ and\ \bibinfo {author} {\bibfnamefont {J.-W.}\ \bibnamefont {Mei}},\
  }\bibfield  {title} {\bibinfo {title} {Effective hamiltonian for nickelate
  oxides
  {${\mathrm{Nd}}_{1\ensuremath{-}x}{\mathrm{Sr}}_{x}{\mathrm{NiO}}_{2}$}},\
  }\href {https://doi.org/10.1103/PhysRevResearch.2.013214} {\bibfield
  {journal} {\bibinfo  {journal} {Phys. Rev. Research}\ }\textbf {\bibinfo
  {volume} {2}},\ \bibinfo {pages} {013214} (\bibinfo {year}
  {2020})}\BibitemShut {NoStop}%
\bibitem [{\citenamefont {Sakakibara}\ \emph {et~al.}(2020)\citenamefont
  {Sakakibara}, \citenamefont {Usui}, \citenamefont {Suzuki}, \citenamefont
  {Kotani}, \citenamefont {Aoki},\ and\ \citenamefont
  {Kuroki}}]{Sakakibara2020}%
  \BibitemOpen
  \bibfield  {author} {\bibinfo {author} {\bibfnamefont {H.}~\bibnamefont
  {Sakakibara}}, \bibinfo {author} {\bibfnamefont {H.}~\bibnamefont {Usui}},
  \bibinfo {author} {\bibfnamefont {K.}~\bibnamefont {Suzuki}}, \bibinfo
  {author} {\bibfnamefont {T.}~\bibnamefont {Kotani}}, \bibinfo {author}
  {\bibfnamefont {H.}~\bibnamefont {Aoki}},\ and\ \bibinfo {author}
  {\bibfnamefont {K.}~\bibnamefont {Kuroki}},\ }\bibfield  {title} {\bibinfo
  {title} {Model construction and a possibility of cupratelike pairing in a new
  ${d}^{9}$ nickelate superconductor
  {$(\mathrm{Nd},\mathrm{Sr}){\mathrm{NiO}}_{2}$}},\ }\href
  {https://doi.org/10.1103/PhysRevLett.125.077003} {\bibfield  {journal}
  {\bibinfo  {journal} {Phys. Rev. Lett.}\ }\textbf {\bibinfo {volume} {125}},\
  \bibinfo {pages} {077003} (\bibinfo {year} {2020})}\BibitemShut {NoStop}%
\bibitem [{\citenamefont {Lechermann}(2020)}]{Lechermann2020}%
  \BibitemOpen
  \bibfield  {author} {\bibinfo {author} {\bibfnamefont {F.}~\bibnamefont
  {Lechermann}},\ }\bibfield  {title} {\bibinfo {title} {Late transition metal
  oxides with infinite-layer structure: Nickelates versus cuprates},\ }\href
  {https://doi.org/10.1103/PhysRevB.101.081110} {\bibfield  {journal} {\bibinfo
   {journal} {Phys. Rev. B}\ }\textbf {\bibinfo {volume} {101}},\ \bibinfo
  {pages} {081110} (\bibinfo {year} {2020})}\BibitemShut {NoStop}%
\bibitem [{\citenamefont {Wang}\ \emph {et~al.}(2020)\citenamefont {Wang},
  \citenamefont {Kang}, \citenamefont {Miao},\ and\ \citenamefont
  {Kotliar}}]{Wang2020E}%
  \BibitemOpen
  \bibfield  {author} {\bibinfo {author} {\bibfnamefont {Y.}~\bibnamefont
  {Wang}}, \bibinfo {author} {\bibfnamefont {C.-J.}\ \bibnamefont {Kang}},
  \bibinfo {author} {\bibfnamefont {H.}~\bibnamefont {Miao}},\ and\ \bibinfo
  {author} {\bibfnamefont {G.}~\bibnamefont {Kotliar}},\ }\bibfield  {title}
  {\bibinfo {title} {Hund's metal physics: From {${\mathrm{SrNiO}}_{2}$ to
  ${\mathrm{LaNiO}}_{2}$}},\ }\href
  {https://doi.org/10.1103/PhysRevB.102.161118} {\bibfield  {journal} {\bibinfo
   {journal} {Phys. Rev. B}\ }\textbf {\bibinfo {volume} {102}},\ \bibinfo
  {pages} {161118} (\bibinfo {year} {2020})}\BibitemShut {NoStop}%
\bibitem [{\citenamefont {Werner}\ and\ \citenamefont
  {Hoshino}(2020)}]{Werner2020}%
  \BibitemOpen
  \bibfield  {author} {\bibinfo {author} {\bibfnamefont {P.}~\bibnamefont
  {Werner}}\ and\ \bibinfo {author} {\bibfnamefont {S.}~\bibnamefont
  {Hoshino}},\ }\bibfield  {title} {\bibinfo {title} {Nickelate
  superconductors: Multiorbital nature and spin freezing},\ }\href
  {https://doi.org/10.1103/PhysRevB.101.041104} {\bibfield  {journal} {\bibinfo
   {journal} {Phys. Rev. B}\ }\textbf {\bibinfo {volume} {101}},\ \bibinfo
  {pages} {041104} (\bibinfo {year} {2020})}\BibitemShut {NoStop}%
\bibitem [{\citenamefont {Karp}\ \emph {et~al.}(2020)\citenamefont {Karp},
  \citenamefont {Botana}, \citenamefont {Norman}, \citenamefont {Park},
  \citenamefont {Zingl},\ and\ \citenamefont {Millis}}]{Karp2020}%
  \BibitemOpen
  \bibfield  {author} {\bibinfo {author} {\bibfnamefont {J.}~\bibnamefont
  {Karp}}, \bibinfo {author} {\bibfnamefont {A.~S.}\ \bibnamefont {Botana}},
  \bibinfo {author} {\bibfnamefont {M.~R.}\ \bibnamefont {Norman}}, \bibinfo
  {author} {\bibfnamefont {H.}~\bibnamefont {Park}}, \bibinfo {author}
  {\bibfnamefont {M.}~\bibnamefont {Zingl}},\ and\ \bibinfo {author}
  {\bibfnamefont {A.}~\bibnamefont {Millis}},\ }\bibfield  {title} {\bibinfo
  {title} {Many-body electronic structure of {${\mathrm{NdNiO}}_{2}$ and
  ${\mathrm{CaCuO}}_{2}$}},\ }\href
  {https://doi.org/10.1103/PhysRevX.10.021061} {\bibfield  {journal} {\bibinfo
  {journal} {Phys. Rev. X}\ }\textbf {\bibinfo {volume} {10}},\ \bibinfo
  {pages} {021061} (\bibinfo {year} {2020})}\BibitemShut {NoStop}%
\bibitem [{\citenamefont {Liu}\ \emph {et~al.}(2021)\citenamefont {Liu},
  \citenamefont {Xu}, \citenamefont {Cao}, \citenamefont {Zhu}, \citenamefont
  {Wang},\ and\ \citenamefont {Yang}}]{Liu2021}%
  \BibitemOpen
  \bibfield  {author} {\bibinfo {author} {\bibfnamefont {Z.}~\bibnamefont
  {Liu}}, \bibinfo {author} {\bibfnamefont {C.}~\bibnamefont {Xu}}, \bibinfo
  {author} {\bibfnamefont {C.}~\bibnamefont {Cao}}, \bibinfo {author}
  {\bibfnamefont {W.}~\bibnamefont {Zhu}}, \bibinfo {author} {\bibfnamefont
  {Z.~F.}\ \bibnamefont {Wang}},\ and\ \bibinfo {author} {\bibfnamefont
  {J.}~\bibnamefont {Yang}},\ }\bibfield  {title} {\bibinfo {title} {Doping
  dependence of electronic structure of infinite-layer
  {${\mathrm{NdNiO}}_{2}$}},\ }\href
  {https://doi.org/10.1103/PhysRevB.103.045103} {\bibfield  {journal} {\bibinfo
   {journal} {Phys. Rev. B}\ }\textbf {\bibinfo {volume} {103}},\ \bibinfo
  {pages} {045103} (\bibinfo {year} {2021})}\BibitemShut {NoStop}%
\bibitem [{\citenamefont {Wan}\ \emph {et~al.}(2021)\citenamefont {Wan},
  \citenamefont {Ivanov}, \citenamefont {Resta}, \citenamefont {Leonov},\ and\
  \citenamefont {Savrasov}}]{Wan2021}%
  \BibitemOpen
  \bibfield  {author} {\bibinfo {author} {\bibfnamefont {X.}~\bibnamefont
  {Wan}}, \bibinfo {author} {\bibfnamefont {V.}~\bibnamefont {Ivanov}},
  \bibinfo {author} {\bibfnamefont {G.}~\bibnamefont {Resta}}, \bibinfo
  {author} {\bibfnamefont {I.}~\bibnamefont {Leonov}},\ and\ \bibinfo {author}
  {\bibfnamefont {S.~Y.}\ \bibnamefont {Savrasov}},\ }\bibfield  {title}
  {\bibinfo {title} {Exchange interactions and sensitivity of the ni two-hole
  spin state to hund's coupling in doped {${\mathrm{NdNiO}}_{2}$}},\ }\href
  {https://doi.org/10.1103/PhysRevB.103.075123} {\bibfield  {journal} {\bibinfo
   {journal} {Phys. Rev. B}\ }\textbf {\bibinfo {volume} {103}},\ \bibinfo
  {pages} {075123} (\bibinfo {year} {2021})}\BibitemShut {NoStop}%
\bibitem [{\citenamefont {Lang}\ \emph {et~al.}(2021)\citenamefont {Lang},
  \citenamefont {Jiang},\ and\ \citenamefont {Ku}}]{Lang2021}%
  \BibitemOpen
  \bibfield  {author} {\bibinfo {author} {\bibfnamefont {Z.-J.}\ \bibnamefont
  {Lang}}, \bibinfo {author} {\bibfnamefont {R.}~\bibnamefont {Jiang}},\ and\
  \bibinfo {author} {\bibfnamefont {W.}~\bibnamefont {Ku}},\ }\bibfield
  {title} {\bibinfo {title} {Strongly correlated doped hole carriers in the
  superconducting nickelates: Their location, local many-body state, and
  low-energy effective hamiltonian},\ }\href
  {https://doi.org/10.1103/PhysRevB.103.L180502} {\bibfield  {journal}
  {\bibinfo  {journal} {Phys. Rev. B}\ }\textbf {\bibinfo {volume} {103}},\
  \bibinfo {pages} {L180502} (\bibinfo {year} {2021})}\BibitemShut {NoStop}%
\bibitem [{\citenamefont {Higashi}\ \emph {et~al.}(2021)\citenamefont
  {Higashi}, \citenamefont {Winder}, \citenamefont {Kune\ifmmode~\check{s}\else
  \v{s}\fi{}},\ and\ \citenamefont {Hariki}}]{Higashi2021}%
  \BibitemOpen
  \bibfield  {author} {\bibinfo {author} {\bibfnamefont {K.}~\bibnamefont
  {Higashi}}, \bibinfo {author} {\bibfnamefont {M.}~\bibnamefont {Winder}},
  \bibinfo {author} {\bibfnamefont {J.}~\bibnamefont
  {Kune\ifmmode~\check{s}\else \v{s}\fi{}}},\ and\ \bibinfo {author}
  {\bibfnamefont {A.}~\bibnamefont {Hariki}},\ }\bibfield  {title} {\bibinfo
  {title} {Core-level x-ray spectroscopy of infinite-layer nickelate:
  $\mathrm{LDA}+\mathrm{DMFT}$ study},\ }\href
  {https://doi.org/10.1103/PhysRevX.11.041009} {\bibfield  {journal} {\bibinfo
  {journal} {Phys. Rev. X}\ }\textbf {\bibinfo {volume} {11}},\ \bibinfo
  {pages} {041009} (\bibinfo {year} {2021})}\BibitemShut {NoStop}%
\bibitem [{\citenamefont {Held}\ \emph {et~al.}(2022)\citenamefont {Held},
  \citenamefont {Si}, \citenamefont {Worm}, \citenamefont {Janson},
  \citenamefont {Arita}, \citenamefont {Zhong}, \citenamefont {Tomczak},\ and\
  \citenamefont {Kitatani}}]{Held2022}%
  \BibitemOpen
  \bibfield  {author} {\bibinfo {author} {\bibfnamefont {K.}~\bibnamefont
  {Held}}, \bibinfo {author} {\bibfnamefont {L.}~\bibnamefont {Si}}, \bibinfo
  {author} {\bibfnamefont {P.}~\bibnamefont {Worm}}, \bibinfo {author}
  {\bibfnamefont {O.}~\bibnamefont {Janson}}, \bibinfo {author} {\bibfnamefont
  {R.}~\bibnamefont {Arita}}, \bibinfo {author} {\bibfnamefont
  {Z.}~\bibnamefont {Zhong}}, \bibinfo {author} {\bibfnamefont
  {J.}~\bibnamefont {Tomczak}},\ and\ \bibinfo {author} {\bibfnamefont
  {M.}~\bibnamefont {Kitatani}},\ }\bibfield  {title} {\bibinfo {title} {Phase
  diagram of nickelate superconductors calculated by dynamical vertex
  approximation},\ }\href
  {https://www.frontiersin.org/articles/10.3389/fphy.2021.810394} {\bibfield
  {journal} {\bibinfo  {journal} {Front. Phys.}\ }\textbf {\bibinfo {volume}
  {9:810394}} (\bibinfo {year} {2022})}\BibitemShut {NoStop}%
\bibitem [{\citenamefont {Chen}\ \emph {et~al.}(2022)\citenamefont {Chen},
  \citenamefont {Hampel}, \citenamefont {Karp}, \citenamefont {Lechermann},\
  and\ \citenamefont {Millis}}]{Chen2022r}%
  \BibitemOpen
  \bibfield  {author} {\bibinfo {author} {\bibfnamefont {H.}~\bibnamefont
  {Chen}}, \bibinfo {author} {\bibfnamefont {A.}~\bibnamefont {Hampel}},
  \bibinfo {author} {\bibfnamefont {J.}~\bibnamefont {Karp}}, \bibinfo {author}
  {\bibfnamefont {F.}~\bibnamefont {Lechermann}},\ and\ \bibinfo {author}
  {\bibfnamefont {A.}~\bibnamefont {Millis}},\ }\bibfield  {title} {\bibinfo
  {title} {Dynamical mean field studies of infinite layer nickelates: Physics
  results and methodological implications},\ }\href
  {https://www.frontiersin.org/articles/10.3389/fphy.2022.835942} {\bibfield
  {journal} {\bibinfo  {journal} {Front. Phys.}\ }\textbf {\bibinfo {volume}
  {10:835942}} (\bibinfo {year} {2022})}\BibitemShut {NoStop}%
\bibitem [{\citenamefont {Hepting}\ \emph {et~al.}(2020)\citenamefont
  {Hepting}, \citenamefont {Li}, \citenamefont {Jia}, \citenamefont {Lu},
  \citenamefont {Paris}, \citenamefont {Tseng}, \citenamefont {Feng},
  \citenamefont {Osada}, \citenamefont {Been}, \citenamefont {Hikita},
  \citenamefont {Chuang}, \citenamefont {Hussain}, \citenamefont {Zhou},
  \citenamefont {Nag}, \citenamefont {Garcia-Fernandez}, \citenamefont {Rossi},
  \citenamefont {Huang}, \citenamefont {Huang}, \citenamefont {Shen},
  \citenamefont {Schmitt}, \citenamefont {Hwang}, \citenamefont {Moritz},
  \citenamefont {Zaanen}, \citenamefont {Devereaux},\ and\ \citenamefont
  {Lee}}]{Hepting2020}%
  \BibitemOpen
  \bibfield  {author} {\bibinfo {author} {\bibfnamefont {M.}~\bibnamefont
  {Hepting}}, \bibinfo {author} {\bibfnamefont {D.}~\bibnamefont {Li}},
  \bibinfo {author} {\bibfnamefont {C.~J.}\ \bibnamefont {Jia}}, \bibinfo
  {author} {\bibfnamefont {H.}~\bibnamefont {Lu}}, \bibinfo {author}
  {\bibfnamefont {E.}~\bibnamefont {Paris}}, \bibinfo {author} {\bibfnamefont
  {Y.}~\bibnamefont {Tseng}}, \bibinfo {author} {\bibfnamefont
  {X.}~\bibnamefont {Feng}}, \bibinfo {author} {\bibfnamefont {M.}~\bibnamefont
  {Osada}}, \bibinfo {author} {\bibfnamefont {E.}~\bibnamefont {Been}},
  \bibinfo {author} {\bibfnamefont {Y.}~\bibnamefont {Hikita}}, \bibinfo
  {author} {\bibfnamefont {Y.-D.}\ \bibnamefont {Chuang}}, \bibinfo {author}
  {\bibfnamefont {Z.}~\bibnamefont {Hussain}}, \bibinfo {author} {\bibfnamefont
  {K.~J.}\ \bibnamefont {Zhou}}, \bibinfo {author} {\bibfnamefont
  {A.}~\bibnamefont {Nag}}, \bibinfo {author} {\bibfnamefont {M.}~\bibnamefont
  {Garcia-Fernandez}}, \bibinfo {author} {\bibfnamefont {M.}~\bibnamefont
  {Rossi}}, \bibinfo {author} {\bibfnamefont {H.~Y.}\ \bibnamefont {Huang}},
  \bibinfo {author} {\bibfnamefont {D.~J.}\ \bibnamefont {Huang}}, \bibinfo
  {author} {\bibfnamefont {Z.~X.}\ \bibnamefont {Shen}}, \bibinfo {author}
  {\bibfnamefont {T.}~\bibnamefont {Schmitt}}, \bibinfo {author} {\bibfnamefont
  {H.~Y.}\ \bibnamefont {Hwang}}, \bibinfo {author} {\bibfnamefont
  {B.}~\bibnamefont {Moritz}}, \bibinfo {author} {\bibfnamefont
  {J.}~\bibnamefont {Zaanen}}, \bibinfo {author} {\bibfnamefont {T.~P.}\
  \bibnamefont {Devereaux}},\ and\ \bibinfo {author} {\bibfnamefont {W.~S.}\
  \bibnamefont {Lee}},\ }\bibfield  {title} {\bibinfo {title} {Electronic
  structure of the parent compound of superconducting infinite-layer
  nickelates},\ }\href {https://doi.org/10.1038/s41563-019-0585-z} {\bibfield
  {journal} {\bibinfo  {journal} {Nat. Mater.}\ }\textbf {\bibinfo {volume}
  {19}},\ \bibinfo {pages} {381} (\bibinfo {year} {2020})}\BibitemShut
  {NoStop}%
\bibitem [{\citenamefont {Goodge}\ \emph {et~al.}(2021)\citenamefont {Goodge},
  \citenamefont {Li}, \citenamefont {Lee}, \citenamefont {Osada}, \citenamefont
  {Wang}, \citenamefont {Sawatzky}, \citenamefont {Hwang},\ and\ \citenamefont
  {Kourkoutis}}]{Goodge2021}%
  \BibitemOpen
  \bibfield  {author} {\bibinfo {author} {\bibfnamefont {B.~H.}\ \bibnamefont
  {Goodge}}, \bibinfo {author} {\bibfnamefont {D.}~\bibnamefont {Li}}, \bibinfo
  {author} {\bibfnamefont {K.}~\bibnamefont {Lee}}, \bibinfo {author}
  {\bibfnamefont {M.}~\bibnamefont {Osada}}, \bibinfo {author} {\bibfnamefont
  {B.~Y.}\ \bibnamefont {Wang}}, \bibinfo {author} {\bibfnamefont {G.~A.}\
  \bibnamefont {Sawatzky}}, \bibinfo {author} {\bibfnamefont {H.~Y.}\
  \bibnamefont {Hwang}},\ and\ \bibinfo {author} {\bibfnamefont {L.~F.}\
  \bibnamefont {Kourkoutis}},\ }\bibfield  {title} {\bibinfo {title} {Doping
  evolution of the mott{\textendash}hubbard landscape in infinite-layer
  nickelates},\ }\href {https://doi.org/10.1073/pnas.2007683118} {\bibfield
  {journal} {\bibinfo  {journal} {PNAS}\ }\textbf {\bibinfo {volume} {118}},\
  \bibinfo {pages} {e2007683118} (\bibinfo {year} {2021})}\BibitemShut
  {NoStop}%
\bibitem [{\citenamefont {Ding}\ \emph {et~al.}(2024)\citenamefont {Ding},
  \citenamefont {Fan}, \citenamefont {Wang}, \citenamefont {Li}, \citenamefont
  {An}, \citenamefont {Ye}, \citenamefont {Tang}, \citenamefont {Lei},
  \citenamefont {Sun}, \citenamefont {Guo}, \citenamefont {Chen}, \citenamefont
  {Sangphet}, \citenamefont {Wang}, \citenamefont {Xu}, \citenamefont {Peng},\
  and\ \citenamefont {Feng}}]{Ding2024}%
  \BibitemOpen
  \bibfield  {author} {\bibinfo {author} {\bibfnamefont {X.}~\bibnamefont
  {Ding}}, \bibinfo {author} {\bibfnamefont {Y.}~\bibnamefont {Fan}}, \bibinfo
  {author} {\bibfnamefont {X.}~\bibnamefont {Wang}}, \bibinfo {author}
  {\bibfnamefont {C.}~\bibnamefont {Li}}, \bibinfo {author} {\bibfnamefont
  {Z.}~\bibnamefont {An}}, \bibinfo {author} {\bibfnamefont {J.}~\bibnamefont
  {Ye}}, \bibinfo {author} {\bibfnamefont {S.}~\bibnamefont {Tang}}, \bibinfo
  {author} {\bibfnamefont {M.}~\bibnamefont {Lei}}, \bibinfo {author}
  {\bibfnamefont {X.}~\bibnamefont {Sun}}, \bibinfo {author} {\bibfnamefont
  {N.}~\bibnamefont {Guo}}, \bibinfo {author} {\bibfnamefont {Z.}~\bibnamefont
  {Chen}}, \bibinfo {author} {\bibfnamefont {S.}~\bibnamefont {Sangphet}},
  \bibinfo {author} {\bibfnamefont {Y.}~\bibnamefont {Wang}}, \bibinfo {author}
  {\bibfnamefont {H.}~\bibnamefont {Xu}}, \bibinfo {author} {\bibfnamefont
  {R.}~\bibnamefont {Peng}},\ and\ \bibinfo {author} {\bibfnamefont
  {D.}~\bibnamefont {Feng}},\ }\bibfield  {title} {\bibinfo {title}
  {{Cuprate-like electronic structures in infinite-layer nickelates with
  substantial hole dopings}},\ }\href {https://doi.org/10.1093/nsr/nwae194}
  {\bibfield  {journal} {\bibinfo  {journal} {Natl. Sci. Rev.}\ }\textbf
  {\bibinfo {volume} {11}},\ \bibinfo {pages} {nwae194} (\bibinfo {year}
  {2024})}\BibitemShut {NoStop}%
\bibitem [{\citenamefont {Sun}\ \emph {et~al.}(2025)\citenamefont {Sun},
  \citenamefont {Jiang}, \citenamefont {Xia}, \citenamefont {Hao},
  \citenamefont {Yan}, \citenamefont {Wang}, \citenamefont {Li}, \citenamefont
  {Liu}, \citenamefont {Ding}, \citenamefont {Liu}, \citenamefont {Liu},
  \citenamefont {Liu}, \citenamefont {Chen}, \citenamefont {Shen},\ and\
  \citenamefont {Nie}}]{Sun2025}%
  \BibitemOpen
  \bibfield  {author} {\bibinfo {author} {\bibfnamefont {W.}~\bibnamefont
  {Sun}}, \bibinfo {author} {\bibfnamefont {Z.}~\bibnamefont {Jiang}}, \bibinfo
  {author} {\bibfnamefont {C.}~\bibnamefont {Xia}}, \bibinfo {author}
  {\bibfnamefont {B.}~\bibnamefont {Hao}}, \bibinfo {author} {\bibfnamefont
  {S.}~\bibnamefont {Yan}}, \bibinfo {author} {\bibfnamefont {M.}~\bibnamefont
  {Wang}}, \bibinfo {author} {\bibfnamefont {Y.}~\bibnamefont {Li}}, \bibinfo
  {author} {\bibfnamefont {H.}~\bibnamefont {Liu}}, \bibinfo {author}
  {\bibfnamefont {J.}~\bibnamefont {Ding}}, \bibinfo {author} {\bibfnamefont
  {J.}~\bibnamefont {Liu}}, \bibinfo {author} {\bibfnamefont {Z.}~\bibnamefont
  {Liu}}, \bibinfo {author} {\bibfnamefont {J.}~\bibnamefont {Liu}}, \bibinfo
  {author} {\bibfnamefont {H.}~\bibnamefont {Chen}}, \bibinfo {author}
  {\bibfnamefont {D.}~\bibnamefont {Shen}},\ and\ \bibinfo {author}
  {\bibfnamefont {Y.}~\bibnamefont {Nie}},\ }\bibfield  {title} {\bibinfo
  {title} {Electronic structure of superconducting infinite-layer lanthanum
  nickelates},\ }\href {https://doi.org/10.1126/sciadv.adr5116} {\bibfield
  {journal} {\bibinfo  {journal} {Sci. Adv.}\ }\textbf {\bibinfo {volume}
  {11}},\ \bibinfo {pages} {eadr5116} (\bibinfo {year} {2025})}\BibitemShut
  {NoStop}%
\bibitem [{\citenamefont {Pintschovius}\ \emph {et~al.}(1991)\citenamefont
  {Pintschovius}, \citenamefont {Pyka}, \citenamefont {Reichardt},
  \citenamefont {Rumiantsev}, \citenamefont {Mitrofanov}, \citenamefont
  {Ivanov}, \citenamefont {Collin},\ and\ \citenamefont
  {Bourges}}]{Pintschovius1991}%
  \BibitemOpen
  \bibfield  {author} {\bibinfo {author} {\bibfnamefont {L.}~\bibnamefont
  {Pintschovius}}, \bibinfo {author} {\bibfnamefont {N.}~\bibnamefont {Pyka}},
  \bibinfo {author} {\bibfnamefont {W.}~\bibnamefont {Reichardt}}, \bibinfo
  {author} {\bibfnamefont {A.}~\bibnamefont {Rumiantsev}}, \bibinfo {author}
  {\bibfnamefont {N.}~\bibnamefont {Mitrofanov}}, \bibinfo {author}
  {\bibfnamefont {A.}~\bibnamefont {Ivanov}}, \bibinfo {author} {\bibfnamefont
  {G.}~\bibnamefont {Collin}},\ and\ \bibinfo {author} {\bibfnamefont
  {P.}~\bibnamefont {Bourges}},\ }\bibfield  {title} {\bibinfo {title}
  {{Lattice dynamical studies of HTSC materials}},\ }\href
  {https://doi.org/https://doi.org/10.1016/0921-4534(91)91965-7} {\bibfield
  {journal} {\bibinfo  {journal} {Physica C: Superconductivity}\ }\textbf
  {\bibinfo {volume} {185-189}},\ \bibinfo {pages} {156} (\bibinfo {year}
  {1991})}\BibitemShut {NoStop}%
\bibitem [{\citenamefont {Fong}\ \emph {et~al.}(1995)\citenamefont {Fong},
  \citenamefont {Keimer}, \citenamefont {Anderson}, \citenamefont {Reznik},
  \citenamefont {Do\ifmmode~\breve{g}\else \u{g}\fi{}an},\ and\ \citenamefont
  {Aksay}}]{Fong1995}%
  \BibitemOpen
  \bibfield  {author} {\bibinfo {author} {\bibfnamefont {H.~F.}\ \bibnamefont
  {Fong}}, \bibinfo {author} {\bibfnamefont {B.}~\bibnamefont {Keimer}},
  \bibinfo {author} {\bibfnamefont {P.~W.}\ \bibnamefont {Anderson}}, \bibinfo
  {author} {\bibfnamefont {D.}~\bibnamefont {Reznik}}, \bibinfo {author}
  {\bibfnamefont {F.}~\bibnamefont {Do\ifmmode~\breve{g}\else \u{g}\fi{}an}},\
  and\ \bibinfo {author} {\bibfnamefont {I.~A.}\ \bibnamefont {Aksay}},\
  }\bibfield  {title} {\bibinfo {title} {{Phonon and Magnetic Neutron
  Scattering at 41 meV in
  YB${\mathrm{a}}_{2}$C${\mathrm{u}}_{3}$${\mathrm{O}}_{7}$}},\ }\href
  {https://doi.org/10.1103/PhysRevLett.75.316} {\bibfield  {journal} {\bibinfo
  {journal} {Phys. Rev. Lett.}\ }\textbf {\bibinfo {volume} {75}},\ \bibinfo
  {pages} {316} (\bibinfo {year} {1995})}\BibitemShut {NoStop}%
\bibitem [{\citenamefont {Reznik}\ \emph
  {et~al.}(2008{\natexlab{a}})\citenamefont {Reznik}, \citenamefont
  {Pintschovius}, \citenamefont {Tranquada}, \citenamefont {Arai},
  \citenamefont {Endoh}, \citenamefont {Masui},\ and\ \citenamefont
  {Tajima}}]{Reznik2008}%
  \BibitemOpen
  \bibfield  {author} {\bibinfo {author} {\bibfnamefont {D.}~\bibnamefont
  {Reznik}}, \bibinfo {author} {\bibfnamefont {L.}~\bibnamefont
  {Pintschovius}}, \bibinfo {author} {\bibfnamefont {J.~M.}\ \bibnamefont
  {Tranquada}}, \bibinfo {author} {\bibfnamefont {M.}~\bibnamefont {Arai}},
  \bibinfo {author} {\bibfnamefont {Y.}~\bibnamefont {Endoh}}, \bibinfo
  {author} {\bibfnamefont {T.}~\bibnamefont {Masui}},\ and\ \bibinfo {author}
  {\bibfnamefont {S.}~\bibnamefont {Tajima}},\ }\bibfield  {title} {\bibinfo
  {title} {{Temperature dependence of the bond-stretching phonon anomaly in
  ${\text{YBa}}_{2}{\text{Cu}}_{3}{\text{O}}_{6.95}$}},\ }\href
  {https://doi.org/10.1103/PhysRevB.78.094507} {\bibfield  {journal} {\bibinfo
  {journal} {Phys. Rev. B}\ }\textbf {\bibinfo {volume} {78}},\ \bibinfo
  {pages} {094507} (\bibinfo {year} {2008}{\natexlab{a}})}\BibitemShut
  {NoStop}%
\bibitem [{\citenamefont {Raichle}\ \emph {et~al.}(2011)\citenamefont
  {Raichle}, \citenamefont {Reznik}, \citenamefont {Lamago}, \citenamefont
  {Heid}, \citenamefont {Li}, \citenamefont {Bakr}, \citenamefont {Ulrich},
  \citenamefont {Hinkov}, \citenamefont {Hradil}, \citenamefont {Lin},\ and\
  \citenamefont {Keimer}}]{Raichle2011}%
  \BibitemOpen
  \bibfield  {author} {\bibinfo {author} {\bibfnamefont {M.}~\bibnamefont
  {Raichle}}, \bibinfo {author} {\bibfnamefont {D.}~\bibnamefont {Reznik}},
  \bibinfo {author} {\bibfnamefont {D.}~\bibnamefont {Lamago}}, \bibinfo
  {author} {\bibfnamefont {R.}~\bibnamefont {Heid}}, \bibinfo {author}
  {\bibfnamefont {Y.}~\bibnamefont {Li}}, \bibinfo {author} {\bibfnamefont
  {M.}~\bibnamefont {Bakr}}, \bibinfo {author} {\bibfnamefont {C.}~\bibnamefont
  {Ulrich}}, \bibinfo {author} {\bibfnamefont {V.}~\bibnamefont {Hinkov}},
  \bibinfo {author} {\bibfnamefont {K.}~\bibnamefont {Hradil}}, \bibinfo
  {author} {\bibfnamefont {C.~T.}\ \bibnamefont {Lin}},\ and\ \bibinfo {author}
  {\bibfnamefont {B.}~\bibnamefont {Keimer}},\ }\bibfield  {title} {\bibinfo
  {title} {{Highly Anisotropic Anomaly in the Dispersion of the Copper-Oxygen
  Bond-Bending Phonon in Superconducting
  ${\mathrm{YBa}}_{2}{\mathrm{Cu}}_{3}{\mathrm{O}}_{7}$ from Inelastic Neutron
  Scattering}},\ }\href {https://doi.org/10.1103/PhysRevLett.107.177004}
  {\bibfield  {journal} {\bibinfo  {journal} {Phys. Rev. Lett.}\ }\textbf
  {\bibinfo {volume} {107}},\ \bibinfo {pages} {177004} (\bibinfo {year}
  {2011})}\BibitemShut {NoStop}%
\bibitem [{\citenamefont {Ahmadova}\ \emph {et~al.}(2020)\citenamefont
  {Ahmadova}, \citenamefont {Sterling}, \citenamefont {Sokolik}, \citenamefont
  {Abernathy}, \citenamefont {Greven},\ and\ \citenamefont
  {Reznik}}]{Ahmadova2020}%
  \BibitemOpen
  \bibfield  {author} {\bibinfo {author} {\bibfnamefont {I.}~\bibnamefont
  {Ahmadova}}, \bibinfo {author} {\bibfnamefont {T.~C.}\ \bibnamefont
  {Sterling}}, \bibinfo {author} {\bibfnamefont {A.~C.}\ \bibnamefont
  {Sokolik}}, \bibinfo {author} {\bibfnamefont {D.~L.}\ \bibnamefont
  {Abernathy}}, \bibinfo {author} {\bibfnamefont {M.}~\bibnamefont {Greven}},\
  and\ \bibinfo {author} {\bibfnamefont {D.}~\bibnamefont {Reznik}},\
  }\bibfield  {title} {\bibinfo {title} {{Phonon spectrum of underdoped
  ${\mathrm{HgBa}}_{2}{\mathrm{CuO}}_{4+\ensuremath{\delta}}$ investigated by
  neutron scattering}},\ }\href {https://doi.org/10.1103/PhysRevB.101.184508}
  {\bibfield  {journal} {\bibinfo  {journal} {Phys. Rev. B}\ }\textbf {\bibinfo
  {volume} {101}},\ \bibinfo {pages} {184508} (\bibinfo {year}
  {2020})}\BibitemShut {NoStop}%
\bibitem [{\citenamefont {Sterling}\ and\ \citenamefont
  {Reznik}(2021)}]{Sterling2021}%
  \BibitemOpen
  \bibfield  {author} {\bibinfo {author} {\bibfnamefont {T.~C.}\ \bibnamefont
  {Sterling}}\ and\ \bibinfo {author} {\bibfnamefont {D.}~\bibnamefont
  {Reznik}},\ }\bibfield  {title} {\bibinfo {title} {{Effect of the electronic
  charge gap on LO bond-stretching phonons in undoped
  ${\mathrm{La}}_{2}{\mathrm{CuO}}_{4}$ calculated using
  $\mathrm{LDA}+\mathrm{U}$}},\ }\href
  {https://doi.org/10.1103/PhysRevB.104.134311} {\bibfield  {journal} {\bibinfo
   {journal} {Phys. Rev. B}\ }\textbf {\bibinfo {volume} {104}},\ \bibinfo
  {pages} {134311} (\bibinfo {year} {2021})}\BibitemShut {NoStop}%
\bibitem [{\citenamefont {Reichardt}\ \emph {et~al.}(1994)\citenamefont
  {Reichardt}, \citenamefont {Pintschovius}, \citenamefont {Pyka},
  \citenamefont {Schwei{\ss}}, \citenamefont {Erb}, \citenamefont {Bourges},
  \citenamefont {Collin}, \citenamefont {Rossat-Mignod}, \citenamefont {Henry},
  \citenamefont {Ivanov}, \citenamefont {Mitrofanov},\ and\ \citenamefont
  {Rumiantsev}}]{Reichardt1994}%
  \BibitemOpen
  \bibfield  {author} {\bibinfo {author} {\bibfnamefont {W.}~\bibnamefont
  {Reichardt}}, \bibinfo {author} {\bibfnamefont {L.}~\bibnamefont
  {Pintschovius}}, \bibinfo {author} {\bibfnamefont {N.}~\bibnamefont {Pyka}},
  \bibinfo {author} {\bibfnamefont {P.}~\bibnamefont {Schwei{\ss}}}, \bibinfo
  {author} {\bibfnamefont {A.}~\bibnamefont {Erb}}, \bibinfo {author}
  {\bibfnamefont {P.}~\bibnamefont {Bourges}}, \bibinfo {author} {\bibfnamefont
  {G.}~\bibnamefont {Collin}}, \bibinfo {author} {\bibfnamefont
  {J.}~\bibnamefont {Rossat-Mignod}}, \bibinfo {author} {\bibfnamefont {I.~Y.}\
  \bibnamefont {Henry}}, \bibinfo {author} {\bibfnamefont {A.~S.}\ \bibnamefont
  {Ivanov}}, \bibinfo {author} {\bibfnamefont {N.~L.}\ \bibnamefont
  {Mitrofanov}},\ and\ \bibinfo {author} {\bibfnamefont {A.~Y.}\ \bibnamefont
  {Rumiantsev}},\ }\bibfield  {title} {\bibinfo {title} {Anharmonicity and
  electron-phonon coupling in cuprate superconductors studied by inelastic
  neutron scattering},\ }\href {https://doi.org/10.1007/BF00724577} {\bibfield
  {journal} {\bibinfo  {journal} {Journal of Superconductivity}\ }\textbf
  {\bibinfo {volume} {7}},\ \bibinfo {pages} {399} (\bibinfo {year}
  {1994})}\BibitemShut {NoStop}%
\bibitem [{\citenamefont {Pintschovius}(2005)}]{Pintschovius2005}%
  \BibitemOpen
  \bibfield  {author} {\bibinfo {author} {\bibfnamefont {L.}~\bibnamefont
  {Pintschovius}},\ }\bibfield  {title} {\bibinfo {title} {Electron–phonon
  coupling effects explored by inelastic neutron scattering},\ }\href
  {https://doi.org/https://doi.org/10.1002/pssb.200404951} {\bibfield
  {journal} {\bibinfo  {journal} {phys. stat. sol. (b)}\ }\textbf {\bibinfo
  {volume} {242}},\ \bibinfo {pages} {30} (\bibinfo {year} {2005})}\BibitemShut
  {NoStop}%
\bibitem [{\citenamefont {Frano}\ \emph {et~al.}(2020)\citenamefont {Frano},
  \citenamefont {Blanco-Canosa}, \citenamefont {Keimer},\ and\ \citenamefont
  {Birgeneau}}]{Frano2020}%
  \BibitemOpen
  \bibfield  {author} {\bibinfo {author} {\bibfnamefont {A.}~\bibnamefont
  {Frano}}, \bibinfo {author} {\bibfnamefont {S.}~\bibnamefont
  {Blanco-Canosa}}, \bibinfo {author} {\bibfnamefont {B.}~\bibnamefont
  {Keimer}},\ and\ \bibinfo {author} {\bibfnamefont {R.~J.}\ \bibnamefont
  {Birgeneau}},\ }\bibfield  {title} {\bibinfo {title} {Charge ordering in
  superconducting copper oxides},\ }\href
  {https://doi.org/10.1088/1361-648X/ab6140} {\bibfield  {journal} {\bibinfo
  {journal} {J. Phys.: Condens. Matter}\ }\textbf {\bibinfo {volume} {32}},\
  \bibinfo {pages} {374005} (\bibinfo {year} {2020})}\BibitemShut {NoStop}%
\bibitem [{\citenamefont {Le~Tacon}\ \emph {et~al.}(2011)\citenamefont
  {Le~Tacon}, \citenamefont {Ghiringhelli}, \citenamefont {Chaloupka},
  \citenamefont {Sala}, \citenamefont {Hinkov}, \citenamefont {Haverkort},
  \citenamefont {Minola}, \citenamefont {Bakr}, \citenamefont {Zhou},
  \citenamefont {Blanco-Canosa}, \citenamefont {Monney}, \citenamefont {Song},
  \citenamefont {Sun}, \citenamefont {Lin}, \citenamefont {De~Luca},
  \citenamefont {Salluzzo}, \citenamefont {Khaliullin}, \citenamefont
  {Schmitt}, \citenamefont {Braicovich},\ and\ \citenamefont
  {Keimer}}]{LeTaconNatPhys2011}%
  \BibitemOpen
  \bibfield  {author} {\bibinfo {author} {\bibfnamefont {M.}~\bibnamefont
  {Le~Tacon}}, \bibinfo {author} {\bibfnamefont {G.}~\bibnamefont
  {Ghiringhelli}}, \bibinfo {author} {\bibfnamefont {J.}~\bibnamefont
  {Chaloupka}}, \bibinfo {author} {\bibfnamefont {M.~M.}\ \bibnamefont {Sala}},
  \bibinfo {author} {\bibfnamefont {V.}~\bibnamefont {Hinkov}}, \bibinfo
  {author} {\bibfnamefont {M.~W.}\ \bibnamefont {Haverkort}}, \bibinfo {author}
  {\bibfnamefont {M.}~\bibnamefont {Minola}}, \bibinfo {author} {\bibfnamefont
  {M.}~\bibnamefont {Bakr}}, \bibinfo {author} {\bibfnamefont {K.~J.}\
  \bibnamefont {Zhou}}, \bibinfo {author} {\bibfnamefont {S.}~\bibnamefont
  {Blanco-Canosa}}, \bibinfo {author} {\bibfnamefont {C.}~\bibnamefont
  {Monney}}, \bibinfo {author} {\bibfnamefont {Y.~T.}\ \bibnamefont {Song}},
  \bibinfo {author} {\bibfnamefont {G.~L.}\ \bibnamefont {Sun}}, \bibinfo
  {author} {\bibfnamefont {C.~T.}\ \bibnamefont {Lin}}, \bibinfo {author}
  {\bibfnamefont {G.~M.}\ \bibnamefont {De~Luca}}, \bibinfo {author}
  {\bibfnamefont {M.}~\bibnamefont {Salluzzo}}, \bibinfo {author}
  {\bibfnamefont {G.}~\bibnamefont {Khaliullin}}, \bibinfo {author}
  {\bibfnamefont {T.}~\bibnamefont {Schmitt}}, \bibinfo {author} {\bibfnamefont
  {L.}~\bibnamefont {Braicovich}},\ and\ \bibinfo {author} {\bibfnamefont
  {B.}~\bibnamefont {Keimer}},\ }\bibfield  {title} {\bibinfo {title} {Intense
  paramagnon excitations in a large family of high-temperature
  superconductors},\ }\href {https://doi.org/10.1038/nphys2041} {\bibfield
  {journal} {\bibinfo  {journal} {Nat. Phys.}\ }\textbf {\bibinfo {volume}
  {7}},\ \bibinfo {pages} {725} (\bibinfo {year} {2011})}\BibitemShut {NoStop}%
\bibitem [{\citenamefont {Dean}\ \emph {et~al.}(2013)\citenamefont {Dean},
  \citenamefont {Dellea}, \citenamefont {Springell}, \citenamefont
  {Yakhou-Harris}, \citenamefont {Kummer}, \citenamefont {Brookes},
  \citenamefont {Liu}, \citenamefont {Sun}, \citenamefont {Strle},
  \citenamefont {Schmitt}, \citenamefont {Braicovich}, \citenamefont
  {Ghiringhelli}, \citenamefont {Bo{\v{z}}ovi{\'{c}}},\ and\ \citenamefont
  {Hill}}]{DeanNatMat2013}%
  \BibitemOpen
  \bibfield  {author} {\bibinfo {author} {\bibfnamefont {M.~P.~M.}\
  \bibnamefont {Dean}}, \bibinfo {author} {\bibfnamefont {G.}~\bibnamefont
  {Dellea}}, \bibinfo {author} {\bibfnamefont {R.~S.}\ \bibnamefont
  {Springell}}, \bibinfo {author} {\bibfnamefont {F.}~\bibnamefont
  {Yakhou-Harris}}, \bibinfo {author} {\bibfnamefont {K.}~\bibnamefont
  {Kummer}}, \bibinfo {author} {\bibfnamefont {N.~B.}\ \bibnamefont {Brookes}},
  \bibinfo {author} {\bibfnamefont {X.}~\bibnamefont {Liu}}, \bibinfo {author}
  {\bibfnamefont {Y.-J.}\ \bibnamefont {Sun}}, \bibinfo {author} {\bibfnamefont
  {J.}~\bibnamefont {Strle}}, \bibinfo {author} {\bibfnamefont
  {T.}~\bibnamefont {Schmitt}}, \bibinfo {author} {\bibfnamefont
  {L.}~\bibnamefont {Braicovich}}, \bibinfo {author} {\bibfnamefont
  {G.}~\bibnamefont {Ghiringhelli}}, \bibinfo {author} {\bibfnamefont
  {I.}~\bibnamefont {Bo{\v{z}}ovi{\'{c}}}},\ and\ \bibinfo {author}
  {\bibfnamefont {J.~P.}\ \bibnamefont {Hill}},\ }\bibfield  {title} {\bibinfo
  {title} {{Persistence of magnetic excitations in La$_{2-x}$Sr$_x$CuO$_4$ from
  the undoped insulator to the heavily overdoped non-superconducting metal}},\
  }\href {https://doi.org/10.1038/nmat3723} {\bibfield  {journal} {\bibinfo
  {journal} {Nat. Mater.}\ }\textbf {\bibinfo {volume} {12}},\ \bibinfo {pages}
  {1019} (\bibinfo {year} {2013})}\BibitemShut {NoStop}%
\bibitem [{\citenamefont {Dai}\ \emph {et~al.}(2001)\citenamefont {Dai},
  \citenamefont {Mook}, \citenamefont {Hunt},\ and\ \citenamefont
  {Do\ifmmode~\breve{g}\else \u{g}\fi{}an}}]{dai01}%
  \BibitemOpen
  \bibfield  {author} {\bibinfo {author} {\bibfnamefont {P.}~\bibnamefont
  {Dai}}, \bibinfo {author} {\bibfnamefont {H.~A.}\ \bibnamefont {Mook}},
  \bibinfo {author} {\bibfnamefont {R.~D.}\ \bibnamefont {Hunt}},\ and\
  \bibinfo {author} {\bibfnamefont {F.}~\bibnamefont {Do\ifmmode~\breve{g}\else
  \u{g}\fi{}an}},\ }\bibfield  {title} {\bibinfo {title} {{Evolution of the
  resonance and incommensurate spin fluctuations in superconducting
  ${\mathrm{YBa}}_{2}{\mathrm{Cu}}_{3}{\mathrm{O}}_{6+x}$}},\ }\href
  {https://doi.org/10.1103/PhysRevB.63.054525} {\bibfield  {journal} {\bibinfo
  {journal} {Phys. Rev. B}\ }\textbf {\bibinfo {volume} {63}},\ \bibinfo
  {pages} {054525} (\bibinfo {year} {2001})}\BibitemShut {NoStop}%
\bibitem [{\citenamefont {Vilardi}\ \emph {et~al.}(2019)\citenamefont
  {Vilardi}, \citenamefont {Taranto},\ and\ \citenamefont
  {Metzner}}]{vilardi19}%
  \BibitemOpen
  \bibfield  {author} {\bibinfo {author} {\bibfnamefont {D.}~\bibnamefont
  {Vilardi}}, \bibinfo {author} {\bibfnamefont {C.}~\bibnamefont {Taranto}},\
  and\ \bibinfo {author} {\bibfnamefont {W.}~\bibnamefont {Metzner}},\
  }\bibfield  {title} {\bibinfo {title} {Antiferromagnetic and $d$-wave pairing
  correlations in the strongly interacting two-dimensional hubbard model from
  the functional renormalization group},\ }\href
  {https://doi.org/10.1103/PhysRevB.99.104501} {\bibfield  {journal} {\bibinfo
  {journal} {Phys. Rev. B}\ }\textbf {\bibinfo {volume} {99}},\ \bibinfo
  {pages} {104501} (\bibinfo {year} {2019})}\BibitemShut {NoStop}%
\bibitem [{\citenamefont {Cuk}\ \emph {et~al.}(2004)\citenamefont {Cuk},
  \citenamefont {Baumberger}, \citenamefont {Lu}, \citenamefont {Ingle},
  \citenamefont {Zhou}, \citenamefont {Eisaki}, \citenamefont {Kaneko},
  \citenamefont {Hussain}, \citenamefont {Devereaux}, \citenamefont {Nagaosa},\
  and\ \citenamefont {Shen}}]{cuk04}%
  \BibitemOpen
  \bibfield  {author} {\bibinfo {author} {\bibfnamefont {T.}~\bibnamefont
  {Cuk}}, \bibinfo {author} {\bibfnamefont {F.}~\bibnamefont {Baumberger}},
  \bibinfo {author} {\bibfnamefont {D.~H.}\ \bibnamefont {Lu}}, \bibinfo
  {author} {\bibfnamefont {N.}~\bibnamefont {Ingle}}, \bibinfo {author}
  {\bibfnamefont {X.~J.}\ \bibnamefont {Zhou}}, \bibinfo {author}
  {\bibfnamefont {H.}~\bibnamefont {Eisaki}}, \bibinfo {author} {\bibfnamefont
  {N.}~\bibnamefont {Kaneko}}, \bibinfo {author} {\bibfnamefont
  {Z.}~\bibnamefont {Hussain}}, \bibinfo {author} {\bibfnamefont {T.~P.}\
  \bibnamefont {Devereaux}}, \bibinfo {author} {\bibfnamefont {N.}~\bibnamefont
  {Nagaosa}},\ and\ \bibinfo {author} {\bibfnamefont {Z.-X.}\ \bibnamefont
  {Shen}},\ }\bibfield  {title} {\bibinfo {title} {{Coupling of the ${B}_{1g}$
  Phonon to the Antinodal Electronic States of
  ${\mathrm{B}\mathrm{i}}_{2}{\mathrm{S}\mathrm{r}}_{2}{\mathrm{C}\mathrm{a}}_{0.92}{\mathrm{Y}}_{0.08}{\mathrm{C}\mathrm{u}}_{2}{\mathrm{O}}_{8+\ensuremath{\delta}}$}},\
  }\href {https://doi.org/10.1103/PhysRevLett.93.117003} {\bibfield  {journal}
  {\bibinfo  {journal} {Phys. Rev. Lett.}\ }\textbf {\bibinfo {volume} {93}},\
  \bibinfo {pages} {117003} (\bibinfo {year} {2004})}\BibitemShut {NoStop}%
\bibitem [{\citenamefont {Heid}\ \emph {et~al.}(2009)\citenamefont {Heid},
  \citenamefont {Zeyher}, \citenamefont {Manske},\ and\ \citenamefont
  {Bohnen}}]{heid09}%
  \BibitemOpen
  \bibfield  {author} {\bibinfo {author} {\bibfnamefont {R.}~\bibnamefont
  {Heid}}, \bibinfo {author} {\bibfnamefont {R.}~\bibnamefont {Zeyher}},
  \bibinfo {author} {\bibfnamefont {D.}~\bibnamefont {Manske}},\ and\ \bibinfo
  {author} {\bibfnamefont {K.-P.}\ \bibnamefont {Bohnen}},\ }\bibfield  {title}
  {\bibinfo {title} {{Phonon-induced pairing interaction in
  ${\text{YBa}}_{2}{\text{Cu}}_{3}{\text{O}}_{7}$ within the local-density
  approximation}},\ }\href {https://doi.org/10.1103/PhysRevB.80.024507}
  {\bibfield  {journal} {\bibinfo  {journal} {Phys. Rev. B}\ }\textbf {\bibinfo
  {volume} {80}},\ \bibinfo {pages} {024507} (\bibinfo {year}
  {2009})}\BibitemShut {NoStop}%
\bibitem [{\citenamefont {Giustino}\ \emph {et~al.}(2008)\citenamefont
  {Giustino}, \citenamefont {Cohen},\ and\ \citenamefont {Louie}}]{giustino08}%
  \BibitemOpen
  \bibfield  {author} {\bibinfo {author} {\bibfnamefont {F.}~\bibnamefont
  {Giustino}}, \bibinfo {author} {\bibfnamefont {M.~L.}\ \bibnamefont
  {Cohen}},\ and\ \bibinfo {author} {\bibfnamefont {S.~G.}\ \bibnamefont
  {Louie}},\ }\bibfield  {title} {\bibinfo {title} {Small phonon contribution
  to the photoemission kink in the copper oxide superconductors},\ }\href
  {https://doi.org/10.1038/nature06874} {\bibfield  {journal} {\bibinfo
  {journal} {Nature}\ }\textbf {\bibinfo {volume} {452}},\ \bibinfo {pages}
  {975} (\bibinfo {year} {2008})}\BibitemShut {NoStop}%
\bibitem [{\citenamefont {Reznik}\ \emph
  {et~al.}(2008{\natexlab{b}})\citenamefont {Reznik}, \citenamefont
  {Sangiovanni}, \citenamefont {Gunnarsson},\ and\ \citenamefont
  {Devereaux}}]{reznik08}%
  \BibitemOpen
  \bibfield  {author} {\bibinfo {author} {\bibfnamefont {D.}~\bibnamefont
  {Reznik}}, \bibinfo {author} {\bibfnamefont {G.}~\bibnamefont {Sangiovanni}},
  \bibinfo {author} {\bibfnamefont {O.}~\bibnamefont {Gunnarsson}},\ and\
  \bibinfo {author} {\bibfnamefont {T.~P.}\ \bibnamefont {Devereaux}},\
  }\bibfield  {title} {\bibinfo {title} {Photoemission kinks and phonons in
  cuprates},\ }\href {https://doi.org/10.1038/nature07364} {\bibfield
  {journal} {\bibinfo  {journal} {Nature}\ }\textbf {\bibinfo {volume} {455}},\
  \bibinfo {pages} {E6} (\bibinfo {year} {2008}{\natexlab{b}})}\BibitemShut
  {NoStop}%
\bibitem [{\citenamefont {Johnston}\ \emph {et~al.}(2010)\citenamefont
  {Johnston}, \citenamefont {Vernay}, \citenamefont {Moritz}, \citenamefont
  {Shen}, \citenamefont {Nagaosa}, \citenamefont {Zaanen},\ and\ \citenamefont
  {Devereaux}}]{johnston10}%
  \BibitemOpen
  \bibfield  {author} {\bibinfo {author} {\bibfnamefont {S.}~\bibnamefont
  {Johnston}}, \bibinfo {author} {\bibfnamefont {F.}~\bibnamefont {Vernay}},
  \bibinfo {author} {\bibfnamefont {B.}~\bibnamefont {Moritz}}, \bibinfo
  {author} {\bibfnamefont {Z.-X.}\ \bibnamefont {Shen}}, \bibinfo {author}
  {\bibfnamefont {N.}~\bibnamefont {Nagaosa}}, \bibinfo {author} {\bibfnamefont
  {J.}~\bibnamefont {Zaanen}},\ and\ \bibinfo {author} {\bibfnamefont {T.~P.}\
  \bibnamefont {Devereaux}},\ }\bibfield  {title} {\bibinfo {title} {Systematic
  study of electron-phonon coupling to oxygen modes across the cuprates},\
  }\href {https://doi.org/10.1103/PhysRevB.82.064513} {\bibfield  {journal}
  {\bibinfo  {journal} {Phys. Rev. B}\ }\textbf {\bibinfo {volume} {82}},\
  \bibinfo {pages} {064513} (\bibinfo {year} {2010})}\BibitemShut {NoStop}%
\bibitem [{\citenamefont {Bauer}\ and\ \citenamefont
  {Falter}(2009)}]{Bauer2009}%
  \BibitemOpen
  \bibfield  {author} {\bibinfo {author} {\bibfnamefont {T.}~\bibnamefont
  {Bauer}}\ and\ \bibinfo {author} {\bibfnamefont {C.}~\bibnamefont {Falter}},\
  }\bibfield  {title} {\bibinfo {title} {{Impact of dynamical screening on the
  phonon dynamics of metallic ${\text{La}}_{2}{\text{CuO}}_{4}$}},\ }\href
  {https://doi.org/10.1103/PhysRevB.80.094525} {\bibfield  {journal} {\bibinfo
  {journal} {Phys. Rev. B}\ }\textbf {\bibinfo {volume} {80}},\ \bibinfo
  {pages} {094525} (\bibinfo {year} {2009})}\BibitemShut {NoStop}%
\bibitem [{\citenamefont {Harvey}\ \emph {et~al.}(2022)\citenamefont {Harvey},
  \citenamefont {Wang}, \citenamefont {Fowlie}, \citenamefont {Osada},
  \citenamefont {Lee}, \citenamefont {Lee}, \citenamefont {Li},\ and\
  \citenamefont {Hwang}}]{Harvey2022}%
  \BibitemOpen
  \bibfield  {author} {\bibinfo {author} {\bibfnamefont {S.~P.}\ \bibnamefont
  {Harvey}}, \bibinfo {author} {\bibfnamefont {B.~Y.}\ \bibnamefont {Wang}},
  \bibinfo {author} {\bibfnamefont {J.}~\bibnamefont {Fowlie}}, \bibinfo
  {author} {\bibfnamefont {M.}~\bibnamefont {Osada}}, \bibinfo {author}
  {\bibfnamefont {K.}~\bibnamefont {Lee}}, \bibinfo {author} {\bibfnamefont
  {Y.}~\bibnamefont {Lee}}, \bibinfo {author} {\bibfnamefont {D.}~\bibnamefont
  {Li}},\ and\ \bibinfo {author} {\bibfnamefont {H.~Y.}\ \bibnamefont
  {Hwang}},\ }\href {https://arxiv.org/abs/2201.12971} {\bibinfo {title}
  {Evidence for nodal superconductivity in infinite-layer nickelates}}
  (\bibinfo {year} {2022}),\ \Eprint {https://arxiv.org/abs/2201.12971}
  {arXiv:2201.12971 [cond-mat.supr-con]} \BibitemShut {NoStop}%
\bibitem [{\citenamefont {Chow}\ \emph {et~al.}(2023)\citenamefont {Chow},
  \citenamefont {Sudheesh}, \citenamefont {Luo}, \citenamefont {Nandi},
  \citenamefont {Heil}, \citenamefont {Deuschle}, \citenamefont {Zeng},
  \citenamefont {Zhang}, \citenamefont {Prakash}, \citenamefont {Du},
  \citenamefont {Lim}, \citenamefont {van Aken}, \citenamefont {Chia},\ and\
  \citenamefont {Ariando}}]{Chow2022Pairing}%
  \BibitemOpen
  \bibfield  {author} {\bibinfo {author} {\bibfnamefont {L.~E.}\ \bibnamefont
  {Chow}}, \bibinfo {author} {\bibfnamefont {S.~K.}\ \bibnamefont {Sudheesh}},
  \bibinfo {author} {\bibfnamefont {Z.~Y.}\ \bibnamefont {Luo}}, \bibinfo
  {author} {\bibfnamefont {P.}~\bibnamefont {Nandi}}, \bibinfo {author}
  {\bibfnamefont {T.}~\bibnamefont {Heil}}, \bibinfo {author} {\bibfnamefont
  {J.}~\bibnamefont {Deuschle}}, \bibinfo {author} {\bibfnamefont {S.~W.}\
  \bibnamefont {Zeng}}, \bibinfo {author} {\bibfnamefont {Z.~T.}\ \bibnamefont
  {Zhang}}, \bibinfo {author} {\bibfnamefont {S.}~\bibnamefont {Prakash}},
  \bibinfo {author} {\bibfnamefont {X.~M.}\ \bibnamefont {Du}}, \bibinfo
  {author} {\bibfnamefont {Z.~S.}\ \bibnamefont {Lim}}, \bibinfo {author}
  {\bibfnamefont {P.~A.}\ \bibnamefont {van Aken}}, \bibinfo {author}
  {\bibfnamefont {E.~E.~M.}\ \bibnamefont {Chia}},\ and\ \bibinfo {author}
  {\bibfnamefont {A.}~\bibnamefont {Ariando}},\ }\href
  {https://arxiv.org/abs/2201.10038} {\bibinfo {title} {Pairing symmetry in
  infinite-layer nickelate superconductor}} (\bibinfo {year} {2023}),\ \Eprint
  {https://arxiv.org/abs/2201.10038} {arXiv:2201.10038 [cond-mat.supr-con]}
  \BibitemShut {NoStop}%
\bibitem [{\citenamefont {Gu}\ \emph {et~al.}(2020)\citenamefont {Gu},
  \citenamefont {Li}, \citenamefont {Wan}, \citenamefont {Li}, \citenamefont
  {Guo}, \citenamefont {Yang}, \citenamefont {Li}, \citenamefont {Zhu},
  \citenamefont {Pan}, \citenamefont {Nie},\ and\ \citenamefont
  {Wen}}]{Gu2020}%
  \BibitemOpen
  \bibfield  {author} {\bibinfo {author} {\bibfnamefont {Q.}~\bibnamefont
  {Gu}}, \bibinfo {author} {\bibfnamefont {Y.}~\bibnamefont {Li}}, \bibinfo
  {author} {\bibfnamefont {S.}~\bibnamefont {Wan}}, \bibinfo {author}
  {\bibfnamefont {H.}~\bibnamefont {Li}}, \bibinfo {author} {\bibfnamefont
  {W.}~\bibnamefont {Guo}}, \bibinfo {author} {\bibfnamefont {H.}~\bibnamefont
  {Yang}}, \bibinfo {author} {\bibfnamefont {Q.}~\bibnamefont {Li}}, \bibinfo
  {author} {\bibfnamefont {X.}~\bibnamefont {Zhu}}, \bibinfo {author}
  {\bibfnamefont {X.}~\bibnamefont {Pan}}, \bibinfo {author} {\bibfnamefont
  {Y.}~\bibnamefont {Nie}},\ and\ \bibinfo {author} {\bibfnamefont {H.-H.}\
  \bibnamefont {Wen}},\ }\bibfield  {title} {\bibinfo {title} {{Single particle
  tunneling spectrum of superconducting Nd$_{1-x}$Sr$_{x}$NiO$_{2}$ thin
  films}},\ }\href {https://doi.org/10.1038/s41467-020-19908-1} {\bibfield
  {journal} {\bibinfo  {journal} {Nat. Commun.}\ }\textbf {\bibinfo {volume}
  {11}},\ \bibinfo {pages} {6027} (\bibinfo {year} {2020})}\BibitemShut
  {NoStop}%
\bibitem [{\citenamefont {Wang}\ \emph {et~al.}(2023)\citenamefont {Wang},
  \citenamefont {Xiong}, \citenamefont {Yan}, \citenamefont {Hu}, \citenamefont
  {Osada}, \citenamefont {Li}, \citenamefont {Hwang}, \citenamefont {Song},
  \citenamefont {Ma},\ and\ \citenamefont {Xue}}]{Wang2023STM}%
  \BibitemOpen
  \bibfield  {author} {\bibinfo {author} {\bibfnamefont {R.-F.}\ \bibnamefont
  {Wang}}, \bibinfo {author} {\bibfnamefont {Y.-L.}\ \bibnamefont {Xiong}},
  \bibinfo {author} {\bibfnamefont {H.}~\bibnamefont {Yan}}, \bibinfo {author}
  {\bibfnamefont {X.}~\bibnamefont {Hu}}, \bibinfo {author} {\bibfnamefont
  {M.}~\bibnamefont {Osada}}, \bibinfo {author} {\bibfnamefont
  {D.}~\bibnamefont {Li}}, \bibinfo {author} {\bibfnamefont {H.~Y.}\
  \bibnamefont {Hwang}}, \bibinfo {author} {\bibfnamefont {C.-L.}\ \bibnamefont
  {Song}}, \bibinfo {author} {\bibfnamefont {X.-C.}\ \bibnamefont {Ma}},\ and\
  \bibinfo {author} {\bibfnamefont {Q.-K.}\ \bibnamefont {Xue}},\ }\bibfield
  {title} {\bibinfo {title} {{Observation of Coulomb blockade and Coulomb
  staircases in superconducting
  ${\mathrm{Pr}}_{0.8}{\mathrm{Sr}}_{0.2}{\mathrm{NiO}}_{2}$ films}},\ }\href
  {https://doi.org/10.1103/PhysRevB.107.115411} {\bibfield  {journal} {\bibinfo
   {journal} {Phys. Rev. B}\ }\textbf {\bibinfo {volume} {107}},\ \bibinfo
  {pages} {115411} (\bibinfo {year} {2023})}\BibitemShut {NoStop}%
\bibitem [{\citenamefont {Choubey}\ and\ \citenamefont
  {Eremin}(2021)}]{Choubey2021}%
  \BibitemOpen
  \bibfield  {author} {\bibinfo {author} {\bibfnamefont {P.}~\bibnamefont
  {Choubey}}\ and\ \bibinfo {author} {\bibfnamefont {I.~M.}\ \bibnamefont
  {Eremin}},\ }\bibfield  {title} {\bibinfo {title} {Electronic theory for
  scanning tunneling microscopy spectra in infinite-layer nickelate
  superconductors},\ }\href {https://doi.org/10.1103/PhysRevB.104.144504}
  {\bibfield  {journal} {\bibinfo  {journal} {Phys. Rev. B}\ }\textbf {\bibinfo
  {volume} {104}},\ \bibinfo {pages} {144504} (\bibinfo {year}
  {2021})}\BibitemShut {NoStop}%
\bibitem [{\citenamefont {Kreisel}\ \emph {et~al.}(2022)\citenamefont
  {Kreisel}, \citenamefont {Andersen}, \citenamefont {R\o{}mer}, \citenamefont
  {Eremin},\ and\ \citenamefont {Lechermann}}]{Kreisel2022}%
  \BibitemOpen
  \bibfield  {author} {\bibinfo {author} {\bibfnamefont {A.}~\bibnamefont
  {Kreisel}}, \bibinfo {author} {\bibfnamefont {B.~M.}\ \bibnamefont
  {Andersen}}, \bibinfo {author} {\bibfnamefont {A.~T.}\ \bibnamefont
  {R\o{}mer}}, \bibinfo {author} {\bibfnamefont {I.~M.}\ \bibnamefont
  {Eremin}},\ and\ \bibinfo {author} {\bibfnamefont {F.}~\bibnamefont
  {Lechermann}},\ }\bibfield  {title} {\bibinfo {title} {Superconducting
  instabilities in strongly correlated infinite-layer nickelates},\ }\href
  {https://doi.org/10.1103/PhysRevLett.129.077002} {\bibfield  {journal}
  {\bibinfo  {journal} {Phys. Rev. Lett.}\ }\textbf {\bibinfo {volume} {129}},\
  \bibinfo {pages} {077002} (\bibinfo {year} {2022})}\BibitemShut {NoStop}%
\bibitem [{\citenamefont {Cheng}\ \emph {et~al.}(2024)\citenamefont {Cheng},
  \citenamefont {Cheng}, \citenamefont {Lee}, \citenamefont {Luo},
  \citenamefont {Chen}, \citenamefont {Lee}, \citenamefont {Wang},
  \citenamefont {Mootz}, \citenamefont {Perakis}, \citenamefont {Shen},
  \citenamefont {Hwang},\ and\ \citenamefont {Wang}}]{Cheng2024}%
  \BibitemOpen
  \bibfield  {author} {\bibinfo {author} {\bibfnamefont {B.}~\bibnamefont
  {Cheng}}, \bibinfo {author} {\bibfnamefont {D.}~\bibnamefont {Cheng}},
  \bibinfo {author} {\bibfnamefont {K.}~\bibnamefont {Lee}}, \bibinfo {author}
  {\bibfnamefont {L.}~\bibnamefont {Luo}}, \bibinfo {author} {\bibfnamefont
  {Z.}~\bibnamefont {Chen}}, \bibinfo {author} {\bibfnamefont {Y.}~\bibnamefont
  {Lee}}, \bibinfo {author} {\bibfnamefont {B.~Y.}\ \bibnamefont {Wang}},
  \bibinfo {author} {\bibfnamefont {M.}~\bibnamefont {Mootz}}, \bibinfo
  {author} {\bibfnamefont {I.~E.}\ \bibnamefont {Perakis}}, \bibinfo {author}
  {\bibfnamefont {Z.-X.}\ \bibnamefont {Shen}}, \bibinfo {author}
  {\bibfnamefont {H.~Y.}\ \bibnamefont {Hwang}},\ and\ \bibinfo {author}
  {\bibfnamefont {J.}~\bibnamefont {Wang}},\ }\bibfield  {title} {\bibinfo
  {title} {{Evidence for d-wave superconductivity of infinite-layer nickelates
  from low-energy electrodynamics}},\ }\href
  {https://doi.org/10.1038/s41563-023-01766-z} {\bibfield  {journal} {\bibinfo
  {journal} {Nat. Mater.}\ }\textbf {\bibinfo {volume} {23}},\ \bibinfo {pages}
  {775} (\bibinfo {year} {2024})}\BibitemShut {NoStop}%
\bibitem [{\citenamefont {Schilling}\ \emph {et~al.}(1993)\citenamefont
  {Schilling}, \citenamefont {Cantoni}, \citenamefont {Guo},\ and\
  \citenamefont {Ott}}]{Schilling1993}%
  \BibitemOpen
  \bibfield  {author} {\bibinfo {author} {\bibfnamefont {A.}~\bibnamefont
  {Schilling}}, \bibinfo {author} {\bibfnamefont {M.}~\bibnamefont {Cantoni}},
  \bibinfo {author} {\bibfnamefont {J.~D.}\ \bibnamefont {Guo}},\ and\ \bibinfo
  {author} {\bibfnamefont {H.~R.}\ \bibnamefont {Ott}},\ }\bibfield  {title}
  {\bibinfo {title} {{Superconductivity above 130 K in the
  Hg–Ba–Ca–Cu–O system}},\ }\href {https://doi.org/10.1038/363056a0}
  {\bibfield  {journal} {\bibinfo  {journal} {Nature}\ }\textbf {\bibinfo
  {volume} {363}},\ \bibinfo {pages} {56} (\bibinfo {year} {1993})}\BibitemShut
  {NoStop}%
\bibitem [{\citenamefont {Lee}\ \emph {et~al.}(2023)\citenamefont {Lee},
  \citenamefont {Wang}, \citenamefont {Osada}, \citenamefont {Goodge},
  \citenamefont {Wang}, \citenamefont {Lee}, \citenamefont {Harvey},
  \citenamefont {Kim}, \citenamefont {Yu}, \citenamefont {Murthy},
  \citenamefont {Raghu}, \citenamefont {Kourkoutis},\ and\ \citenamefont
  {Hwang}}]{Lee2023}%
  \BibitemOpen
  \bibfield  {author} {\bibinfo {author} {\bibfnamefont {K.}~\bibnamefont
  {Lee}}, \bibinfo {author} {\bibfnamefont {B.~Y.}\ \bibnamefont {Wang}},
  \bibinfo {author} {\bibfnamefont {M.}~\bibnamefont {Osada}}, \bibinfo
  {author} {\bibfnamefont {B.~H.}\ \bibnamefont {Goodge}}, \bibinfo {author}
  {\bibfnamefont {T.~C.}\ \bibnamefont {Wang}}, \bibinfo {author}
  {\bibfnamefont {Y.}~\bibnamefont {Lee}}, \bibinfo {author} {\bibfnamefont
  {S.}~\bibnamefont {Harvey}}, \bibinfo {author} {\bibfnamefont {W.~J.}\
  \bibnamefont {Kim}}, \bibinfo {author} {\bibfnamefont {Y.}~\bibnamefont
  {Yu}}, \bibinfo {author} {\bibfnamefont {C.}~\bibnamefont {Murthy}}, \bibinfo
  {author} {\bibfnamefont {S.}~\bibnamefont {Raghu}}, \bibinfo {author}
  {\bibfnamefont {L.~F.}\ \bibnamefont {Kourkoutis}},\ and\ \bibinfo {author}
  {\bibfnamefont {H.~Y.}\ \bibnamefont {Hwang}},\ }\bibfield  {title} {\bibinfo
  {title} {{Linear-in-temperature resistivity for optimally superconducting
  (Nd,Sr)NiO$_{2}$}},\ }\href {https://doi.org/10.1038/s41586-023-06129-x}
  {\bibfield  {journal} {\bibinfo  {journal} {Nature}\ }\textbf {\bibinfo
  {volume} {619}},\ \bibinfo {pages} {288} (\bibinfo {year}
  {2023})}\BibitemShut {NoStop}%
\bibitem [{\citenamefont {Xia}\ \emph {et~al.}(2022)\citenamefont {Xia},
  \citenamefont {Wu}, \citenamefont {Chen},\ and\ \citenamefont
  {Chen}}]{Xia2022}%
  \BibitemOpen
  \bibfield  {author} {\bibinfo {author} {\bibfnamefont {C.}~\bibnamefont
  {Xia}}, \bibinfo {author} {\bibfnamefont {J.}~\bibnamefont {Wu}}, \bibinfo
  {author} {\bibfnamefont {Y.}~\bibnamefont {Chen}},\ and\ \bibinfo {author}
  {\bibfnamefont {H.}~\bibnamefont {Chen}},\ }\bibfield  {title} {\bibinfo
  {title} {Dynamical structural instability and its implications for the
  physical properties of infinite-layer nickelates},\ }\href
  {https://doi.org/10.1103/PhysRevB.105.115134} {\bibfield  {journal} {\bibinfo
   {journal} {Phys. Rev. B}\ }\textbf {\bibinfo {volume} {105}},\ \bibinfo
  {pages} {115134} (\bibinfo {year} {2022})}\BibitemShut {NoStop}%
\bibitem [{\citenamefont {Carrasco~\'Alvarez}\ \emph
  {et~al.}(2022)\citenamefont {Carrasco~\'Alvarez}, \citenamefont {Petit},
  \citenamefont {Iglesias}, \citenamefont {Prellier}, \citenamefont {Bibes},\
  and\ \citenamefont {Varignon}}]{Carrasco2022}%
  \BibitemOpen
  \bibfield  {author} {\bibinfo {author} {\bibfnamefont {A.~A.}\ \bibnamefont
  {Carrasco~\'Alvarez}}, \bibinfo {author} {\bibfnamefont {S.}~\bibnamefont
  {Petit}}, \bibinfo {author} {\bibfnamefont {L.}~\bibnamefont {Iglesias}},
  \bibinfo {author} {\bibfnamefont {W.}~\bibnamefont {Prellier}}, \bibinfo
  {author} {\bibfnamefont {M.}~\bibnamefont {Bibes}},\ and\ \bibinfo {author}
  {\bibfnamefont {J.}~\bibnamefont {Varignon}},\ }\bibfield  {title} {\bibinfo
  {title} {Structural instabilities of infinite-layer nickelates from
  first-principles simulations},\ }\href
  {https://doi.org/10.1103/PhysRevResearch.4.023064} {\bibfield  {journal}
  {\bibinfo  {journal} {Phys. Rev. Res.}\ }\textbf {\bibinfo {volume} {4}},\
  \bibinfo {pages} {023064} (\bibinfo {year} {2022})}\BibitemShut {NoStop}%
\bibitem [{\citenamefont {Zhang}\ \emph {et~al.}(2023)\citenamefont {Zhang},
  \citenamefont {Zhang}, \citenamefont {Li}, \citenamefont {Sahoo},
  \citenamefont {He}, \citenamefont {Wang},\ and\ \citenamefont
  {Ghosez}}]{Zhang2023Phonon}%
  \BibitemOpen
  \bibfield  {author} {\bibinfo {author} {\bibfnamefont {Y.}~\bibnamefont
  {Zhang}}, \bibinfo {author} {\bibfnamefont {J.}~\bibnamefont {Zhang}},
  \bibinfo {author} {\bibfnamefont {J.}~\bibnamefont {Li}}, \bibinfo {author}
  {\bibfnamefont {M.~P.~K.}\ \bibnamefont {Sahoo}}, \bibinfo {author}
  {\bibfnamefont {X.}~\bibnamefont {He}}, \bibinfo {author} {\bibfnamefont
  {J.}~\bibnamefont {Wang}},\ and\ \bibinfo {author} {\bibfnamefont
  {P.}~\bibnamefont {Ghosez}},\ }\bibfield  {title} {\bibinfo {title}
  {Displacive phase transitions in infinite-layer nickelates from first- and
  second-principles calculations},\ }\href
  {https://doi.org/10.1103/PhysRevB.108.165117} {\bibfield  {journal} {\bibinfo
   {journal} {Phys. Rev. B}\ }\textbf {\bibinfo {volume} {108}},\ \bibinfo
  {pages} {165117} (\bibinfo {year} {2023})}\BibitemShut {NoStop}%
\bibitem [{\citenamefont {Meier}\ \emph {et~al.}(2024)\citenamefont {Meier},
  \citenamefont {de~Vaulx}, \citenamefont {Bernardini}, \citenamefont {Botana},
  \citenamefont {Blase}, \citenamefont {Olevano},\ and\ \citenamefont
  {Cano}}]{Meier2024}%
  \BibitemOpen
  \bibfield  {author} {\bibinfo {author} {\bibfnamefont {Q.~N.}\ \bibnamefont
  {Meier}}, \bibinfo {author} {\bibfnamefont {J.~B.}\ \bibnamefont {de~Vaulx}},
  \bibinfo {author} {\bibfnamefont {F.}~\bibnamefont {Bernardini}}, \bibinfo
  {author} {\bibfnamefont {A.~S.}\ \bibnamefont {Botana}}, \bibinfo {author}
  {\bibfnamefont {X.}~\bibnamefont {Blase}}, \bibinfo {author} {\bibfnamefont
  {V.}~\bibnamefont {Olevano}},\ and\ \bibinfo {author} {\bibfnamefont
  {A.}~\bibnamefont {Cano}},\ }\bibfield  {title} {\bibinfo {title} {Preempted
  phonon-mediated superconductivity in the infinite-layer nickelates},\ }\href
  {https://doi.org/10.1103/PhysRevB.109.184505} {\bibfield  {journal} {\bibinfo
   {journal} {Phys. Rev. B}\ }\textbf {\bibinfo {volume} {109}},\ \bibinfo
  {pages} {184505} (\bibinfo {year} {2024})}\BibitemShut {NoStop}%
\bibitem [{\citenamefont {Sakakibara}\ \emph {et~al.}(2025)\citenamefont
  {Sakakibara}, \citenamefont {Mizuno}, \citenamefont {Ochi}, \citenamefont
  {Usui},\ and\ \citenamefont {Kuroki}}]{Sakakibara2024}%
  \BibitemOpen
  \bibfield  {author} {\bibinfo {author} {\bibfnamefont {H.}~\bibnamefont
  {Sakakibara}}, \bibinfo {author} {\bibfnamefont {R.}~\bibnamefont {Mizuno}},
  \bibinfo {author} {\bibfnamefont {M.}~\bibnamefont {Ochi}}, \bibinfo {author}
  {\bibfnamefont {H.}~\bibnamefont {Usui}},\ and\ \bibinfo {author}
  {\bibfnamefont {K.}~\bibnamefont {Kuroki}},\ }\bibfield  {title} {\bibinfo
  {title} {{Theoretical study on the possibility of high Tc
  s\textsuperscript{\textpm}-wave superconductivity in the heavily hole-doped
  infinite layer nickelates}},\ }\href {https://doi.org/10.1103/nl4m-n5dx}
  {\bibfield  {journal} {\bibinfo  {journal} {Phys. Rev. B}\ }\textbf {\bibinfo
  {volume} {111}},\ \bibinfo {pages} {224511} (\bibinfo {year}
  {2025})}\BibitemShut {NoStop}%
\bibitem [{\citenamefont {Zhang}\ \emph {et~al.}(2024)\citenamefont {Zhang},
  \citenamefont {Lane}, \citenamefont {Nokelainen}, \citenamefont {Singh},
  \citenamefont {Barbiellini}, \citenamefont {Markiewicz}, \citenamefont
  {Bansil},\ and\ \citenamefont {Sun}}]{Zhang2024EPC}%
  \BibitemOpen
  \bibfield  {author} {\bibinfo {author} {\bibfnamefont {R.}~\bibnamefont
  {Zhang}}, \bibinfo {author} {\bibfnamefont {C.}~\bibnamefont {Lane}},
  \bibinfo {author} {\bibfnamefont {J.}~\bibnamefont {Nokelainen}}, \bibinfo
  {author} {\bibfnamefont {B.}~\bibnamefont {Singh}}, \bibinfo {author}
  {\bibfnamefont {B.}~\bibnamefont {Barbiellini}}, \bibinfo {author}
  {\bibfnamefont {R.~S.}\ \bibnamefont {Markiewicz}}, \bibinfo {author}
  {\bibfnamefont {A.}~\bibnamefont {Bansil}},\ and\ \bibinfo {author}
  {\bibfnamefont {J.}~\bibnamefont {Sun}},\ }\bibfield  {title} {\bibinfo
  {title} {{Emergence of Competing Stripe Phases in Undoped Infinite-Layer
  Nickelates}},\ }\href {https://doi.org/10.1103/PhysRevLett.133.066401}
  {\bibfield  {journal} {\bibinfo  {journal} {Phys. Rev. Lett.}\ }\textbf
  {\bibinfo {volume} {133}},\ \bibinfo {pages} {066401} (\bibinfo {year}
  {2024})}\BibitemShut {NoStop}%
\bibitem [{\citenamefont {Li}\ and\ \citenamefont
  {Louie}(2024)}]{Li2024Phonon}%
  \BibitemOpen
  \bibfield  {author} {\bibinfo {author} {\bibfnamefont {Z.}~\bibnamefont
  {Li}}\ and\ \bibinfo {author} {\bibfnamefont {S.~G.}\ \bibnamefont {Louie}},\
  }\bibfield  {title} {\bibinfo {title} {{Two-Gap Superconductivity and the
  Decisive Role of Rare-Earth $d$ Electrons in Infinite-Layer Nickelates}},\
  }\href {https://doi.org/10.1103/PhysRevLett.133.126401} {\bibfield  {journal}
  {\bibinfo  {journal} {Phys. Rev. Lett.}\ }\textbf {\bibinfo {volume} {133}},\
  \bibinfo {pages} {126401} (\bibinfo {year} {2024})}\BibitemShut {NoStop}%
\bibitem [{\citenamefont {Carrasco~\'Alvarez}\ \emph
  {et~al.}(2024)\citenamefont {Carrasco~\'Alvarez}, \citenamefont {Iglesias},
  \citenamefont {Petit}, \citenamefont {Prellier}, \citenamefont {Bibes},\ and\
  \citenamefont {Varignon}}]{Alvarez2024}%
  \BibitemOpen
  \bibfield  {author} {\bibinfo {author} {\bibfnamefont {A.~A.}\ \bibnamefont
  {Carrasco~\'Alvarez}}, \bibinfo {author} {\bibfnamefont {L.}~\bibnamefont
  {Iglesias}}, \bibinfo {author} {\bibfnamefont {S.}~\bibnamefont {Petit}},
  \bibinfo {author} {\bibfnamefont {W.}~\bibnamefont {Prellier}}, \bibinfo
  {author} {\bibfnamefont {M.}~\bibnamefont {Bibes}},\ and\ \bibinfo {author}
  {\bibfnamefont {J.}~\bibnamefont {Varignon}},\ }\bibfield  {title} {\bibinfo
  {title} {{Charge ordering as the driving mechanism for superconductivity in
  rare-earth nickel oxides}},\ }\href
  {https://doi.org/10.1103/PhysRevMaterials.8.064801} {\bibfield  {journal}
  {\bibinfo  {journal} {Phys. Rev. Mater.}\ }\textbf {\bibinfo {volume} {8}},\
  \bibinfo {pages} {064801} (\bibinfo {year} {2024})}\BibitemShut {NoStop}%
\bibitem [{\citenamefont {Sui}\ \emph {et~al.}(2023)\citenamefont {Sui},
  \citenamefont {Wang}, \citenamefont {Chen}, \citenamefont {Ding},
  \citenamefont {Zhou}, \citenamefont {Cao}, \citenamefont {Qiao},
  \citenamefont {Lin},\ and\ \citenamefont {Huang}}]{Sui2023}%
  \BibitemOpen
  \bibfield  {author} {\bibinfo {author} {\bibfnamefont {X.}~\bibnamefont
  {Sui}}, \bibinfo {author} {\bibfnamefont {J.}~\bibnamefont {Wang}}, \bibinfo
  {author} {\bibfnamefont {C.}~\bibnamefont {Chen}}, \bibinfo {author}
  {\bibfnamefont {X.}~\bibnamefont {Ding}}, \bibinfo {author} {\bibfnamefont
  {K.-J.}\ \bibnamefont {Zhou}}, \bibinfo {author} {\bibfnamefont
  {C.}~\bibnamefont {Cao}}, \bibinfo {author} {\bibfnamefont {L.}~\bibnamefont
  {Qiao}}, \bibinfo {author} {\bibfnamefont {H.}~\bibnamefont {Lin}},\ and\
  \bibinfo {author} {\bibfnamefont {B.}~\bibnamefont {Huang}},\ }\bibfield
  {title} {\bibinfo {title} {Hole doping dependent electronic instability and
  electron-phonon coupling in infinite-layer nickelates},\ }\href
  {https://doi.org/10.1103/PhysRevB.107.075159} {\bibfield  {journal} {\bibinfo
   {journal} {Phys. Rev. B}\ }\textbf {\bibinfo {volume} {107}},\ \bibinfo
  {pages} {075159} (\bibinfo {year} {2023})}\BibitemShut {NoStop}%
\bibitem [{\citenamefont {Craco}(2025)}]{Craco2025}%
  \BibitemOpen
  \bibfield  {author} {\bibinfo {author} {\bibfnamefont {L.}~\bibnamefont
  {Craco}},\ }\bibfield  {title} {\bibinfo {title} {{Orbital-Nematic and
  Two-Fluid Superconductivity in Hole-Doped NdNiO$_2$}},\ }\bibfield  {journal}
  {\bibinfo  {journal} {Condens. Matter}\ }\textbf {\bibinfo {volume} {10}},\
  \href {https://doi.org/10.3390/condmat10010018} {10.3390/condmat10010018}
  (\bibinfo {year} {2025})\BibitemShut {NoStop}%
\bibitem [{\citenamefont {Di~Cataldo}\ \emph {et~al.}(2023)\citenamefont
  {Di~Cataldo}, \citenamefont {Worm}, \citenamefont {Si},\ and\ \citenamefont
  {Held}}]{DiCataldo2023}%
  \BibitemOpen
  \bibfield  {author} {\bibinfo {author} {\bibfnamefont {S.}~\bibnamefont
  {Di~Cataldo}}, \bibinfo {author} {\bibfnamefont {P.}~\bibnamefont {Worm}},
  \bibinfo {author} {\bibfnamefont {L.}~\bibnamefont {Si}},\ and\ \bibinfo
  {author} {\bibfnamefont {K.}~\bibnamefont {Held}},\ }\bibfield  {title}
  {\bibinfo {title} {Absence of electron-phonon-mediated superconductivity in
  hydrogen-intercalated nickelates},\ }\href
  {https://doi.org/10.1103/PhysRevB.108.174512} {\bibfield  {journal} {\bibinfo
   {journal} {Phys. Rev. B}\ }\textbf {\bibinfo {volume} {108}},\ \bibinfo
  {pages} {174512} (\bibinfo {year} {2023})}\BibitemShut {NoStop}%
\bibitem [{\citenamefont {Kitatani}\ \emph {et~al.}(2023)\citenamefont
  {Kitatani}, \citenamefont {Si}, \citenamefont {Worm}, \citenamefont
  {Tomczak}, \citenamefont {Arita},\ and\ \citenamefont {Held}}]{Kitatani2023}%
  \BibitemOpen
  \bibfield  {author} {\bibinfo {author} {\bibfnamefont {M.}~\bibnamefont
  {Kitatani}}, \bibinfo {author} {\bibfnamefont {L.}~\bibnamefont {Si}},
  \bibinfo {author} {\bibfnamefont {P.}~\bibnamefont {Worm}}, \bibinfo {author}
  {\bibfnamefont {J.~M.}\ \bibnamefont {Tomczak}}, \bibinfo {author}
  {\bibfnamefont {R.}~\bibnamefont {Arita}},\ and\ \bibinfo {author}
  {\bibfnamefont {K.}~\bibnamefont {Held}},\ }\bibfield  {title} {\bibinfo
  {title} {{Optimizing Superconductivity: From Cuprates via Nickelates to
  Palladates}},\ }\href {https://doi.org/10.1103/PhysRevLett.130.166002}
  {\bibfield  {journal} {\bibinfo  {journal} {Phys. Rev. Lett.}\ }\textbf
  {\bibinfo {volume} {130}},\ \bibinfo {pages} {166002} (\bibinfo {year}
  {2023})}\BibitemShut {NoStop}%
\bibitem [{\citenamefont {Adhikary}\ \emph {et~al.}(2020)\citenamefont
  {Adhikary}, \citenamefont {Bandyopadhyay}, \citenamefont {Das}, \citenamefont
  {Dasgupta},\ and\ \citenamefont {Saha-Dasgupta}}]{Adhikary2020}%
  \BibitemOpen
  \bibfield  {author} {\bibinfo {author} {\bibfnamefont {P.}~\bibnamefont
  {Adhikary}}, \bibinfo {author} {\bibfnamefont {S.}~\bibnamefont
  {Bandyopadhyay}}, \bibinfo {author} {\bibfnamefont {T.}~\bibnamefont {Das}},
  \bibinfo {author} {\bibfnamefont {I.}~\bibnamefont {Dasgupta}},\ and\
  \bibinfo {author} {\bibfnamefont {T.}~\bibnamefont {Saha-Dasgupta}},\
  }\bibfield  {title} {\bibinfo {title} {Orbital-selective superconductivity in
  a two-band model of infinite-layer nickelates},\ }\href
  {https://doi.org/10.1103/PhysRevB.102.100501} {\bibfield  {journal} {\bibinfo
   {journal} {Phys. Rev. B}\ }\textbf {\bibinfo {volume} {102}},\ \bibinfo
  {pages} {100501} (\bibinfo {year} {2020})}\BibitemShut {NoStop}%
\bibitem [{\citenamefont {Si}\ \emph {et~al.}(2023)\citenamefont {Si},
  \citenamefont {Worm}, \citenamefont {Chen},\ and\ \citenamefont
  {Held}}]{Si2023}%
  \BibitemOpen
  \bibfield  {author} {\bibinfo {author} {\bibfnamefont {L.}~\bibnamefont
  {Si}}, \bibinfo {author} {\bibfnamefont {P.}~\bibnamefont {Worm}}, \bibinfo
  {author} {\bibfnamefont {D.}~\bibnamefont {Chen}},\ and\ \bibinfo {author}
  {\bibfnamefont {K.}~\bibnamefont {Held}},\ }\bibfield  {title} {\bibinfo
  {title} {{Topotactic hydrogen forms chains in $AB{\mathrm{O}}_{2}$ nickelate
  superconductors}},\ }\href {https://doi.org/10.1103/PhysRevB.107.165116}
  {\bibfield  {journal} {\bibinfo  {journal} {Phys. Rev. B}\ }\textbf {\bibinfo
  {volume} {107}},\ \bibinfo {pages} {165116} (\bibinfo {year}
  {2023})}\BibitemShut {NoStop}%
\bibitem [{\citenamefont {Sharma}\ \emph {et~al.}(2024)\citenamefont {Sharma},
  \citenamefont {Jung}, \citenamefont {LaBollita},\ and\ \citenamefont
  {Botana}}]{Sharma2024}%
  \BibitemOpen
  \bibfield  {author} {\bibinfo {author} {\bibfnamefont {S.}~\bibnamefont
  {Sharma}}, \bibinfo {author} {\bibfnamefont {M.-C.}\ \bibnamefont {Jung}},
  \bibinfo {author} {\bibfnamefont {H.}~\bibnamefont {LaBollita}},\ and\
  \bibinfo {author} {\bibfnamefont {A.~S.}\ \bibnamefont {Botana}},\ }\bibfield
   {title} {\bibinfo {title} {Pressure effects on the electronic structure and
  magnetic properties of infinite-layer nickelates},\ }\href
  {https://doi.org/10.1103/PhysRevB.110.155156} {\bibfield  {journal} {\bibinfo
   {journal} {Phys. Rev. B}\ }\textbf {\bibinfo {volume} {110}},\ \bibinfo
  {pages} {155156} (\bibinfo {year} {2024})}\BibitemShut {NoStop}%
\bibitem [{\citenamefont {Wang}(2025)}]{Wang2025}%
  \BibitemOpen
  \bibfield  {author} {\bibinfo {author} {\bibfnamefont {Y.}~\bibnamefont
  {Wang}},\ }\bibfield  {title} {\bibinfo {title} {{Hardening of Ni-O
  bond-stretching phonons in ${\mathrm{LaNiO}}_{2}$}},\ }\href
  {https://doi.org/10.1103/PhysRevB.111.085117} {\bibfield  {journal} {\bibinfo
   {journal} {Phys. Rev. B}\ }\textbf {\bibinfo {volume} {111}},\ \bibinfo
  {pages} {085117} (\bibinfo {year} {2025})}\BibitemShut {NoStop}%
\bibitem [{\citenamefont {Zhang}\ \emph {et~al.}(2025)\citenamefont {Zhang},
  \citenamefont {Wang}, \citenamefont {Engel}, \citenamefont {Lane},
  \citenamefont {Miranda}, \citenamefont {Hou}, \citenamefont {Chowdhury},
  \citenamefont {Singh}, \citenamefont {Barbiellini}, \citenamefont {Zhu},
  \citenamefont {Markiewicz}, \citenamefont {Gross}, \citenamefont {Kresse},
  \citenamefont {Bansil},\ and\ \citenamefont {Sun}}]{Zhang2025EPC}%
  \BibitemOpen
  \bibfield  {author} {\bibinfo {author} {\bibfnamefont {R.}~\bibnamefont
  {Zhang}}, \bibinfo {author} {\bibfnamefont {Y.}~\bibnamefont {Wang}},
  \bibinfo {author} {\bibfnamefont {M.}~\bibnamefont {Engel}}, \bibinfo
  {author} {\bibfnamefont {C.}~\bibnamefont {Lane}}, \bibinfo {author}
  {\bibfnamefont {H.}~\bibnamefont {Miranda}}, \bibinfo {author} {\bibfnamefont
  {L.}~\bibnamefont {Hou}}, \bibinfo {author} {\bibfnamefont {S.}~\bibnamefont
  {Chowdhury}}, \bibinfo {author} {\bibfnamefont {B.}~\bibnamefont {Singh}},
  \bibinfo {author} {\bibfnamefont {B.}~\bibnamefont {Barbiellini}}, \bibinfo
  {author} {\bibfnamefont {J.-X.}\ \bibnamefont {Zhu}}, \bibinfo {author}
  {\bibfnamefont {R.~S.}\ \bibnamefont {Markiewicz}}, \bibinfo {author}
  {\bibfnamefont {E.~K.~U.}\ \bibnamefont {Gross}}, \bibinfo {author}
  {\bibfnamefont {G.}~\bibnamefont {Kresse}}, \bibinfo {author} {\bibfnamefont
  {A.}~\bibnamefont {Bansil}},\ and\ \bibinfo {author} {\bibfnamefont
  {J.}~\bibnamefont {Sun}},\ }\href {https://arxiv.org/abs/2504.13025}
  {\bibinfo {title} {{Magnetism-Enhanced Strong Electron-Phonon Coupling in
  Infinite-Layer Nickelate}}} (\bibinfo {year} {2025}),\ \Eprint
  {https://arxiv.org/abs/2504.13025} {arXiv:2504.13025 [cond-mat.str-el]}
  \BibitemShut {NoStop}%
\bibitem [{\citenamefont {Cervasio}\ \emph {et~al.}(2023)\citenamefont
  {Cervasio}, \citenamefont {Tomarchio}, \citenamefont {Verseils},
  \citenamefont {Brubach}, \citenamefont {Macis}, \citenamefont {Zeng},
  \citenamefont {Ariando}, \citenamefont {Roy},\ and\ \citenamefont
  {Lupi}}]{Cervasio2023}%
  \BibitemOpen
  \bibfield  {author} {\bibinfo {author} {\bibfnamefont {R.}~\bibnamefont
  {Cervasio}}, \bibinfo {author} {\bibfnamefont {L.}~\bibnamefont {Tomarchio}},
  \bibinfo {author} {\bibfnamefont {M.}~\bibnamefont {Verseils}}, \bibinfo
  {author} {\bibfnamefont {J.-B.}\ \bibnamefont {Brubach}}, \bibinfo {author}
  {\bibfnamefont {S.}~\bibnamefont {Macis}}, \bibinfo {author} {\bibfnamefont
  {S.}~\bibnamefont {Zeng}}, \bibinfo {author} {\bibfnamefont {A.}~\bibnamefont
  {Ariando}}, \bibinfo {author} {\bibfnamefont {P.}~\bibnamefont {Roy}},\ and\
  \bibinfo {author} {\bibfnamefont {S.}~\bibnamefont {Lupi}},\ }\bibfield
  {title} {\bibinfo {title} {{Optical Properties of Superconducting
  Nd$_{0.8}$Sr$_{0.2}$NiO$_{2}$ Nickelate}},\ }\href
  {https://doi.org/10.1021/acsaelm.3c00506} {\bibfield  {journal} {\bibinfo
  {journal} {ACS Appl. Electron. Mater.}\ }\textbf {\bibinfo {volume} {5}},\
  \bibinfo {pages} {4770} (\bibinfo {year} {2023})}\BibitemShut {NoStop}%
\bibitem [{\citenamefont {Burns}\ \emph {et~al.}(1989)\citenamefont {Burns},
  \citenamefont {Crawford}, \citenamefont {Dacol}, \citenamefont {McCarron},\
  and\ \citenamefont {Shaw}}]{Burns1989}%
  \BibitemOpen
  \bibfield  {author} {\bibinfo {author} {\bibfnamefont {G.}~\bibnamefont
  {Burns}}, \bibinfo {author} {\bibfnamefont {M.~K.}\ \bibnamefont {Crawford}},
  \bibinfo {author} {\bibfnamefont {F.~H.}\ \bibnamefont {Dacol}}, \bibinfo
  {author} {\bibfnamefont {E.~M.}\ \bibnamefont {McCarron}},\ and\ \bibinfo
  {author} {\bibfnamefont {T.~M.}\ \bibnamefont {Shaw}},\ }\bibfield  {title}
  {\bibinfo {title} {{Phonons in ${\mathrm{CaCuO}}_{2}$}},\ }\href
  {https://doi.org/10.1103/PhysRevB.40.6717} {\bibfield  {journal} {\bibinfo
  {journal} {Phys. Rev. B}\ }\textbf {\bibinfo {volume} {40}},\ \bibinfo
  {pages} {6717} (\bibinfo {year} {1989})}\BibitemShut {NoStop}%
\bibitem [{\citenamefont {Puphal}\ \emph {et~al.}(2021)\citenamefont {Puphal},
  \citenamefont {Wu}, \citenamefont {F{\"{u}}rsich}, \citenamefont {Lee},
  \citenamefont {Pakdaman}, \citenamefont {Bruin}, \citenamefont {Nuss},
  \citenamefont {Suyolcu}, \citenamefont {van Aken}, \citenamefont {Keimer},
  \citenamefont {Isobe},\ and\ \citenamefont {Hepting}}]{Puphal2021}%
  \BibitemOpen
  \bibfield  {author} {\bibinfo {author} {\bibfnamefont {P.}~\bibnamefont
  {Puphal}}, \bibinfo {author} {\bibfnamefont {Y.-M.}\ \bibnamefont {Wu}},
  \bibinfo {author} {\bibfnamefont {K.}~\bibnamefont {F{\"{u}}rsich}}, \bibinfo
  {author} {\bibfnamefont {H.}~\bibnamefont {Lee}}, \bibinfo {author}
  {\bibfnamefont {M.}~\bibnamefont {Pakdaman}}, \bibinfo {author}
  {\bibfnamefont {J.~A.~N.}\ \bibnamefont {Bruin}}, \bibinfo {author}
  {\bibfnamefont {J.}~\bibnamefont {Nuss}}, \bibinfo {author} {\bibfnamefont
  {Y.~E.}\ \bibnamefont {Suyolcu}}, \bibinfo {author} {\bibfnamefont {P.~A.}\
  \bibnamefont {van Aken}}, \bibinfo {author} {\bibfnamefont {B.}~\bibnamefont
  {Keimer}}, \bibinfo {author} {\bibfnamefont {M.}~\bibnamefont {Isobe}},\ and\
  \bibinfo {author} {\bibfnamefont {M.}~\bibnamefont {Hepting}},\ }\bibfield
  {title} {\bibinfo {title} {{Topotactic transformation of single crystals:
  From perovskite to infinite-layer nickelates}},\ }\href
  {https://doi.org/10.1126/sciadv.abl8091} {\bibfield  {journal} {\bibinfo
  {journal} {Sci. Adv.}\ }\textbf {\bibinfo {volume} {7}},\ \bibinfo {pages}
  {eabl8091} (\bibinfo {year} {2021})}\BibitemShut {NoStop}%
\bibitem [{\citenamefont {Baroni}\ \emph {et~al.}(2001)\citenamefont {Baroni},
  \citenamefont {de~Gironcoli}, \citenamefont {Dal~Corso},\ and\ \citenamefont
  {Giannozzi}}]{RevModPhys.73.515}%
  \BibitemOpen
  \bibfield  {author} {\bibinfo {author} {\bibfnamefont {S.}~\bibnamefont
  {Baroni}}, \bibinfo {author} {\bibfnamefont {S.}~\bibnamefont
  {de~Gironcoli}}, \bibinfo {author} {\bibfnamefont {A.}~\bibnamefont
  {Dal~Corso}},\ and\ \bibinfo {author} {\bibfnamefont {P.}~\bibnamefont
  {Giannozzi}},\ }\bibfield  {title} {\bibinfo {title} {Phonons and related
  crystal properties from density-functional perturbation theory},\ }\href
  {https://doi.org/10.1103/RevModPhys.73.515} {\bibfield  {journal} {\bibinfo
  {journal} {Rev. Mod. Phys.}\ }\textbf {\bibinfo {volume} {73}},\ \bibinfo
  {pages} {515} (\bibinfo {year} {2001})}\BibitemShut {NoStop}%
\bibitem [{\citenamefont {Puphal}\ \emph
  {et~al.}(2023{\natexlab{a}})\citenamefont {Puphal}, \citenamefont {Wehinger},
  \citenamefont {Nuss}, \citenamefont {K\"uster}, \citenamefont {Starke},
  \citenamefont {Garbarino}, \citenamefont {Keimer}, \citenamefont {Isobe},\
  and\ \citenamefont {Hepting}}]{Puphal2023}%
  \BibitemOpen
  \bibfield  {author} {\bibinfo {author} {\bibfnamefont {P.}~\bibnamefont
  {Puphal}}, \bibinfo {author} {\bibfnamefont {B.}~\bibnamefont {Wehinger}},
  \bibinfo {author} {\bibfnamefont {J.}~\bibnamefont {Nuss}}, \bibinfo {author}
  {\bibfnamefont {K.}~\bibnamefont {K\"uster}}, \bibinfo {author}
  {\bibfnamefont {U.}~\bibnamefont {Starke}}, \bibinfo {author} {\bibfnamefont
  {G.}~\bibnamefont {Garbarino}}, \bibinfo {author} {\bibfnamefont
  {B.}~\bibnamefont {Keimer}}, \bibinfo {author} {\bibfnamefont
  {M.}~\bibnamefont {Isobe}},\ and\ \bibinfo {author} {\bibfnamefont
  {M.}~\bibnamefont {Hepting}},\ }\bibfield  {title} {\bibinfo {title}
  {{Synthesis and physical properties of ${\mathrm{LaNiO}}_{2}$ crystals}},\
  }\href {https://doi.org/10.1103/PhysRevMaterials.7.014804} {\bibfield
  {journal} {\bibinfo  {journal} {Phys. Rev. Mater.}\ }\textbf {\bibinfo
  {volume} {7}},\ \bibinfo {pages} {014804} (\bibinfo {year}
  {2023}{\natexlab{a}})}\BibitemShut {NoStop}%
\bibitem [{\citenamefont {Puphal}\ \emph
  {et~al.}(2023{\natexlab{b}})\citenamefont {Puphal}, \citenamefont
  {Sundaramurthy}, \citenamefont {Zimmermann}, \citenamefont {Küster},
  \citenamefont {Starke}, \citenamefont {Isobe}, \citenamefont {Keimer},\ and\
  \citenamefont {Hepting}}]{Puphal2023a}%
  \BibitemOpen
  \bibfield  {author} {\bibinfo {author} {\bibfnamefont {P.}~\bibnamefont
  {Puphal}}, \bibinfo {author} {\bibfnamefont {V.}~\bibnamefont
  {Sundaramurthy}}, \bibinfo {author} {\bibfnamefont {V.}~\bibnamefont
  {Zimmermann}}, \bibinfo {author} {\bibfnamefont {K.}~\bibnamefont {Küster}},
  \bibinfo {author} {\bibfnamefont {U.}~\bibnamefont {Starke}}, \bibinfo
  {author} {\bibfnamefont {M.}~\bibnamefont {Isobe}}, \bibinfo {author}
  {\bibfnamefont {B.}~\bibnamefont {Keimer}},\ and\ \bibinfo {author}
  {\bibfnamefont {M.}~\bibnamefont {Hepting}},\ }\bibfield  {title} {\bibinfo
  {title} {{Phase formation in hole- and electron-doped rare-earth nickelate
  single crystals}},\ }\href {https://doi.org/10.1063/5.0160912} {\bibfield
  {journal} {\bibinfo  {journal} {APL Mater.}\ }\textbf {\bibinfo {volume}
  {11}},\ \bibinfo {pages} {081107} (\bibinfo {year}
  {2023}{\natexlab{b}})}\BibitemShut {NoStop}%
\bibitem [{\citenamefont {Hayashida}\ \emph {et~al.}(2024)\citenamefont
  {Hayashida}, \citenamefont {Sundaramurthy}, \citenamefont {Puphal},
  \citenamefont {Garcia-Fernandez}, \citenamefont {Zhou}, \citenamefont {Fenk},
  \citenamefont {Isobe}, \citenamefont {Minola}, \citenamefont {Wu},
  \citenamefont {Suyolcu}, \citenamefont {van Aken}, \citenamefont {Keimer},\
  and\ \citenamefont {Hepting}}]{Hayashida2024}%
  \BibitemOpen
  \bibfield  {author} {\bibinfo {author} {\bibfnamefont {S.}~\bibnamefont
  {Hayashida}}, \bibinfo {author} {\bibfnamefont {V.}~\bibnamefont
  {Sundaramurthy}}, \bibinfo {author} {\bibfnamefont {P.}~\bibnamefont
  {Puphal}}, \bibinfo {author} {\bibfnamefont {M.}~\bibnamefont
  {Garcia-Fernandez}}, \bibinfo {author} {\bibfnamefont {K.-J.}\ \bibnamefont
  {Zhou}}, \bibinfo {author} {\bibfnamefont {B.}~\bibnamefont {Fenk}}, \bibinfo
  {author} {\bibfnamefont {M.}~\bibnamefont {Isobe}}, \bibinfo {author}
  {\bibfnamefont {M.}~\bibnamefont {Minola}}, \bibinfo {author} {\bibfnamefont
  {Y.-M.}\ \bibnamefont {Wu}}, \bibinfo {author} {\bibfnamefont {Y.~E.}\
  \bibnamefont {Suyolcu}}, \bibinfo {author} {\bibfnamefont {P.~A.}\
  \bibnamefont {van Aken}}, \bibinfo {author} {\bibfnamefont {B.}~\bibnamefont
  {Keimer}},\ and\ \bibinfo {author} {\bibfnamefont {M.}~\bibnamefont
  {Hepting}},\ }\bibfield  {title} {\bibinfo {title} {{Investigation of spin
  excitations and charge order in bulk crystals of the infinite-layer nickelate
  ${\mathrm{LaNiO}}_{2}$}},\ }\href
  {https://doi.org/10.1103/PhysRevB.109.235106} {\bibfield  {journal} {\bibinfo
   {journal} {Phys. Rev. B}\ }\textbf {\bibinfo {volume} {109}},\ \bibinfo
  {pages} {235106} (\bibinfo {year} {2024})}\BibitemShut {NoStop}%
\bibitem [{SM()}]{SM}%
  \BibitemOpen
  \href@noop {} {}\bibinfo {note} {See Supplemental Material for the
  complementary inelastic neutron scattering spectra of LaNiO$_{2}$ crystals.
  The Supplemental Material also contains
  Refs.~\cite{Puphal2023,Wu2024,Hayashida2024,EuphonicWeb,EuphonicFair,LuScience2021,Xu2013,InternationalTable,togo2023implementation,PhysRevB.72.035105,PhysRevB.73.045112}.}\BibitemShut
  {Stop}%
\bibitem [{\citenamefont {F{\aa}k}\ \emph {et~al.}(2022)\citenamefont
  {F{\aa}k}, \citenamefont {Rols}, \citenamefont {Manzin},\ and\ \citenamefont
  {Meulien}}]{Fak2022}%
  \BibitemOpen
  \bibfield  {author} {\bibinfo {author} {\bibfnamefont {B.}~\bibnamefont
  {F{\aa}k}}, \bibinfo {author} {\bibfnamefont {S.}~\bibnamefont {Rols}},
  \bibinfo {author} {\bibfnamefont {G.}~\bibnamefont {Manzin}},\ and\ \bibinfo
  {author} {\bibfnamefont {O.}~\bibnamefont {Meulien}},\ }\bibfield  {title}
  {\bibinfo {title} {{Panther --- the new thermal neutron time-of-flight
  spectrometer at the ILL}},\ }\href
  {https://doi.org/10.1051/epjconf/202227202001} {\bibfield  {journal}
  {\bibinfo  {journal} {EPJ Web Conf.}\ }\textbf {\bibinfo {volume} {272}},\
  \bibinfo {pages} {7} (\bibinfo {year} {2022})}\BibitemShut {NoStop}%
\bibitem [{\citenamefont {Arnold}\ \emph {et~al.}(2014)\citenamefont {Arnold},
  \citenamefont {Bilheux}, \citenamefont {Borreguero}, \citenamefont {Buts},
  \citenamefont {Campbell}, \citenamefont {Chapon}, \citenamefont {Doucet},
  \citenamefont {Draper}, \citenamefont {{Ferraz Leal}}, \citenamefont {Gigg},
  \citenamefont {Lynch}, \citenamefont {Markvardsen}, \citenamefont
  {Mikkelson}, \citenamefont {Mikkelson}, \citenamefont {Miller}, \citenamefont
  {Palmen}, \citenamefont {Parker}, \citenamefont {Passos}, \citenamefont
  {Perring}, \citenamefont {Peterson}, \citenamefont {Ren}, \citenamefont
  {Reuter}, \citenamefont {Savici}, \citenamefont {Taylor}, \citenamefont
  {Taylor}, \citenamefont {Tolchenov}, \citenamefont {Zhou},\ and\
  \citenamefont {Zikovsky}}]{Mantid}%
  \BibitemOpen
  \bibfield  {author} {\bibinfo {author} {\bibfnamefont {O.}~\bibnamefont
  {Arnold}}, \bibinfo {author} {\bibfnamefont {J.}~\bibnamefont {Bilheux}},
  \bibinfo {author} {\bibfnamefont {J.}~\bibnamefont {Borreguero}}, \bibinfo
  {author} {\bibfnamefont {A.}~\bibnamefont {Buts}}, \bibinfo {author}
  {\bibfnamefont {S.}~\bibnamefont {Campbell}}, \bibinfo {author}
  {\bibfnamefont {L.}~\bibnamefont {Chapon}}, \bibinfo {author} {\bibfnamefont
  {M.}~\bibnamefont {Doucet}}, \bibinfo {author} {\bibfnamefont
  {N.}~\bibnamefont {Draper}}, \bibinfo {author} {\bibfnamefont
  {R.}~\bibnamefont {{Ferraz Leal}}}, \bibinfo {author} {\bibfnamefont
  {M.}~\bibnamefont {Gigg}}, \bibinfo {author} {\bibfnamefont {V.}~\bibnamefont
  {Lynch}}, \bibinfo {author} {\bibfnamefont {A.}~\bibnamefont {Markvardsen}},
  \bibinfo {author} {\bibfnamefont {D.}~\bibnamefont {Mikkelson}}, \bibinfo
  {author} {\bibfnamefont {R.}~\bibnamefont {Mikkelson}}, \bibinfo {author}
  {\bibfnamefont {R.}~\bibnamefont {Miller}}, \bibinfo {author} {\bibfnamefont
  {K.}~\bibnamefont {Palmen}}, \bibinfo {author} {\bibfnamefont
  {P.}~\bibnamefont {Parker}}, \bibinfo {author} {\bibfnamefont
  {G.}~\bibnamefont {Passos}}, \bibinfo {author} {\bibfnamefont
  {T.}~\bibnamefont {Perring}}, \bibinfo {author} {\bibfnamefont
  {P.}~\bibnamefont {Peterson}}, \bibinfo {author} {\bibfnamefont
  {S.}~\bibnamefont {Ren}}, \bibinfo {author} {\bibfnamefont {M.}~\bibnamefont
  {Reuter}}, \bibinfo {author} {\bibfnamefont {A.}~\bibnamefont {Savici}},
  \bibinfo {author} {\bibfnamefont {J.}~\bibnamefont {Taylor}}, \bibinfo
  {author} {\bibfnamefont {R.}~\bibnamefont {Taylor}}, \bibinfo {author}
  {\bibfnamefont {R.}~\bibnamefont {Tolchenov}}, \bibinfo {author}
  {\bibfnamefont {W.}~\bibnamefont {Zhou}},\ and\ \bibinfo {author}
  {\bibfnamefont {J.}~\bibnamefont {Zikovsky}},\ }\bibfield  {title} {\bibinfo
  {title} {{Mantid—Data analysis and visualization package for neutron
  scattering and $\mu$SR experiments}},\ }\href
  {https://doi.org/https://doi.org/10.1016/j.nima.2014.07.029} {\bibfield
  {journal} {\bibinfo  {journal} {Nucl. Instrum. Methods Phys. Res. A}\
  }\textbf {\bibinfo {volume} {764}},\ \bibinfo {pages} {156} (\bibinfo {year}
  {2014})}\BibitemShut {NoStop}%
\bibitem [{\citenamefont {Ewings}\ \emph {et~al.}(2016)\citenamefont {Ewings},
  \citenamefont {Buts}, \citenamefont {Le}, \citenamefont {{van Duijn}},
  \citenamefont {Bustinduy},\ and\ \citenamefont {Perring}}]{Horace}%
  \BibitemOpen
  \bibfield  {author} {\bibinfo {author} {\bibfnamefont {R.}~\bibnamefont
  {Ewings}}, \bibinfo {author} {\bibfnamefont {A.}~\bibnamefont {Buts}},
  \bibinfo {author} {\bibfnamefont {M.}~\bibnamefont {Le}}, \bibinfo {author}
  {\bibfnamefont {J.}~\bibnamefont {{van Duijn}}}, \bibinfo {author}
  {\bibfnamefont {I.}~\bibnamefont {Bustinduy}},\ and\ \bibinfo {author}
  {\bibfnamefont {T.}~\bibnamefont {Perring}},\ }\bibfield  {title} {\bibinfo
  {title} {{Horace: Software for the analysis of data from single crystal
  spectroscopy experiments at time-of-flight neutron instruments}},\ }\href
  {https://doi.org/https://doi.org/10.1016/j.nima.2016.07.036} {\bibfield
  {journal} {\bibinfo  {journal} {Nucl. Instrum. Methods Phys. Res. A}\
  }\textbf {\bibinfo {volume} {834}},\ \bibinfo {pages} {132} (\bibinfo {year}
  {2016})}\BibitemShut {NoStop}%
\bibitem [{\citenamefont {Hohenberg}\ and\ \citenamefont
  {Kohn}(1964)}]{PhysRev.136.B864}%
  \BibitemOpen
  \bibfield  {author} {\bibinfo {author} {\bibfnamefont {P.}~\bibnamefont
  {Hohenberg}}\ and\ \bibinfo {author} {\bibfnamefont {W.}~\bibnamefont
  {Kohn}},\ }\bibfield  {title} {\bibinfo {title} {Inhomogeneous electron
  gas},\ }\href {http://link.aps.org/doi/10.1103/PhysRev.136.B864} {\bibfield
  {journal} {\bibinfo  {journal} {Phys. Rev.}\ }\textbf {\bibinfo {volume}
  {136}},\ \bibinfo {pages} {B864} (\bibinfo {year} {1964})}\BibitemShut
  {NoStop}%
\bibitem [{\citenamefont {Kohn}\ and\ \citenamefont
  {Sham}(1965)}]{PhysRev.140.A1133}%
  \BibitemOpen
  \bibfield  {author} {\bibinfo {author} {\bibfnamefont {W.}~\bibnamefont
  {Kohn}}\ and\ \bibinfo {author} {\bibfnamefont {L.~J.}\ \bibnamefont
  {Sham}},\ }\bibfield  {title} {\bibinfo {title} {Self-consistent equations
  including exchange and correlation effects},\ }\href
  {http://link.aps.org/doi/10.1103/PhysRev.140.A1133} {\bibfield  {journal}
  {\bibinfo  {journal} {Phys. Rev.}\ }\textbf {\bibinfo {volume} {140}},\
  \bibinfo {pages} {A1133} (\bibinfo {year} {1965})}\BibitemShut {NoStop}%
\bibitem [{\citenamefont {Kresse}\ and\ \citenamefont
  {F{\"u}rthm{\"u}ller}(1996)}]{kresse1996efficiency}%
  \BibitemOpen
  \bibfield  {author} {\bibinfo {author} {\bibfnamefont {G.}~\bibnamefont
  {Kresse}}\ and\ \bibinfo {author} {\bibfnamefont {J.}~\bibnamefont
  {F{\"u}rthm{\"u}ller}},\ }\bibfield  {title} {\bibinfo {title} {Efficiency of
  ab-initio total energy calculations for metals and semiconductors using a
  plane-wave basis set},\ }\href {https://doi.org/10.1016/0927-0256(96)00008-0}
  {\bibfield  {journal} {\bibinfo  {journal} {Comput. Mater. Sci.}\ }\textbf
  {\bibinfo {volume} {6}},\ \bibinfo {pages} {15} (\bibinfo {year}
  {1996})}\BibitemShut {NoStop}%
\bibitem [{\citenamefont {Kresse}\ and\ \citenamefont
  {Furthm\"uller}(1996)}]{PhysRevB.54.11169}%
  \BibitemOpen
  \bibfield  {author} {\bibinfo {author} {\bibfnamefont {G.}~\bibnamefont
  {Kresse}}\ and\ \bibinfo {author} {\bibfnamefont {J.}~\bibnamefont
  {Furthm\"uller}},\ }\bibfield  {title} {\bibinfo {title} {Efficient iterative
  schemes for ab initio total-energy calculations using a plane-wave basis
  set},\ }\href {https://link.aps.org/doi/10.1103/PhysRevB.54.11169} {\bibfield
   {journal} {\bibinfo  {journal} {Phys. Rev. B}\ }\textbf {\bibinfo {volume}
  {54}},\ \bibinfo {pages} {11169} (\bibinfo {year} {1996})}\BibitemShut
  {NoStop}%
\bibitem [{\citenamefont {Kresse}\ and\ \citenamefont
  {Joubert}(1999)}]{PhysRevB.59.1758}%
  \BibitemOpen
  \bibfield  {author} {\bibinfo {author} {\bibfnamefont {G.}~\bibnamefont
  {Kresse}}\ and\ \bibinfo {author} {\bibfnamefont {D.}~\bibnamefont
  {Joubert}},\ }\bibfield  {title} {\bibinfo {title} {From ultrasoft
  pseudopotentials to the projector augmented-wave method},\ }\href
  {https://doi.org/10.1103/PhysRevB.59.1758} {\bibfield  {journal} {\bibinfo
  {journal} {Phys. Rev. B}\ }\textbf {\bibinfo {volume} {59}},\ \bibinfo
  {pages} {1758} (\bibinfo {year} {1999})}\BibitemShut {NoStop}%
\bibitem [{\citenamefont {Perdew}\ \emph {et~al.}(1996)\citenamefont {Perdew},
  \citenamefont {Burke},\ and\ \citenamefont
  {Ernzerhof}}]{PhysRevLett.77.3865}%
  \BibitemOpen
  \bibfield  {author} {\bibinfo {author} {\bibfnamefont {J.~P.}\ \bibnamefont
  {Perdew}}, \bibinfo {author} {\bibfnamefont {K.}~\bibnamefont {Burke}},\ and\
  \bibinfo {author} {\bibfnamefont {M.}~\bibnamefont {Ernzerhof}},\ }\bibfield
  {title} {\bibinfo {title} {Generalized gradient approximation made simple},\
  }\href {https://link.aps.org/doi/10.1103/PhysRevLett.77.3865} {\bibfield
  {journal} {\bibinfo  {journal} {Phys. Rev. Lett.}\ }\textbf {\bibinfo
  {volume} {77}},\ \bibinfo {pages} {3865} (\bibinfo {year}
  {1996})}\BibitemShut {NoStop}%
\bibitem [{\citenamefont {Togo}\ and\ \citenamefont
  {Tanaka}(2015)}]{togo2015first}%
  \BibitemOpen
  \bibfield  {author} {\bibinfo {author} {\bibfnamefont {A.}~\bibnamefont
  {Togo}}\ and\ \bibinfo {author} {\bibfnamefont {I.}~\bibnamefont {Tanaka}},\
  }\bibfield  {title} {\bibinfo {title} {{First principles phonon calculations
  in materials science}},\ }\href
  {https://www.sciencedirect.com/science/article/pii/S1359646215003127}
  {\bibfield  {journal} {\bibinfo  {journal} {Scripta Materialia}\ }\textbf
  {\bibinfo {volume} {108}},\ \bibinfo {pages} {1} (\bibinfo {year}
  {2015})}\BibitemShut {NoStop}%
\bibitem [{\citenamefont {Fair}\ \emph {et~al.}(2025)\citenamefont {Fair},
  \citenamefont {Farmer}, \citenamefont {Jackson}, \citenamefont {King},
  \citenamefont {Le}, \citenamefont {Perring}, \citenamefont {Pettitt},
  \citenamefont {Refson}, \citenamefont {Tucker}, \citenamefont {Voneshen},\
  and\ \citenamefont {Wilkins}}]{EuphonicWeb}%
  \BibitemOpen
  \bibfield  {author} {\bibinfo {author} {\bibfnamefont {R.~L.}\ \bibnamefont
  {Fair}}, \bibinfo {author} {\bibfnamefont {J.~L.}\ \bibnamefont {Farmer}},
  \bibinfo {author} {\bibfnamefont {A.~J.}\ \bibnamefont {Jackson}}, \bibinfo
  {author} {\bibfnamefont {J.~C.}\ \bibnamefont {King}}, \bibinfo {author}
  {\bibfnamefont {M.~D.}\ \bibnamefont {Le}}, \bibinfo {author} {\bibfnamefont
  {T.~G.}\ \bibnamefont {Perring}}, \bibinfo {author} {\bibfnamefont
  {C.}~\bibnamefont {Pettitt}}, \bibinfo {author} {\bibfnamefont
  {K.}~\bibnamefont {Refson}}, \bibinfo {author} {\bibfnamefont {G.~S.}\
  \bibnamefont {Tucker}}, \bibinfo {author} {\bibfnamefont {D.~J.}\
  \bibnamefont {Voneshen}},\ and\ \bibinfo {author} {\bibfnamefont {J.~S.}\
  \bibnamefont {Wilkins}},\ }\href {https://doi.org/10.5286/SOFTWARE/EUPHONIC}
  {\bibinfo {title} {Euphonic}} (\bibinfo {year} {2025})\BibitemShut {NoStop}%
\bibitem [{\citenamefont {Fair}\ \emph {et~al.}(2022)\citenamefont {Fair},
  \citenamefont {Jackson}, \citenamefont {Voneshen}, \citenamefont {Jochym},
  \citenamefont {Le}, \citenamefont {Refson},\ and\ \citenamefont
  {Perring}}]{EuphonicFair}%
  \BibitemOpen
  \bibfield  {author} {\bibinfo {author} {\bibfnamefont {R.}~\bibnamefont
  {Fair}}, \bibinfo {author} {\bibfnamefont {A.}~\bibnamefont {Jackson}},
  \bibinfo {author} {\bibfnamefont {D.}~\bibnamefont {Voneshen}}, \bibinfo
  {author} {\bibfnamefont {D.}~\bibnamefont {Jochym}}, \bibinfo {author}
  {\bibfnamefont {D.}~\bibnamefont {Le}}, \bibinfo {author} {\bibfnamefont
  {K.}~\bibnamefont {Refson}},\ and\ \bibinfo {author} {\bibfnamefont
  {T.}~\bibnamefont {Perring}},\ }\bibfield  {title} {\bibinfo {title} {{{{\it
  Euphonic}: inelastic neutron scattering simulations from force constants and
  visualization tools for phonon properties}}},\ }\href
  {https://doi.org/10.1107/S1600576722009256} {\bibfield  {journal} {\bibinfo
  {journal} {J. Appl. Cryst.}\ }\textbf {\bibinfo {volume} {55}},\ \bibinfo
  {pages} {1689} (\bibinfo {year} {2022})}\BibitemShut {NoStop}%
\bibitem [{\citenamefont {Klenner}\ \emph {et~al.}(1994)\citenamefont
  {Klenner}, \citenamefont {Falter},\ and\ \citenamefont {Chen}}]{klenner94}%
  \BibitemOpen
  \bibfield  {author} {\bibinfo {author} {\bibfnamefont {M.}~\bibnamefont
  {Klenner}}, \bibinfo {author} {\bibfnamefont {C.}~\bibnamefont {Falter}},\
  and\ \bibinfo {author} {\bibfnamefont {Q.}~\bibnamefont {Chen}},\ }\bibfield
  {title} {\bibinfo {title} {Calculated phonon dispersion of infinite-layer
  compounds and the effects of charge fluctuations},\ }\href
  {https://doi.org/10.1007/BF01313348} {\bibfield  {journal} {\bibinfo
  {journal} {Z. Phys. B}\ }\textbf {\bibinfo {volume} {95}},\ \bibinfo {pages}
  {417} (\bibinfo {year} {1994})}\BibitemShut {NoStop}%
\bibitem [{\citenamefont {Falter}\ \emph {et~al.}(1993)\citenamefont {Falter},
  \citenamefont {Klenner},\ and\ \citenamefont {Ludwig}}]{Falter1992}%
  \BibitemOpen
  \bibfield  {author} {\bibinfo {author} {\bibfnamefont {C.}~\bibnamefont
  {Falter}}, \bibinfo {author} {\bibfnamefont {M.}~\bibnamefont {Klenner}},\
  and\ \bibinfo {author} {\bibfnamefont {W.}~\bibnamefont {Ludwig}},\
  }\bibfield  {title} {\bibinfo {title} {{Effect of charge fluctuations on the
  phonon dispersion and electron-phonon interaction in
  ${\mathrm{La}}_{2}$${\mathrm{CuO}}_{4}$}},\ }\href
  {https://doi.org/10.1103/PhysRevB.47.5390} {\bibfield  {journal} {\bibinfo
  {journal} {Phys. Rev. B}\ }\textbf {\bibinfo {volume} {47}},\ \bibinfo
  {pages} {5390} (\bibinfo {year} {1993})}\BibitemShut {NoStop}%
\bibitem [{\citenamefont {Falter}\ \emph {et~al.}(1997)\citenamefont {Falter},
  \citenamefont {Klenner}, \citenamefont {Hoffmann},\ and\ \citenamefont
  {Chen}}]{Falter1997}%
  \BibitemOpen
  \bibfield  {author} {\bibinfo {author} {\bibfnamefont {C.}~\bibnamefont
  {Falter}}, \bibinfo {author} {\bibfnamefont {M.}~\bibnamefont {Klenner}},
  \bibinfo {author} {\bibfnamefont {G.~A.}\ \bibnamefont {Hoffmann}},\ and\
  \bibinfo {author} {\bibfnamefont {Q.}~\bibnamefont {Chen}},\ }\bibfield
  {title} {\bibinfo {title} {{Origin of phonon anomalies in
  ${\mathrm{La}}_{2}$${\mathrm{CuO}}_{4}$}},\ }\href
  {https://doi.org/10.1103/PhysRevB.55.3308} {\bibfield  {journal} {\bibinfo
  {journal} {Phys. Rev. B}\ }\textbf {\bibinfo {volume} {55}},\ \bibinfo
  {pages} {3308} (\bibinfo {year} {1997})}\BibitemShut {NoStop}%
\bibitem [{\citenamefont {Falter}\ \emph {et~al.}(2006)\citenamefont {Falter},
  \citenamefont {Bauer},\ and\ \citenamefont {Schnetg\"oke}}]{Falter2006}%
  \BibitemOpen
  \bibfield  {author} {\bibinfo {author} {\bibfnamefont {C.}~\bibnamefont
  {Falter}}, \bibinfo {author} {\bibfnamefont {T.}~\bibnamefont {Bauer}},\ and\
  \bibinfo {author} {\bibfnamefont {F.}~\bibnamefont {Schnetg\"oke}},\
  }\bibfield  {title} {\bibinfo {title} {{Modeling the electronic state of the
  high-${T}_{c}$ superconductor $\mathrm{LaCuO}$: Phonon dynamics and charge
  response}},\ }\href {https://doi.org/10.1103/PhysRevB.73.224502} {\bibfield
  {journal} {\bibinfo  {journal} {Phys. Rev. B}\ }\textbf {\bibinfo {volume}
  {73}},\ \bibinfo {pages} {224502} (\bibinfo {year} {2006})}\BibitemShut
  {NoStop}%
\bibitem [{\citenamefont {J.~Birgeneau}\ \emph {et~al.}(2006)\citenamefont
  {J.~Birgeneau}, \citenamefont {Stock}, \citenamefont {M.~Tranquada},\ and\
  \citenamefont {Yamada}}]{Birgeneau2006}%
  \BibitemOpen
  \bibfield  {author} {\bibinfo {author} {\bibfnamefont {R.}~\bibnamefont
  {J.~Birgeneau}}, \bibinfo {author} {\bibfnamefont {C.}~\bibnamefont {Stock}},
  \bibinfo {author} {\bibfnamefont {J.}~\bibnamefont {M.~Tranquada}},\ and\
  \bibinfo {author} {\bibfnamefont {K.}~\bibnamefont {Yamada}},\ }\bibfield
  {title} {\bibinfo {title} {Magnetic neutron scattering in hole-doped cuprate
  superconductors},\ }\href {https://doi.org/10.1143/JPSJ.75.111003} {\bibfield
   {journal} {\bibinfo  {journal} {J. Phys. Soc. Jpn.}\ }\textbf {\bibinfo
  {volume} {75}},\ \bibinfo {pages} {111003} (\bibinfo {year}
  {2006})}\BibitemShut {NoStop}%
\bibitem [{\citenamefont {Lu}\ \emph {et~al.}(2021)\citenamefont {Lu},
  \citenamefont {Rossi}, \citenamefont {Nag}, \citenamefont {Osada},
  \citenamefont {Li}, \citenamefont {Lee}, \citenamefont {Wang}, \citenamefont
  {Garcia-Fernandez}, \citenamefont {Agrestini}, \citenamefont {Shen},
  \citenamefont {Been}, \citenamefont {Moritz}, \citenamefont {Devereaux},
  \citenamefont {Zaanen}, \citenamefont {Hwang}, \citenamefont {Zhou},\ and\
  \citenamefont {Lee}}]{LuScience2021}%
  \BibitemOpen
  \bibfield  {author} {\bibinfo {author} {\bibfnamefont {H.}~\bibnamefont
  {Lu}}, \bibinfo {author} {\bibfnamefont {M.}~\bibnamefont {Rossi}}, \bibinfo
  {author} {\bibfnamefont {A.}~\bibnamefont {Nag}}, \bibinfo {author}
  {\bibfnamefont {M.}~\bibnamefont {Osada}}, \bibinfo {author} {\bibfnamefont
  {D.~F.}\ \bibnamefont {Li}}, \bibinfo {author} {\bibfnamefont
  {K.}~\bibnamefont {Lee}}, \bibinfo {author} {\bibfnamefont {B.~Y.}\
  \bibnamefont {Wang}}, \bibinfo {author} {\bibfnamefont {M.}~\bibnamefont
  {Garcia-Fernandez}}, \bibinfo {author} {\bibfnamefont {S.}~\bibnamefont
  {Agrestini}}, \bibinfo {author} {\bibfnamefont {Z.~X.}\ \bibnamefont {Shen}},
  \bibinfo {author} {\bibfnamefont {E.~M.}\ \bibnamefont {Been}}, \bibinfo
  {author} {\bibfnamefont {B.}~\bibnamefont {Moritz}}, \bibinfo {author}
  {\bibfnamefont {T.~P.}\ \bibnamefont {Devereaux}}, \bibinfo {author}
  {\bibfnamefont {J.}~\bibnamefont {Zaanen}}, \bibinfo {author} {\bibfnamefont
  {H.~Y.}\ \bibnamefont {Hwang}}, \bibinfo {author} {\bibfnamefont {K.-J.}\
  \bibnamefont {Zhou}},\ and\ \bibinfo {author} {\bibfnamefont {W.~S.}\
  \bibnamefont {Lee}},\ }\bibfield  {title} {\bibinfo {title} {Magnetic
  excitations in infinite-layer nickelates},\ }\href
  {https://doi.org/10.1126/science.abd7726} {\bibfield  {journal} {\bibinfo
  {journal} {Science}\ }\textbf {\bibinfo {volume} {373}},\ \bibinfo {pages}
  {213} (\bibinfo {year} {2021})}\BibitemShut {NoStop}%
\bibitem [{\citenamefont {Tam}\ \emph {et~al.}(2022)\citenamefont {Tam},
  \citenamefont {Choi}, \citenamefont {Ding}, \citenamefont {Agrestini},
  \citenamefont {Nag}, \citenamefont {Wu}, \citenamefont {Huang}, \citenamefont
  {Luo}, \citenamefont {Gao}, \citenamefont {Garc{\'i}a-Fern{\'a}ndez},
  \citenamefont {Qiao},\ and\ \citenamefont {Zhou}}]{TamNatMat2022}%
  \BibitemOpen
  \bibfield  {author} {\bibinfo {author} {\bibfnamefont {C.~C.}\ \bibnamefont
  {Tam}}, \bibinfo {author} {\bibfnamefont {J.}~\bibnamefont {Choi}}, \bibinfo
  {author} {\bibfnamefont {X.}~\bibnamefont {Ding}}, \bibinfo {author}
  {\bibfnamefont {S.}~\bibnamefont {Agrestini}}, \bibinfo {author}
  {\bibfnamefont {A.}~\bibnamefont {Nag}}, \bibinfo {author} {\bibfnamefont
  {M.}~\bibnamefont {Wu}}, \bibinfo {author} {\bibfnamefont {B.}~\bibnamefont
  {Huang}}, \bibinfo {author} {\bibfnamefont {H.}~\bibnamefont {Luo}}, \bibinfo
  {author} {\bibfnamefont {P.}~\bibnamefont {Gao}}, \bibinfo {author}
  {\bibfnamefont {M.}~\bibnamefont {Garc{\'i}a-Fern{\'a}ndez}}, \bibinfo
  {author} {\bibfnamefont {L.}~\bibnamefont {Qiao}},\ and\ \bibinfo {author}
  {\bibfnamefont {K.-J.}\ \bibnamefont {Zhou}},\ }\bibfield  {title} {\bibinfo
  {title} {{Charge density waves in infinite-layer NdNiO$_2$ nickelates}},\
  }\href {https://doi.org/10.1038/s41563-022-01330-1} {\bibfield  {journal}
  {\bibinfo  {journal} {Nat. Mater.}\ }\textbf {\bibinfo {volume} {21}},\
  \bibinfo {pages} {1116} (\bibinfo {year} {2022})}\BibitemShut {NoStop}%
\bibitem [{\citenamefont {Krieger}\ \emph {et~al.}(2022)\citenamefont
  {Krieger}, \citenamefont {Martinelli}, \citenamefont {Zeng}, \citenamefont
  {Chow}, \citenamefont {Kummer}, \citenamefont {Arpaia}, \citenamefont
  {Moretti~Sala}, \citenamefont {Brookes}, \citenamefont {Ariando},
  \citenamefont {Viart}, \citenamefont {Salluzzo}, \citenamefont
  {Ghiringhelli},\ and\ \citenamefont {Preziosi}}]{KriegerPRL2022}%
  \BibitemOpen
  \bibfield  {author} {\bibinfo {author} {\bibfnamefont {G.}~\bibnamefont
  {Krieger}}, \bibinfo {author} {\bibfnamefont {L.}~\bibnamefont {Martinelli}},
  \bibinfo {author} {\bibfnamefont {S.}~\bibnamefont {Zeng}}, \bibinfo {author}
  {\bibfnamefont {L.~E.}\ \bibnamefont {Chow}}, \bibinfo {author}
  {\bibfnamefont {K.}~\bibnamefont {Kummer}}, \bibinfo {author} {\bibfnamefont
  {R.}~\bibnamefont {Arpaia}}, \bibinfo {author} {\bibfnamefont
  {M.}~\bibnamefont {Moretti~Sala}}, \bibinfo {author} {\bibfnamefont {N.~B.}\
  \bibnamefont {Brookes}}, \bibinfo {author} {\bibfnamefont {A.}~\bibnamefont
  {Ariando}}, \bibinfo {author} {\bibfnamefont {N.}~\bibnamefont {Viart}},
  \bibinfo {author} {\bibfnamefont {M.}~\bibnamefont {Salluzzo}}, \bibinfo
  {author} {\bibfnamefont {G.}~\bibnamefont {Ghiringhelli}},\ and\ \bibinfo
  {author} {\bibfnamefont {D.}~\bibnamefont {Preziosi}},\ }\bibfield  {title}
  {\bibinfo {title} {{Charge and Spin Order Dichotomy in ${\mathrm{NdNiO}}_{2}$
  Driven by the Capping Layer}},\ }\href
  {https://doi.org/10.1103/PhysRevLett.129.027002} {\bibfield  {journal}
  {\bibinfo  {journal} {Phys. Rev. Lett.}\ }\textbf {\bibinfo {volume} {129}},\
  \bibinfo {pages} {027002} (\bibinfo {year} {2022})}\BibitemShut {NoStop}%
\bibitem [{\citenamefont {Zhang}\ \emph {et~al.}(2007)\citenamefont {Zhang},
  \citenamefont {Louie},\ and\ \citenamefont {Cohen}}]{Zhang2007}%
  \BibitemOpen
  \bibfield  {author} {\bibinfo {author} {\bibfnamefont {P.}~\bibnamefont
  {Zhang}}, \bibinfo {author} {\bibfnamefont {S.~G.}\ \bibnamefont {Louie}},\
  and\ \bibinfo {author} {\bibfnamefont {M.~L.}\ \bibnamefont {Cohen}},\
  }\bibfield  {title} {\bibinfo {title} {{Electron-Phonon Renormalization in
  Cuprate Superconductors}},\ }\href
  {https://doi.org/10.1103/PhysRevLett.98.067005} {\bibfield  {journal}
  {\bibinfo  {journal} {Phys. Rev. Lett.}\ }\textbf {\bibinfo {volume} {98}},\
  \bibinfo {pages} {067005} (\bibinfo {year} {2007})}\BibitemShut {NoStop}%
\bibitem [{\citenamefont {Chow}\ \emph {et~al.}(2025)\citenamefont {Chow},
  \citenamefont {Luo},\ and\ \citenamefont {Ariando}}]{Chow2025}%
  \BibitemOpen
  \bibfield  {author} {\bibinfo {author} {\bibfnamefont {S.~L.~E.}\
  \bibnamefont {Chow}}, \bibinfo {author} {\bibfnamefont {Z.}~\bibnamefont
  {Luo}},\ and\ \bibinfo {author} {\bibfnamefont {A.}~\bibnamefont {Ariando}},\
  }\bibfield  {title} {\bibinfo {title} {{Bulk superconductivity near 40 K in
  hole-doped SmNiO$_{2}$ at ambient pressure}},\ }\href
  {https://doi.org/10.1038/s41586-025-08893-4} {\bibfield  {journal} {\bibinfo
  {journal} {Nature}\ } (\bibinfo {year} {2025})}\BibitemShut {NoStop}%
\bibitem [{\citenamefont {Yang}\ \emph {et~al.}(2025)\citenamefont {Yang},
  \citenamefont {Wang}, \citenamefont {Tang}, \citenamefont {Luo},
  \citenamefont {Wu}, \citenamefont {Xu}, \citenamefont {Wang}, \citenamefont
  {Wu}, \citenamefont {Mao}, \citenamefont {Wang}, \citenamefont {Pei},
  \citenamefont {Zhou}, \citenamefont {Dong}, \citenamefont {Feng},
  \citenamefont {Shi}, \citenamefont {Meng}, \citenamefont {Xi}, \citenamefont
  {Pi}, \citenamefont {Lu}, \citenamefont {Okamoto}, \citenamefont {Huang},
  \citenamefont {Huang}, \citenamefont {Huang}, \citenamefont {Wang},
  \citenamefont {Gao}, \citenamefont {Chen},\ and\ \citenamefont
  {Li}}]{Yang2025}%
  \BibitemOpen
  \bibfield  {author} {\bibinfo {author} {\bibfnamefont {M.}~\bibnamefont
  {Yang}}, \bibinfo {author} {\bibfnamefont {H.}~\bibnamefont {Wang}}, \bibinfo
  {author} {\bibfnamefont {J.}~\bibnamefont {Tang}}, \bibinfo {author}
  {\bibfnamefont {J.}~\bibnamefont {Luo}}, \bibinfo {author} {\bibfnamefont
  {X.}~\bibnamefont {Wu}}, \bibinfo {author} {\bibfnamefont {W.}~\bibnamefont
  {Xu}}, \bibinfo {author} {\bibfnamefont {A.}~\bibnamefont {Wang}}, \bibinfo
  {author} {\bibfnamefont {Y.}~\bibnamefont {Wu}}, \bibinfo {author}
  {\bibfnamefont {R.}~\bibnamefont {Mao}}, \bibinfo {author} {\bibfnamefont
  {Z.}~\bibnamefont {Wang}}, \bibinfo {author} {\bibfnamefont {Z.}~\bibnamefont
  {Pei}}, \bibinfo {author} {\bibfnamefont {G.}~\bibnamefont {Zhou}}, \bibinfo
  {author} {\bibfnamefont {Z.}~\bibnamefont {Dong}}, \bibinfo {author}
  {\bibfnamefont {B.}~\bibnamefont {Feng}}, \bibinfo {author} {\bibfnamefont
  {L.}~\bibnamefont {Shi}}, \bibinfo {author} {\bibfnamefont {W.}~\bibnamefont
  {Meng}}, \bibinfo {author} {\bibfnamefont {C.}~\bibnamefont {Xi}}, \bibinfo
  {author} {\bibfnamefont {L.}~\bibnamefont {Pi}}, \bibinfo {author}
  {\bibfnamefont {Q.}~\bibnamefont {Lu}}, \bibinfo {author} {\bibfnamefont
  {J.}~\bibnamefont {Okamoto}}, \bibinfo {author} {\bibfnamefont {H.-Y.}\
  \bibnamefont {Huang}}, \bibinfo {author} {\bibfnamefont {D.-J.}\ \bibnamefont
  {Huang}}, \bibinfo {author} {\bibfnamefont {H.}~\bibnamefont {Huang}},
  \bibinfo {author} {\bibfnamefont {Q.}~\bibnamefont {Wang}}, \bibinfo {author}
  {\bibfnamefont {P.}~\bibnamefont {Gao}}, \bibinfo {author} {\bibfnamefont
  {Z.}~\bibnamefont {Chen}},\ and\ \bibinfo {author} {\bibfnamefont
  {D.}~\bibnamefont {Li}},\ }\href {https://arxiv.org/abs/2503.18346} {\bibinfo
  {title} {{Enhanced Superconductivity and Mixed-dimensional Behaviour in
  Infinite-layer Samarium Nickelate Thin Films}}} (\bibinfo {year} {2025}),\
  \Eprint {https://arxiv.org/abs/2503.18346} {arXiv:2503.18346
  [cond-mat.supr-con]} \BibitemShut {NoStop}%
\bibitem [{\citenamefont {Hayashida}\ \emph {et~al.}()\citenamefont
  {Hayashida}, \citenamefont {F{\AA}K}, \citenamefont {Hepting}, \citenamefont
  {Keimer}, \citenamefont {Keller}, \citenamefont {Krajewska}, \citenamefont
  {Puphal},\ and\ \citenamefont {Sundaramurthy}}]{ILLdata}%
  \BibitemOpen
  \bibfield  {author} {\bibinfo {author} {\bibfnamefont {S.}~\bibnamefont
  {Hayashida}}, \bibinfo {author} {\bibfnamefont {B.}~\bibnamefont {F{\AA}K}},
  \bibinfo {author} {\bibfnamefont {M.}~\bibnamefont {Hepting}}, \bibinfo
  {author} {\bibfnamefont {B.}~\bibnamefont {Keimer}}, \bibinfo {author}
  {\bibfnamefont {T.}~\bibnamefont {Keller}}, \bibinfo {author} {\bibfnamefont
  {A.}~\bibnamefont {Krajewska}}, \bibinfo {author} {\bibfnamefont
  {P.}~\bibnamefont {Puphal}},\ and\ \bibinfo {author} {\bibfnamefont
  {V.}~\bibnamefont {Sundaramurthy}},\ }\bibfield  {title} {\bibinfo {title}
  {{Spin dynamics in single-crystals of the infinite-layer-nickelate
  LaNiO$_2$}},\ }\bibfield  {journal} {\bibinfo  {journal} {Institut
  Laue-Langevin (ILL)}\ }\href {https://doi.org/doi:10.5291/ILL-DATA.4-02-634}
  {doi:10.5291/ILL-DATA.4-02-634}\BibitemShut {NoStop}%
\bibitem [{\citenamefont {Yokoo}\ \emph {et~al.}(2013)\citenamefont {Yokoo},
  \citenamefont {Ohoyama}, \citenamefont {Itoh}, \citenamefont {Suzuki},
  \citenamefont {Iwasa}, \citenamefont {Sato}, \citenamefont {Kira},
  \citenamefont {Sakaguchi}, \citenamefont {Ino}, \citenamefont {Oku},
  \citenamefont {Tomiyasu}, \citenamefont {Matsuura}, \citenamefont {Hiraka},
  \citenamefont {Fujita}, \citenamefont {Kimura}, \citenamefont {Sato},
  \citenamefont {Takeda}, \citenamefont {Kaneko}, \citenamefont {Hino},\ and\
  \citenamefont {Muto}}]{Yokoo2013}%
  \BibitemOpen
  \bibfield  {author} {\bibinfo {author} {\bibfnamefont {T.}~\bibnamefont
  {Yokoo}}, \bibinfo {author} {\bibfnamefont {K.}~\bibnamefont {Ohoyama}},
  \bibinfo {author} {\bibfnamefont {S.}~\bibnamefont {Itoh}}, \bibinfo {author}
  {\bibfnamefont {J.}~\bibnamefont {Suzuki}}, \bibinfo {author} {\bibfnamefont
  {K.}~\bibnamefont {Iwasa}}, \bibinfo {author} {\bibfnamefont {T.~J.}\
  \bibnamefont {Sato}}, \bibinfo {author} {\bibfnamefont {H.}~\bibnamefont
  {Kira}}, \bibinfo {author} {\bibfnamefont {Y.}~\bibnamefont {Sakaguchi}},
  \bibinfo {author} {\bibfnamefont {T.}~\bibnamefont {Ino}}, \bibinfo {author}
  {\bibfnamefont {T.}~\bibnamefont {Oku}}, \bibinfo {author} {\bibfnamefont
  {K.}~\bibnamefont {Tomiyasu}}, \bibinfo {author} {\bibfnamefont
  {M.}~\bibnamefont {Matsuura}}, \bibinfo {author} {\bibfnamefont
  {H.}~\bibnamefont {Hiraka}}, \bibinfo {author} {\bibfnamefont
  {M.}~\bibnamefont {Fujita}}, \bibinfo {author} {\bibfnamefont
  {H.}~\bibnamefont {Kimura}}, \bibinfo {author} {\bibfnamefont
  {T.}~\bibnamefont {Sato}}, \bibinfo {author} {\bibfnamefont {M.}~\bibnamefont
  {Takeda}}, \bibinfo {author} {\bibfnamefont {K.}~\bibnamefont {Kaneko}},
  \bibinfo {author} {\bibfnamefont {M.}~\bibnamefont {Hino}},\ and\ \bibinfo
  {author} {\bibfnamefont {S.}~\bibnamefont {Muto}},\ }\bibfield  {title}
  {\bibinfo {title} {{Newly Proposed Inelastic Neutron Spectrometer POLANO}},\
  }\href {https://doi.org/10.7566/JPSJS.82SA.SA035} {\bibfield  {journal}
  {\bibinfo  {journal} {J. Phys. Soc. Jpn.}\ }\textbf {\bibinfo {volume}
  {82}},\ \bibinfo {pages} {SA035} (\bibinfo {year} {2013})}\BibitemShut
  {NoStop}%
\bibitem [{\citenamefont {Wu}\ \emph {et~al.}(2024)\citenamefont {Wu},
  \citenamefont {Puphal}, \citenamefont {Isobe}, \citenamefont {Keimer},
  \citenamefont {Hepting}, \citenamefont {Suyolcu},\ and\ \citenamefont {van
  Aken}}]{Wu2024}%
  \BibitemOpen
  \bibfield  {author} {\bibinfo {author} {\bibfnamefont {Y.-M.}\ \bibnamefont
  {Wu}}, \bibinfo {author} {\bibfnamefont {P.}~\bibnamefont {Puphal}}, \bibinfo
  {author} {\bibfnamefont {M.}~\bibnamefont {Isobe}}, \bibinfo {author}
  {\bibfnamefont {B.}~\bibnamefont {Keimer}}, \bibinfo {author} {\bibfnamefont
  {M.}~\bibnamefont {Hepting}}, \bibinfo {author} {\bibfnamefont {Y.~E.}\
  \bibnamefont {Suyolcu}},\ and\ \bibinfo {author} {\bibfnamefont {P.~A.}\
  \bibnamefont {van Aken}},\ }\bibfield  {title} {\bibinfo {title} {{Unraveling
  nano-scale effects of topotactic reduction in LaNiO$_2$ crystals}},\ }\href
  {https://doi.org/10.1063/5.0227732} {\bibfield  {journal} {\bibinfo
  {journal} {APL Materials}\ }\textbf {\bibinfo {volume} {12}},\ \bibinfo
  {pages} {091119} (\bibinfo {year} {2024})}\BibitemShut {NoStop}%
\bibitem [{\citenamefont {Xu}\ \emph {et~al.}(2013)\citenamefont {Xu},
  \citenamefont {Xu},\ and\ \citenamefont {Tranquada}}]{Xu2013}%
  \BibitemOpen
  \bibfield  {author} {\bibinfo {author} {\bibfnamefont {G.}~\bibnamefont
  {Xu}}, \bibinfo {author} {\bibfnamefont {Z.}~\bibnamefont {Xu}},\ and\
  \bibinfo {author} {\bibfnamefont {J.~M.}\ \bibnamefont {Tranquada}},\
  }\bibfield  {title} {\bibinfo {title} {Absolute cross-section normalization
  of magnetic neutron scattering data},\ }\href
  {https://doi.org/10.1063/1.4818323} {\bibfield  {journal} {\bibinfo
  {journal} {Rev. Sci. Instrum.}\ }\textbf {\bibinfo {volume} {84}},\ \bibinfo
  {pages} {083906} (\bibinfo {year} {2013})}\BibitemShut {NoStop}%
\bibitem [{\citenamefont {Brown}(2004)}]{InternationalTable}%
  \BibitemOpen
  \bibfield  {author} {\bibinfo {author} {\bibfnamefont {P.~J.}\ \bibnamefont
  {Brown}},\ }\href {https://doi.org/10.1107/97809553602060000103} {\emph
  {\bibinfo {title} {International Tables for Crystallography}}},\ edited by\
  \bibinfo {editor} {\bibfnamefont {E.}~\bibnamefont {Prince}},\ Vol.~\bibinfo
  {volume} {C}\ (\bibinfo  {publisher} {Kluwer Academic Publishers,
  Dordrecht/Boston/London},\ \bibinfo {year} {2004})\ Chap.\ \bibinfo {chapter}
  {4.4}, p.\ \bibinfo {pages} {454}\BibitemShut {NoStop}%
\bibitem [{\citenamefont {Togo}\ \emph {et~al.}(2023)\citenamefont {Togo},
  \citenamefont {Chaput}, \citenamefont {Tadano},\ and\ \citenamefont
  {Tanaka}}]{togo2023implementation}%
  \BibitemOpen
  \bibfield  {author} {\bibinfo {author} {\bibfnamefont {A.}~\bibnamefont
  {Togo}}, \bibinfo {author} {\bibfnamefont {L.}~\bibnamefont {Chaput}},
  \bibinfo {author} {\bibfnamefont {T.}~\bibnamefont {Tadano}},\ and\ \bibinfo
  {author} {\bibfnamefont {I.}~\bibnamefont {Tanaka}},\ }\bibfield  {title}
  {\bibinfo {title} {{Implementation strategies in phonopy and phono3py}},\
  }\href {https://doi.org/10.1088/1361-648X/acd831} {\bibfield  {journal}
  {\bibinfo  {journal} {Journal of Physics: Condensed Matter}\ }\textbf
  {\bibinfo {volume} {35}},\ \bibinfo {pages} {353001} (\bibinfo {year}
  {2023})}\BibitemShut {NoStop}%
\bibitem [{\citenamefont {Wu}\ \emph {et~al.}(2005)\citenamefont {Wu},
  \citenamefont {Vanderbilt},\ and\ \citenamefont
  {Hamann}}]{PhysRevB.72.035105}%
  \BibitemOpen
  \bibfield  {author} {\bibinfo {author} {\bibfnamefont {X.}~\bibnamefont
  {Wu}}, \bibinfo {author} {\bibfnamefont {D.}~\bibnamefont {Vanderbilt}},\
  and\ \bibinfo {author} {\bibfnamefont {D.~R.}\ \bibnamefont {Hamann}},\
  }\bibfield  {title} {\bibinfo {title} {Systematic treatment of displacements,
  strains, and electric fields in density-functional perturbation theory},\
  }\href {https://doi.org/10.1103/PhysRevB.72.035105} {\bibfield  {journal}
  {\bibinfo  {journal} {Phys. Rev. B}\ }\textbf {\bibinfo {volume} {72}},\
  \bibinfo {pages} {035105} (\bibinfo {year} {2005})}\BibitemShut {NoStop}%
\bibitem [{\citenamefont {Gajdo\ifmmode~\check{s}\else \v{s}\fi{}}\ \emph
  {et~al.}(2006)\citenamefont {Gajdo\ifmmode~\check{s}\else \v{s}\fi{}},
  \citenamefont {Hummer}, \citenamefont {Kresse}, \citenamefont
  {Furthm\"uller},\ and\ \citenamefont {Bechstedt}}]{PhysRevB.73.045112}%
  \BibitemOpen
  \bibfield  {author} {\bibinfo {author} {\bibfnamefont {M.}~\bibnamefont
  {Gajdo\ifmmode~\check{s}\else \v{s}\fi{}}}, \bibinfo {author} {\bibfnamefont
  {K.}~\bibnamefont {Hummer}}, \bibinfo {author} {\bibfnamefont
  {G.}~\bibnamefont {Kresse}}, \bibinfo {author} {\bibfnamefont
  {J.}~\bibnamefont {Furthm\"uller}},\ and\ \bibinfo {author} {\bibfnamefont
  {F.}~\bibnamefont {Bechstedt}},\ }\bibfield  {title} {\bibinfo {title}
  {{Linear optical properties in the projector-augmented wave methodology}},\
  }\href {https://doi.org/10.1103/PhysRevB.73.045112} {\bibfield  {journal}
  {\bibinfo  {journal} {Phys. Rev. B}\ }\textbf {\bibinfo {volume} {73}},\
  \bibinfo {pages} {045112} (\bibinfo {year} {2006})}\BibitemShut {NoStop}%
\end{thebibliography}%


\begin{thebibliography}{11}%
\makeatletter
\providecommand \@ifxundefined [1]{%
 \@ifx{#1\undefined}
}%
\providecommand \@ifnum [1]{%
 \ifnum #1\expandafter \@firstoftwo
 \else \expandafter \@secondoftwo
 \fi
}%
\providecommand \@ifx [1]{%
 \ifx #1\expandafter \@firstoftwo
 \else \expandafter \@secondoftwo
 \fi
}%
\providecommand \natexlab [1]{#1}%
\providecommand \enquote  [1]{``#1''}%
\providecommand \bibnamefont  [1]{#1}%
\providecommand \bibfnamefont [1]{#1}%
\providecommand \citenamefont [1]{#1}%
\providecommand \href@noop [0]{\@secondoftwo}%
\providecommand \href [0]{\begingroup \@sanitize@url \@href}%
\providecommand \@href[1]{\@@startlink{#1}\@@href}%
\providecommand \@@href[1]{\endgroup#1\@@endlink}%
\providecommand \@sanitize@url [0]{\catcode `\\12\catcode `\$12\catcode
  `\&12\catcode `\#12\catcode `\^12\catcode `\_12\catcode `\%12\relax}%
\providecommand \@@startlink[1]{}%
\providecommand \@@endlink[0]{}%
\providecommand \url  [0]{\begingroup\@sanitize@url \@url }%
\providecommand \@url [1]{\endgroup\@href {#1}{\urlprefix }}%
\providecommand \urlprefix  [0]{URL }%
\providecommand \Eprint [0]{\href }%
\providecommand \doibase [0]{https://doi.org/}%
\providecommand \selectlanguage [0]{\@gobble}%
\providecommand \bibinfo  [0]{\@secondoftwo}%
\providecommand \bibfield  [0]{\@secondoftwo}%
\providecommand \translation [1]{[#1]}%
\providecommand \BibitemOpen [0]{}%
\providecommand \bibitemStop [0]{}%
\providecommand \bibitemNoStop [0]{.\EOS\space}%
\providecommand \EOS [0]{\spacefactor3000\relax}%
\providecommand \BibitemShut  [1]{\csname bibitem#1\endcsname}%
\let\auto@bib@innerbib\@empty
\bibitem [{\citenamefont {Puphal}\ \emph {et~al.}(2023)\citenamefont {Puphal},
  \citenamefont {Wehinger}, \citenamefont {Nuss}, \citenamefont {K\"uster},
  \citenamefont {Starke}, \citenamefont {Garbarino}, \citenamefont {Keimer},
  \citenamefont {Isobe},\ and\ \citenamefont {Hepting}}]{Puphal2023}%
  \BibitemOpen
  \bibfield  {author} {\bibinfo {author} {\bibfnamefont {P.}~\bibnamefont
  {Puphal}}, \bibinfo {author} {\bibfnamefont {B.}~\bibnamefont {Wehinger}},
  \bibinfo {author} {\bibfnamefont {J.}~\bibnamefont {Nuss}}, \bibinfo {author}
  {\bibfnamefont {K.}~\bibnamefont {K\"uster}}, \bibinfo {author}
  {\bibfnamefont {U.}~\bibnamefont {Starke}}, \bibinfo {author} {\bibfnamefont
  {G.}~\bibnamefont {Garbarino}}, \bibinfo {author} {\bibfnamefont
  {B.}~\bibnamefont {Keimer}}, \bibinfo {author} {\bibfnamefont
  {M.}~\bibnamefont {Isobe}},\ and\ \bibinfo {author} {\bibfnamefont
  {M.}~\bibnamefont {Hepting}},\ }\bibfield  {title} {\bibinfo {title}
  {{Synthesis and physical properties of ${\mathrm{LaNiO}}_{2}$ crystals}},\
  }\href {https://doi.org/10.1103/PhysRevMaterials.7.014804} {\bibfield
  {journal} {\bibinfo  {journal} {Phys. Rev. Mater.}\ }\textbf {\bibinfo
  {volume} {7}},\ \bibinfo {pages} {014804} (\bibinfo {year}
  {2023})}\BibitemShut {NoStop}%
\bibitem [{\citenamefont {Wu}\ \emph {et~al.}(2024)\citenamefont {Wu},
  \citenamefont {Puphal}, \citenamefont {Isobe}, \citenamefont {Keimer},
  \citenamefont {Hepting}, \citenamefont {Suyolcu},\ and\ \citenamefont {van
  Aken}}]{Wu2024}%
  \BibitemOpen
  \bibfield  {author} {\bibinfo {author} {\bibfnamefont {Y.-M.}\ \bibnamefont
  {Wu}}, \bibinfo {author} {\bibfnamefont {P.}~\bibnamefont {Puphal}}, \bibinfo
  {author} {\bibfnamefont {M.}~\bibnamefont {Isobe}}, \bibinfo {author}
  {\bibfnamefont {B.}~\bibnamefont {Keimer}}, \bibinfo {author} {\bibfnamefont
  {M.}~\bibnamefont {Hepting}}, \bibinfo {author} {\bibfnamefont {Y.~E.}\
  \bibnamefont {Suyolcu}},\ and\ \bibinfo {author} {\bibfnamefont {P.~A.}\
  \bibnamefont {van Aken}},\ }\bibfield  {title} {\bibinfo {title} {{Unraveling
  nano-scale effects of topotactic reduction in LaNiO$_2$ crystals}},\ }\href
  {https://doi.org/10.1063/5.0227732} {\bibfield  {journal} {\bibinfo
  {journal} {APL Materials}\ }\textbf {\bibinfo {volume} {12}},\ \bibinfo
  {pages} {091119} (\bibinfo {year} {2024})}\BibitemShut {NoStop}%
\bibitem [{\citenamefont {Fair}\ \emph {et~al.}(2025)\citenamefont {Fair},
  \citenamefont {Farmer}, \citenamefont {Jackson}, \citenamefont {King},
  \citenamefont {Le}, \citenamefont {Perring}, \citenamefont {Pettitt},
  \citenamefont {Refson}, \citenamefont {Tucker}, \citenamefont {Voneshen},\
  and\ \citenamefont {Wilkins}}]{EuphonicWeb}%
  \BibitemOpen
  \bibfield  {author} {\bibinfo {author} {\bibfnamefont {R.~L.}\ \bibnamefont
  {Fair}}, \bibinfo {author} {\bibfnamefont {J.~L.}\ \bibnamefont {Farmer}},
  \bibinfo {author} {\bibfnamefont {A.~J.}\ \bibnamefont {Jackson}}, \bibinfo
  {author} {\bibfnamefont {J.~C.}\ \bibnamefont {King}}, \bibinfo {author}
  {\bibfnamefont {M.~D.}\ \bibnamefont {Le}}, \bibinfo {author} {\bibfnamefont
  {T.~G.}\ \bibnamefont {Perring}}, \bibinfo {author} {\bibfnamefont
  {C.}~\bibnamefont {Pettitt}}, \bibinfo {author} {\bibfnamefont
  {K.}~\bibnamefont {Refson}}, \bibinfo {author} {\bibfnamefont {G.~S.}\
  \bibnamefont {Tucker}}, \bibinfo {author} {\bibfnamefont {D.~J.}\
  \bibnamefont {Voneshen}},\ and\ \bibinfo {author} {\bibfnamefont {J.~S.}\
  \bibnamefont {Wilkins}},\ }\href {https://doi.org/10.5286/SOFTWARE/EUPHONIC}
  {\bibinfo {title} {Euphonic}} (\bibinfo {year} {2025})\BibitemShut {NoStop}%
\bibitem [{\citenamefont {Fair}\ \emph {et~al.}(2022)\citenamefont {Fair},
  \citenamefont {Jackson}, \citenamefont {Voneshen}, \citenamefont {Jochym},
  \citenamefont {Le}, \citenamefont {Refson},\ and\ \citenamefont
  {Perring}}]{EuphonicFair}%
  \BibitemOpen
  \bibfield  {author} {\bibinfo {author} {\bibfnamefont {R.}~\bibnamefont
  {Fair}}, \bibinfo {author} {\bibfnamefont {A.}~\bibnamefont {Jackson}},
  \bibinfo {author} {\bibfnamefont {D.}~\bibnamefont {Voneshen}}, \bibinfo
  {author} {\bibfnamefont {D.}~\bibnamefont {Jochym}}, \bibinfo {author}
  {\bibfnamefont {D.}~\bibnamefont {Le}}, \bibinfo {author} {\bibfnamefont
  {K.}~\bibnamefont {Refson}},\ and\ \bibinfo {author} {\bibfnamefont
  {T.}~\bibnamefont {Perring}},\ }\bibfield  {title} {\bibinfo {title} {{{{\it
  Euphonic}: inelastic neutron scattering simulations from force constants and
  visualization tools for phonon properties}}},\ }\href
  {https://doi.org/10.1107/S1600576722009256} {\bibfield  {journal} {\bibinfo
  {journal} {J. Appl. Cryst.}\ }\textbf {\bibinfo {volume} {55}},\ \bibinfo
  {pages} {1689} (\bibinfo {year} {2022})}\BibitemShut {NoStop}%
\bibitem [{\citenamefont {Lu}\ \emph {et~al.}(2021)\citenamefont {Lu},
  \citenamefont {Rossi}, \citenamefont {Nag}, \citenamefont {Osada},
  \citenamefont {Li}, \citenamefont {Lee}, \citenamefont {Wang}, \citenamefont
  {Garcia-Fernandez}, \citenamefont {Agrestini}, \citenamefont {Shen},
  \citenamefont {Been}, \citenamefont {Moritz}, \citenamefont {Devereaux},
  \citenamefont {Zaanen}, \citenamefont {Hwang}, \citenamefont {Zhou},\ and\
  \citenamefont {Lee}}]{LuScience2021}%
  \BibitemOpen
  \bibfield  {author} {\bibinfo {author} {\bibfnamefont {H.}~\bibnamefont
  {Lu}}, \bibinfo {author} {\bibfnamefont {M.}~\bibnamefont {Rossi}}, \bibinfo
  {author} {\bibfnamefont {A.}~\bibnamefont {Nag}}, \bibinfo {author}
  {\bibfnamefont {M.}~\bibnamefont {Osada}}, \bibinfo {author} {\bibfnamefont
  {D.~F.}\ \bibnamefont {Li}}, \bibinfo {author} {\bibfnamefont
  {K.}~\bibnamefont {Lee}}, \bibinfo {author} {\bibfnamefont {B.~Y.}\
  \bibnamefont {Wang}}, \bibinfo {author} {\bibfnamefont {M.}~\bibnamefont
  {Garcia-Fernandez}}, \bibinfo {author} {\bibfnamefont {S.}~\bibnamefont
  {Agrestini}}, \bibinfo {author} {\bibfnamefont {Z.~X.}\ \bibnamefont {Shen}},
  \bibinfo {author} {\bibfnamefont {E.~M.}\ \bibnamefont {Been}}, \bibinfo
  {author} {\bibfnamefont {B.}~\bibnamefont {Moritz}}, \bibinfo {author}
  {\bibfnamefont {T.~P.}\ \bibnamefont {Devereaux}}, \bibinfo {author}
  {\bibfnamefont {J.}~\bibnamefont {Zaanen}}, \bibinfo {author} {\bibfnamefont
  {H.~Y.}\ \bibnamefont {Hwang}}, \bibinfo {author} {\bibfnamefont {K.-J.}\
  \bibnamefont {Zhou}},\ and\ \bibinfo {author} {\bibfnamefont {W.~S.}\
  \bibnamefont {Lee}},\ }\bibfield  {title} {\bibinfo {title} {Magnetic
  excitations in infinite-layer nickelates},\ }\href
  {https://doi.org/10.1126/science.abd7726} {\bibfield  {journal} {\bibinfo
  {journal} {Science}\ }\textbf {\bibinfo {volume} {373}},\ \bibinfo {pages}
  {213} (\bibinfo {year} {2021})}\BibitemShut {NoStop}%
\bibitem [{\citenamefont {Hayashida}\ \emph {et~al.}(2024)\citenamefont
  {Hayashida}, \citenamefont {Sundaramurthy}, \citenamefont {Puphal},
  \citenamefont {Garcia-Fernandez}, \citenamefont {Zhou}, \citenamefont {Fenk},
  \citenamefont {Isobe}, \citenamefont {Minola}, \citenamefont {Wu},
  \citenamefont {Suyolcu}, \citenamefont {van Aken}, \citenamefont {Keimer},\
  and\ \citenamefont {Hepting}}]{Hayashida2024}%
  \BibitemOpen
  \bibfield  {author} {\bibinfo {author} {\bibfnamefont {S.}~\bibnamefont
  {Hayashida}}, \bibinfo {author} {\bibfnamefont {V.}~\bibnamefont
  {Sundaramurthy}}, \bibinfo {author} {\bibfnamefont {P.}~\bibnamefont
  {Puphal}}, \bibinfo {author} {\bibfnamefont {M.}~\bibnamefont
  {Garcia-Fernandez}}, \bibinfo {author} {\bibfnamefont {K.-J.}\ \bibnamefont
  {Zhou}}, \bibinfo {author} {\bibfnamefont {B.}~\bibnamefont {Fenk}}, \bibinfo
  {author} {\bibfnamefont {M.}~\bibnamefont {Isobe}}, \bibinfo {author}
  {\bibfnamefont {M.}~\bibnamefont {Minola}}, \bibinfo {author} {\bibfnamefont
  {Y.-M.}\ \bibnamefont {Wu}}, \bibinfo {author} {\bibfnamefont {Y.~E.}\
  \bibnamefont {Suyolcu}}, \bibinfo {author} {\bibfnamefont {P.~A.}\
  \bibnamefont {van Aken}}, \bibinfo {author} {\bibfnamefont {B.}~\bibnamefont
  {Keimer}},\ and\ \bibinfo {author} {\bibfnamefont {M.}~\bibnamefont
  {Hepting}},\ }\bibfield  {title} {\bibinfo {title} {{Investigation of spin
  excitations and charge order in bulk crystals of the infinite-layer nickelate
  ${\mathrm{LaNiO}}_{2}$}},\ }\href
  {https://doi.org/10.1103/PhysRevB.109.235106} {\bibfield  {journal} {\bibinfo
   {journal} {Phys. Rev. B}\ }\textbf {\bibinfo {volume} {109}},\ \bibinfo
  {pages} {235106} (\bibinfo {year} {2024})}\BibitemShut {NoStop}%
\bibitem [{\citenamefont {Xu}\ \emph {et~al.}(2013)\citenamefont {Xu},
  \citenamefont {Xu},\ and\ \citenamefont {Tranquada}}]{Xu2013}%
  \BibitemOpen
  \bibfield  {author} {\bibinfo {author} {\bibfnamefont {G.}~\bibnamefont
  {Xu}}, \bibinfo {author} {\bibfnamefont {Z.}~\bibnamefont {Xu}},\ and\
  \bibinfo {author} {\bibfnamefont {J.~M.}\ \bibnamefont {Tranquada}},\
  }\bibfield  {title} {\bibinfo {title} {Absolute cross-section normalization
  of magnetic neutron scattering data},\ }\href
  {https://doi.org/10.1063/1.4818323} {\bibfield  {journal} {\bibinfo
  {journal} {Rev. Sci. Instrum.}\ }\textbf {\bibinfo {volume} {84}},\ \bibinfo
  {pages} {083906} (\bibinfo {year} {2013})}\BibitemShut {NoStop}%
\bibitem [{\citenamefont {Brown}(2004)}]{InternationalTable}%
  \BibitemOpen
  \bibfield  {author} {\bibinfo {author} {\bibfnamefont {P.~J.}\ \bibnamefont
  {Brown}},\ }\href {https://doi.org/10.1107/97809553602060000103} {\emph
  {\bibinfo {title} {International Tables for Crystallography}}},\ edited by\
  \bibinfo {editor} {\bibfnamefont {E.}~\bibnamefont {Prince}},\ Vol.~\bibinfo
  {volume} {C}\ (\bibinfo  {publisher} {Kluwer Academic Publishers,
  Dordrecht/Boston/London},\ \bibinfo {year} {2004})\ Chap.\ \bibinfo {chapter}
  {4.4}, p.\ \bibinfo {pages} {454}\BibitemShut {NoStop}%
\bibitem [{\citenamefont {Togo}\ \emph {et~al.}(2023)\citenamefont {Togo},
  \citenamefont {Chaput}, \citenamefont {Tadano},\ and\ \citenamefont
  {Tanaka}}]{togo2023implementation}%
  \BibitemOpen
  \bibfield  {author} {\bibinfo {author} {\bibfnamefont {A.}~\bibnamefont
  {Togo}}, \bibinfo {author} {\bibfnamefont {L.}~\bibnamefont {Chaput}},
  \bibinfo {author} {\bibfnamefont {T.}~\bibnamefont {Tadano}},\ and\ \bibinfo
  {author} {\bibfnamefont {I.}~\bibnamefont {Tanaka}},\ }\bibfield  {title}
  {\bibinfo {title} {{Implementation strategies in phonopy and phono3py}},\
  }\href {https://doi.org/10.1088/1361-648X/acd831} {\bibfield  {journal}
  {\bibinfo  {journal} {Journal of Physics: Condensed Matter}\ }\textbf
  {\bibinfo {volume} {35}},\ \bibinfo {pages} {353001} (\bibinfo {year}
  {2023})}\BibitemShut {NoStop}%
\bibitem [{\citenamefont {Wu}\ \emph {et~al.}(2005)\citenamefont {Wu},
  \citenamefont {Vanderbilt},\ and\ \citenamefont
  {Hamann}}]{PhysRevB.72.035105}%
  \BibitemOpen
  \bibfield  {author} {\bibinfo {author} {\bibfnamefont {X.}~\bibnamefont
  {Wu}}, \bibinfo {author} {\bibfnamefont {D.}~\bibnamefont {Vanderbilt}},\
  and\ \bibinfo {author} {\bibfnamefont {D.~R.}\ \bibnamefont {Hamann}},\
  }\bibfield  {title} {\bibinfo {title} {Systematic treatment of displacements,
  strains, and electric fields in density-functional perturbation theory},\
  }\href {https://doi.org/10.1103/PhysRevB.72.035105} {\bibfield  {journal}
  {\bibinfo  {journal} {Phys. Rev. B}\ }\textbf {\bibinfo {volume} {72}},\
  \bibinfo {pages} {035105} (\bibinfo {year} {2005})}\BibitemShut {NoStop}%
\bibitem [{\citenamefont {Gajdo\ifmmode~\check{s}\else \v{s}\fi{}}\ \emph
  {et~al.}(2006)\citenamefont {Gajdo\ifmmode~\check{s}\else \v{s}\fi{}},
  \citenamefont {Hummer}, \citenamefont {Kresse}, \citenamefont
  {Furthm\"uller},\ and\ \citenamefont {Bechstedt}}]{PhysRevB.73.045112}%
  \BibitemOpen
  \bibfield  {author} {\bibinfo {author} {\bibfnamefont {M.}~\bibnamefont
  {Gajdo\ifmmode~\check{s}\else \v{s}\fi{}}}, \bibinfo {author} {\bibfnamefont
  {K.}~\bibnamefont {Hummer}}, \bibinfo {author} {\bibfnamefont
  {G.}~\bibnamefont {Kresse}}, \bibinfo {author} {\bibfnamefont
  {J.}~\bibnamefont {Furthm\"uller}},\ and\ \bibinfo {author} {\bibfnamefont
  {F.}~\bibnamefont {Bechstedt}},\ }\bibfield  {title} {\bibinfo {title}
  {{Linear optical properties in the projector-augmented wave methodology}},\
  }\href {https://doi.org/10.1103/PhysRevB.73.045112} {\bibfield  {journal}
  {\bibinfo  {journal} {Phys. Rev. B}\ }\textbf {\bibinfo {volume} {73}},\
  \bibinfo {pages} {045112} (\bibinfo {year} {2006})}\BibitemShut {NoStop}%
\end{thebibliography}%

\end{document}


\title{Supplementary information for ``Lattice dynamics of the infinite-layer nickelate LaNiO$_2$"}

\author{Shohei~Hayashida}
\email[]{s\_hayashida@cross.or.jp}
\affiliation{Neutron Science and Technology Center, Comprehensive Research Organization for Science and Society (CROSS), Tokai, Ibaraki 319-1106, Japan}
\affiliation{Max-Planck-Institute for Solid State Research, Heisenbergstra{\ss}e 1, 70569 Stuttgart, Germany}
\author{Vignesh~Sundaramurthy}
\affiliation{Max-Planck-Institute for Solid State Research, Heisenbergstra{\ss}e 1, 70569 Stuttgart, Germany}
\author{Wenfeng~Wu}
\affiliation{Key Laboratory of Materials Physics, Institute of Solid State Physics, HFIPS, Chinese Academy of Sciences, Hefei 230031, China}
\affiliation{Science Island Branch of Graduate School, University of Science and Technology of China, Hefei 230026, China}
\author{Pascal~Puphal}
\affiliation{Max-Planck-Institute for Solid State Research, Heisenbergstra{\ss}e 1, 70569 Stuttgart, Germany}
\author{Thomas~Keller}
\affiliation{Max-Planck-Institute for Solid State Research, Heisenbergstra{\ss}e 1, 70569 Stuttgart, Germany}
\affiliation{Max Planck Society Outstation at the Heinz Maier-Leibnitz Zentrum (MLZ), Lichtenbergstra{\ss}e 1, 85748 Garching, Germany}
\author{Björn~F{\aa}k}
\affiliation{Institut Laue-Langevin, 71 Avenue des Martyrs, 38042 Grenoble Cedex 9, France}
\author{Masahiko~Isobe}
\affiliation{Max-Planck-Institute for Solid State Research, Heisenbergstra{\ss}e 1, 70569 Stuttgart, Germany}
\author{Bernhard~Keimer}
\affiliation{Max-Planck-Institute for Solid State Research, Heisenbergstra{\ss}e 1, 70569 Stuttgart, Germany}
\author{Karsten~Held}
\affiliation{Institut für Festkörperphysik, Technische Universität Wien, 1040 Wien, Austria}
\author{Liang~Si}
\email[]{si@ifp.tuwien.ac.at}
\affiliation{School of Physics, Northwest University, Xi’an 710127, China}
\affiliation{Institut für Festkörperphysik, Technische Universität Wien, 1040 Wien, Austria}
\author{Matthias~Hepting}
\email[]{hepting@fkf.mpg.de}
\affiliation{Max-Planck-Institute for Solid State Research, Heisenbergstra{\ss}e 1, 70569 Stuttgart, Germany}
\affiliation{Max Planck Society Outstation at the Heinz Maier-Leibnitz Zentrum (MLZ), Lichtenbergstra{\ss}e 1, 85748 Garching, Germany}

\maketitle


\section{Experimental details and complementary neutron scattering data}
Inelastic neutron scattering (INS) measurements on LaNiO$_2$ crystals were conducted at $T=1.5$~K with incident neutron energies of $E_{i}=76$ and 30~meV using the PANTHER spectrometer. The lattice constants determined from the nuclear Bragg peaks $(220)$ and $(002)$ using $E_{i}=19$~meV are: $a = b = 3.90(14)$~{\AA}, and $c = 3.3575(78)$~{\AA} for the tetragonal $P4/mmm$ unit cell of LaNiO$_2$.

In our neutron scattering maps of the $(H,K,0)$ plane [Figs. 2(a) and 3(a) of the main text], the main Bragg peaks exhibit broad intensity profiles along with satellite peaks. These features are comparable to those observed in single crystal x-ray diffraction (XRD) maps from individual LaNiO$_{2}$ crystals in Ref.~\cite{Puphal2023}, which were characterized by a poor crystal mosaicity and the presence of twin domains. Nevertheless, electron microscopic investigations revealed that the phase purity and crystalline quality in LaNiO$_{2}$ crystals was very high on a local scale of  tens of nanometers. Grain boundary-like regions and secondary phase NiO$_{1-x}$ inclusions occur on larger length scales of several micrometers \cite{Wu2024}, resulting in a poor macroscopic mosaicity. 

\begin{figure}[tb]
\includegraphics[scale=1]{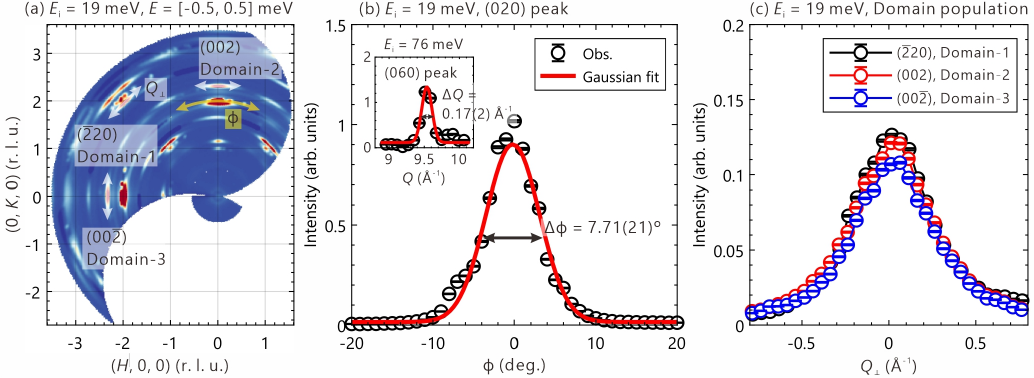}
\caption{(a) Constant energy slice of the INS spectrum in the $(H,K,0)$ plane for incident neutron energy $E_{\rm i}=19$~meV. The transferred energies are integrated over $-0.5 \leq E \leq 0.5$~meV. The integration range along the out-of-plane direction is $\pm 0.37$~{\AA}$^{-1}$. (b) $\phi$ scan around $(020)$ peak, as illustrated by the yellow arrow in (a). The peak is fitted with a Gaussian profile, and the FWHM of 7.71(21)$^{\circ}$ is indicated. The inset shows a scan along the $[010]$ direction around the (060) peak, acquired with $E_{\rm i}=76$~meV, integrated over energy transfers of $-2 \leq E \leq 2$~meV. The integration of the momentum transfer along the orthogonal directions is $\pm0.2$~{\AA}$^{-1}$. (c) Intensity profiles of the $(\bar{2}20)$, $(002)$, and $(00\bar{2})$ Bragg peaks, acquired with $E_{\rm i}=19$~meV, with scan directions indicated by the white arrows in (a). The integration of the momentum transfer along the orthogonal directions is $\pm 16$~{\AA}$^{-1}$. For the $(\bar{2}20)$ peak, the in-plane integration range is restricted to $[-0.16, 0.05]$~{\AA}$^{-1}$ to avoid contributions from different peaks.
}
\label{fig:suppl_Bragg_peaks}
\end{figure}

We quantify the mosaicity of our crystal array via an azimuthal ($\phi$) scan of the $(020)$ nuclear Bragg peak measured with $E_{\rm i}=19$~meV [Fig.~\ref{fig:suppl_Bragg_peaks}(a)]. A Gaussian fit to the peak yields a full width at half maximum (FWHM) of 7.71(21)$^{\circ}$ [Fig.~\ref{fig:suppl_Bragg_peaks}(b)], corresponding to a measure of the mosaicity.
The corresponding $Q$ width perpendicular to the $[010]$ direction around the $(060)$ peak is 1.29~{\AA}$^{-1}$. 
The $Q$ width along the $[010]$ direction around the $(060)$ peak is 0.17~{\AA}$^{-1}$ [see inset of Fig.~\ref{fig:suppl_Bragg_peaks}(b)]. For the Gaussian broadening of the calculated spectrum in Fig.~3(e) of the main text, we use the averaged momentum linewidth, which is 0.92~{\AA}$^{-1}$.


We estimate an equal population of the twin domains in our crystal array from the Bragg peak intensities. At 1.5~K, where the Debye-Waller factor is negligible, the structure factors of the $(220)$ and $(002)$ peaks are expected to be equal for the tetragonal $P4/mmm$ unit cell of LaNiO$_2$.  
Figure~\ref{fig:suppl_Bragg_peaks}(c) shows the measured Bragg peak intensities from each domain. A comparable intensity of the peaks suggests that the three domains are nearly equally populated.

\begin{figure}[tb]
\includegraphics[scale=1]{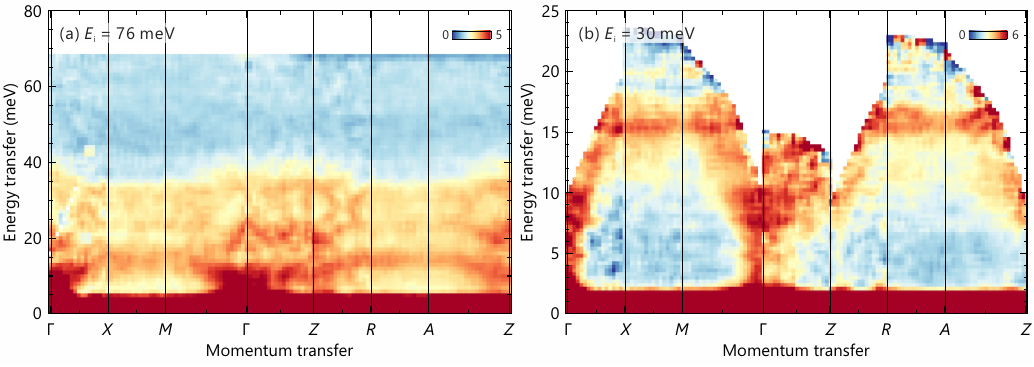}
\caption{INS maps along high-symmetry points from folded data around the (040) Bragg peak, acquired with (a) $E_{\rm i}=76$ and (b) $30$~meV. The momentum integration range in the orthogonal direction is $\pm0.2$~{\AA}$^{-1}$. 
}
\label{fig:suppl_INS_spaghetti}
\end{figure}

Next, we compare INS data from the proximity of different Bragg peaks. For comparison with the INS map in Fig.~3(d) of the main text, which was generated from folded data around $(0 6 0)$ and $(6 0 0)$ [Figs. 3(a) and 3(b) of the main text], we present in Fig.~\ref{fig:suppl_INS_spaghetti}(a) the map resulting from folded data around $(040)$ and $(400)$. While the energy transfer coverage up to high energies is more comprehensive in Fig.~\ref{fig:suppl_INS_spaghetti}(a), the phonon intensity is lower and the individual branches below $\sim$38~meV are less resolved in this map generated from lower-$\textbf{Q}$ data.

For comprehensiveness, we show in Fig.~\ref{fig:suppl_INS_spaghetti}(b) an INS intensity map along $\Gamma$-$X$-$M$-$\Gamma$-$Z$-$R$-$A$-$Z$ acquired with $E_{\rm i}=30$~meV. The map results from folded data around $(040)$. Within the covered low-energy region in this higher-resolution map, acoustic phonon branches emanate from $\Gamma$, consistent with the INS maps acquired with $E_{\rm i}=76$~meV [Fig.~\ref{fig:suppl_INS_spaghetti}(a) and Fig.~3(d) of the main text].

\section{Simulation of the inelastic neutron scattering intensity}

\begin{figure}[tb]
\includegraphics[scale=1]{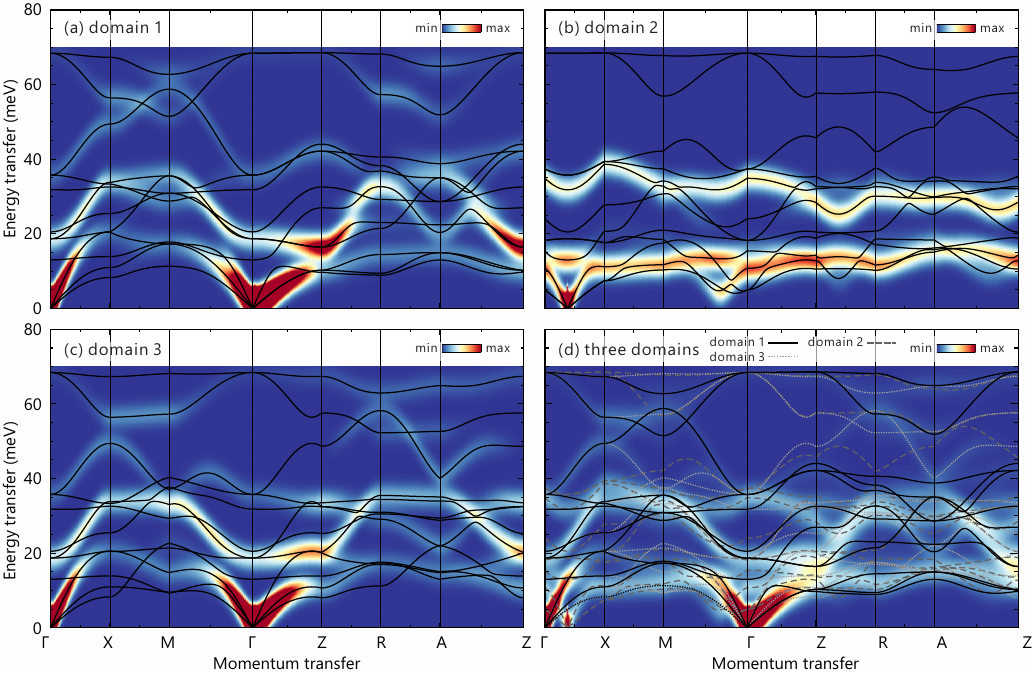}
\caption{Calculated phonon intensities of different twin domains of LaNiO$_{2}$, obtained with the Euphonic software package~\cite{EuphonicWeb,EuphonicFair}, using the DFPT phonons as input. The intensities are convoluted along the energy transfer direction by a Gaussian function representing the experimental energy resolution $\Delta E=4.9$~meV at the elastic line for $E_{\rm i}=76$~meV. (a) Calculated phonon intensities for a high-symmetry path around $(060)$ for domain-1. (b) Calculated phonon intensities for domain-2. The same indexing for the high-symmetry path as for domain-1 is used in the map, although the corresponding reciprocal space coordinates for domain-2 are different (see Table~\ref{tab:BZ_path}). (c) Calculated phonon intensities for domain-3, with the high-symmetry path indexed according to domain-1. (d) Superposition of the calculated phonon intensities with equal weight for all three domains, with the high-symmetry path indexed according to domain-1.
}
\label{fig:suppl_euphonic}
\end{figure}

The Euphonic software package~\cite{EuphonicWeb,EuphonicFair} with the eigenenergies and eigenvectors resulting from the DFPT calculations as an input was used to simulate the phonon intensities in INS. Figures~\ref{fig:suppl_euphonic}(a)-\ref{fig:suppl_euphonic}(c) show the maps of the simulated phonon intensities along the high-symmetry paths around $(060)$ for the three tetragonal twin domains of LaNiO$_{2}$. The labels $\Gamma$-$X$-$M$-$\Gamma$-$Z$-$R$-$A$-$Z$ correspond to the BZ of domain 1 [Fig.~\ref{fig:suppl_euphonic}(a)], while the intensities of domain 2 and 3 are projected onto the same BZ [Figs.~\ref{fig:suppl_euphonic}(b) and \ref{fig:suppl_euphonic}(c)], reflecting the presence of the two other domains in our sample and accounting for the experimental scattering geometry. The corresponding $(H,K,0)$ coordinates for each domain are given in Table~\ref{tab:BZ_path}. The superposition of the simulated INS intensity of the three domains is displayed in Fig.~\ref{fig:suppl_euphonic}(d). The simulation with equal contribution from the three domains corresponds to our assumption of the presence of a large number of domains in our sample, which are in average equally populated.

Gaussian broadening of the calculated phonon intensity maps both along the energy and the momentum transfer directions was applied to match the experimental conditions as follows: The FWHM value used as input for the applied broadening was given by the experimental energy resolution $\Delta E=4.9$~meV, obtained as incoherent scattering at the elastic line for $E_{\rm i}=76$~meV. For the momentum resolution,  we use the averaged momentum linewidth of 0.92~{\AA}$^{-1}$ of the (060) peak (see above).

\begin{table}[htb]
    \centering
    \begin{tabular}{l|l|l|l}
         & Domain-1 & Domain-2 & Domain-3 \\ \hline
       Path  & $(H,K,L)$ & $(H,K,L)$ & $(H,K,L)$ \\ \hline
  $\Gamma$  & $(0,6,0)$ & $(0,0,5.165)$ & $(6,0,0)$ \\
  $X$  & $(0,5.5,0)$ & $(0,0,4.735)$ & $(5.5,0,0)$ \\
  $M$  & $(0.5,5.5,0)$ & $(0.5,0,4.735)$ & $(5.5,0,0.430)$ \\
  $\Gamma$  & $(0,6,0)$ & $(0,0,5.165)$ & $(6,0,0)$ \\
  $Z$  & $(0,6,0.5)$ & $(0,0.581,5.165)$ & $(6,0.581,0)$ \\
  $R$  & $(0,5.5,0.5)$ & $(0,0.581,4.735)$ & $(5.5,0.581,0)$ \\
  $A$  & $(0.5,5.5,0.5)$ & $(0.5,0.581,4.735)$ & $(5.5,0.581,0.430)$ \\
  $Z$  & $(0,6,0.5)$ & $(0,0.581,5.165)$ & $(6,0.581,0)$ \\ \hline
    \end{tabular}
    \caption{Mapping of the $\Gamma$-$X$-$M$-$\Gamma$-$Z$-$R$-$A$-$Z$ high-symmetry path and the $(H,K,L)$ coordinates of domain-1 onto the other two domains. The indexing in the text and the figures refers to domain-1 and the Brillouin zone around $(0, 6, 0)$. Given an orthogonal orientation of domain-2 and domain-3 with respect to domain-1, the $\Gamma$-point of domain-1 with coordinates $(0, 6, 0)$ corresponds to $(0, 0, 5.165)$ in the reference frame of domain-2 and $(6, 0, 0)$ in domain-3. The correspondence of the other high-symmetry points is also given in the table.}
    \label{tab:BZ_path}
\end{table}

\section{Estimation of the spin wave intensity}


Evidence for spin waves was not observed in our INS experiment. Given that prior resonant inelastic x-ray scattering (RIXS) studies have detected paramagnon excitations both in IL nickelate thin films \cite{LuScience2021} and in LaNiO$_{2}$ crystals \cite{Hayashida2024} from the same batch, we estimate in the following whether in principle signatures of magnetic excitations should have been detectable in our INS experiment. To this end, we evaluate the INS cross-section in absolute units (b/meV) from the observed acoustic phonon intensity and compare our observed background signal around $\mathbf{Q}=(1/2,1/2,0)$ (antiferromagnetic zone center) with the spin wave intensity calculated by linear spin wave theory.

The momentum ($\mathbf{Q}$) integrated INS intensity for acoustic phonons at small momenta $\mathbf{q} = \mathbf{Q} - \mathbf{G}$ can be approximated as~\cite{Xu2013}
\begin{equation}
\int I(\mathbf{Q},\omega)d\mathbf{q}
= \frac{1}{d\omega /dq}\frac{n_{q}}{\hbar\omega(q)}\frac{(\hbar\mathbf{Q})^{2}}{2m}
\frac{m}{M}\cos^{2}\beta \left|F_{\rm N}(\mathbf{G})\right|^{2}e^{-2W}
NR_{0},
\label{eq:phonon_normalization}
\end{equation} 
where $d\omega/dq$ is the phonon group velocity, $n_{q}$ the Bose factor, $\hbar\omega(q)$ the phonon energy,  $m$ the neutron mass, $M$ the sum of the masses of the atoms in the LaNiO$_{2}$ unit cell involved in the phonon, $\beta$ the phonon polarization angle, $\mathbf{G}$ the Bragg wave vector near which the acoustic phonon is measured, $F_{\rm N}(\mathbf{G})$ the acoustic phonon structure factor which is the same as the Bragg structure factor at $\mathbf{G}$, $e^{-2W}$ the Debye-Waller factor, and $NR_{0}$ the resolution volume which contains information about the sample and instrument (number of unit cells and instrument resolution).

\begin{figure}[tb]
\includegraphics[scale=1]{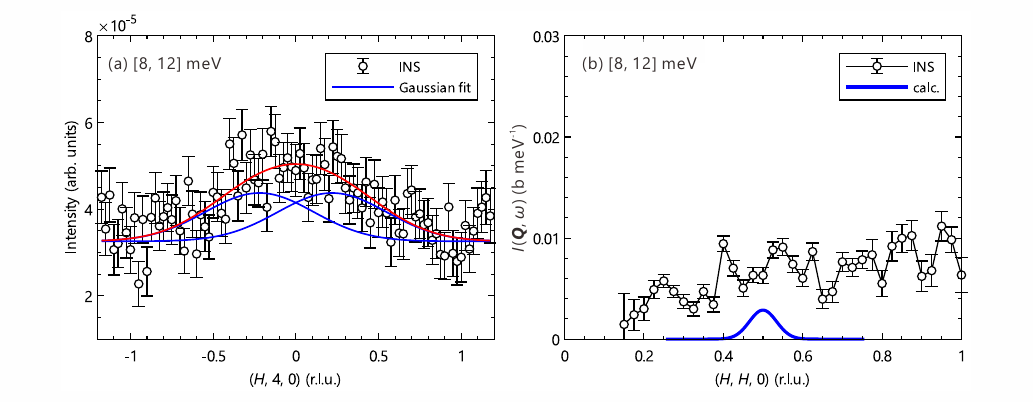}
\caption{Constant-energy cuts ($E=10$~meV) of the INS intensities along (a) the $(H,4,0)$ and (b) the $(H,H,0)$ direction, acquired with $E_{\rm i}=76$~meV. The $Q$ integration width is $\Delta Q=0.4$~{\AA}$^{-1}$, and the energy integration range is $8\leq E \leq 12$~meV. 
The blue solid curves in (a) are Gaussian fits of the acoustic phonon intensities, with the Gaussian profiles symmetric to $H=0$. The red line is the sum of the two Gaussian profiles. The INS intensity in (b) is normalized by the factor $NR_{0}$ (see text). The blue solid curve in (b) is the calculated spin wave intensity (see text), divided by a factor of 3 to account for the three equally populated twin domains in our sample. 
}
\label{fig:suppl_abs_intensity}
\end{figure}

Figure~\ref{fig:suppl_abs_intensity}(a) shows a constant energy scan ($\hbar\omega=10$~meV, integration range $8\leq E \leq 12$~meV) around $\mathbf{Q}=(0,4,0)$ at $T=1.5$~K.
By integrating over the acoustic phonon peak intensity, we obtain 
\begin{equation}
    NR_{0}= 1.3(7)\times 10^{-3}~{\rm meV}/{\rm b}.
\end{equation}
The absolute neutron scattering intensity, $I(\mathbf{Q},\omega)$, in units of ($\mathrm{b}/\mathrm{meV}^{-1}$) is determined by normalizing the measured INS intensity by $NR_{0}$.

On the other hand, the absolute neutron scattering intensity of spin wave excitations is given by 
\begin{equation}
    I(\mathbf{Q},\omega) = \left(\dfrac{\gamma r_{0}}{2}\right)^{2} g^{2}\left|f(\mathbf{Q})\right|^{2}e^{-2W}S(\mathbf{Q},\omega),
\end{equation}
where $\left(\dfrac{\gamma r_{0}}{2}\right)^{2}$ is a constant equal to $7.26\times 10^{-2}$~b, with $\gamma$ and $r_{0}$ being the neutron gyromagnetic ratio and the classical electron radius, respectively.
$g$ is the $g$-factor of the magnetic Ni ion and is equal to 2, $f(\mathbf{Q})$ is the magnetic form factor of the Ni ion taken from Ref.~\cite{InternationalTable}, and $S(\mathbf{Q},\omega)$ is the dynamical structure factor derived from spin-wave excitations in a spin $1/2$ square lattice model based on the linear spin wave theory, as detailed in Ref.~\cite{LuScience2021,Hayashida2024}.

Figure~\ref{fig:suppl_abs_intensity}(b) compares the measured INS intensity with the calculated spin wave intensity---both in absolute units---around $\mathbf{Q}=(1/2,1/2,0)$ where the spin wave intensity in IL nickelates is expected to be strongest~\cite{LuScience2021,Hayashida2024}.
Notably, the calculated spin wave intensity [see blue solid line in Fig.~\ref{fig:suppl_abs_intensity}(b)] lies below the background level of the measured INS intensity, suggesting that such an excitation may be below the detection threshold of our experiment. Moreover, due to the absence of long-range magnetic order and the metallic character of LaNiO$_2$, the spin excitation spectrum is likely to be further broadened and damped compared to the calculated peak profile in Fig.~\ref{fig:suppl_abs_intensity}(b), which may further hinder its observation.


\section{Complementary phonon calculations}

\begin{figure}[tb]
\includegraphics[scale=0.32]{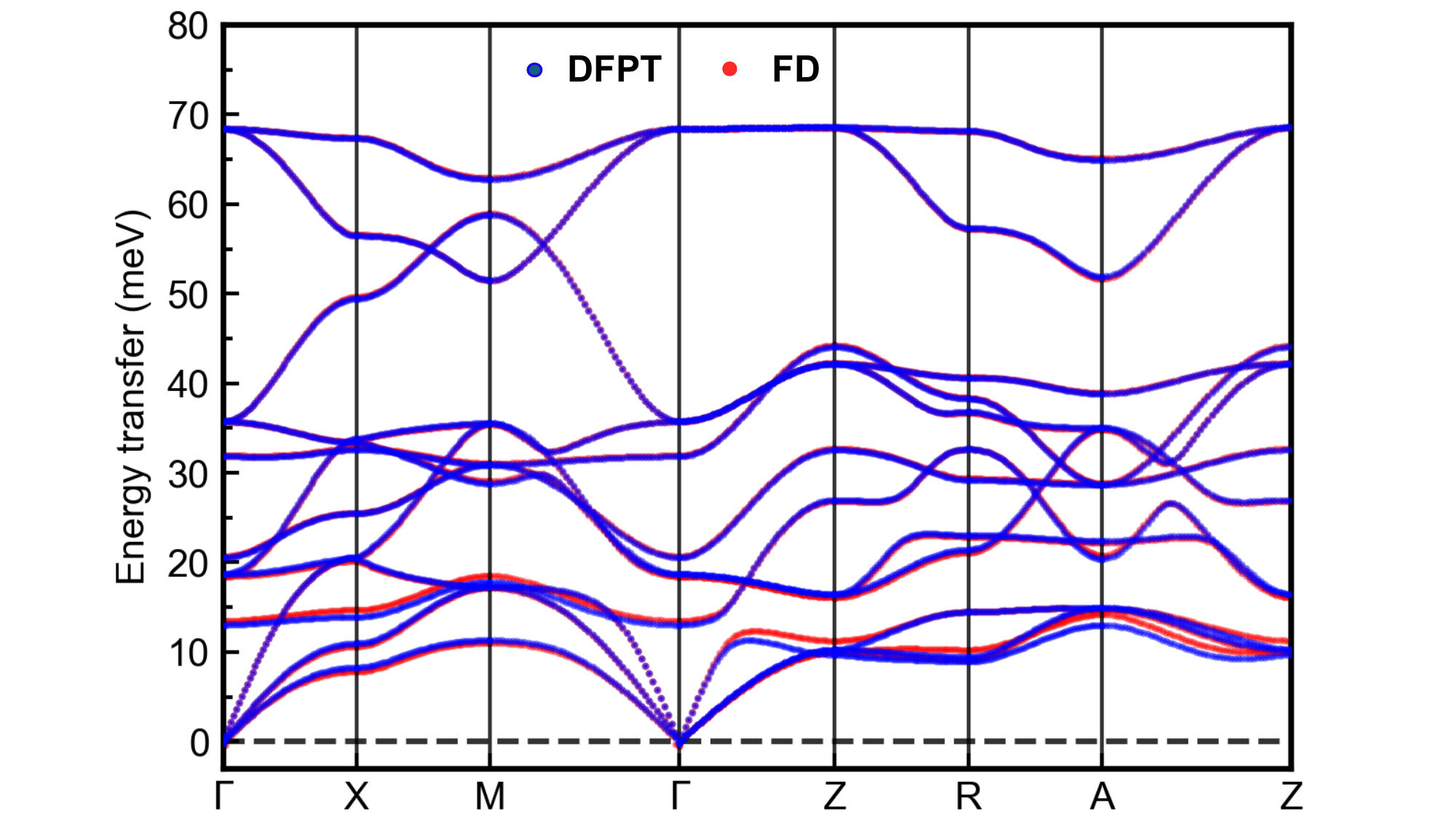}
\caption{Phonon dispersion of LaNiO$_{2}$ calculated using the DFPT (blue) and FD (red) methods.}
\label{fig:suppl_comparison}
\end{figure}

In this study, we computed the phonon dispersion of bulk LaNiO$_2$ using the finite displacement (FD) method \cite{togo2023implementation} and compared the results with those obtained from density functional perturbation theory (DFPT) \cite{PhysRevB.72.035105,PhysRevB.73.045112}. The FD calculations were performed on a 2$\times$2$\times$2 supercell with a 6$\times$6$\times$8 $k$-mesh.
As illustrated in Fig.~\ref{fig:suppl_comparison}, both methods yield nearly identical phonon spectra above 20\,meV. However, discrepancies emerge around 15\,meV, corresponding to long-wavelength phonon modes. These modes are particularly sensitive to long-range interatomic interactions. DFPT captures these effects by explicitly accounting for the linear response of the electronic system to atomic displacements, thereby incorporating effects such as electronic screening and electron-phonon coupling with higher accuracy. In contrast, the FD approach, constrained by the finite supercell size, folds long-wavelength phonons into shorter wavevectors, introducing artificial hardening and limiting its ability to accurately describe long-range dynamical interactions. Given these considerations, we employed DFPT for subsequent calculations due to its superior treatment of long-range effects and improved numerical accuracy.

\bibliography{LaNiO2_INS}

\newpage